\documentclass[traditabstract,longauth]{aa}
\pdfoutput=1
\usepackage{txfonts}
\usepackage{graphicx}
\usepackage[breaklinks, colorlinks, citecolor=blue]{hyperref}
\usepackage{fixltx2e}
\usepackage{natbib}
\usepackage[flushleft]{threeparttable}
\bibpunct{(}{)}{;}{a}{}{,} 

\def\setsymbol#1#2{\expandafter\def\csname #1\endcsname{#2}}
\def\getsymbol#1{\csname #1\endcsname}

\def\Planck{\textit{Planck}}

\newbox\tablebox    \newdimen\tablewidth
\def\leaderfil{\leaders\hbox to 5pt{\hss.\hss}\hfil}
\def\endPlancktable{\tablewidth=\columnwidth 
    $$\hss\copy\tablebox\hss$$
    \vskip-\lastskip\vskip -2pt}
\def\endPlancktablewide{\tablewidth=\textwidth 
    $$\hss\copy\tablebox\hss$$
    \vskip-\lastskip\vskip -2pt}
\def\tablenote#1 #2\par{\begingroup \parindent=0.8em
    \abovedisplayshortskip=0pt\belowdisplayshortskip=0pt
    \noindent
    $$\hss\vbox{\hsize\tablewidth \hangindent=\parindent \hangafter=1 \noindent
    \hbox to \parindent{$^#1$\hss}\strut#2\strut\par}\hss$$
    \endgroup}
\def\doubleline{\vskip 3pt\hrule \vskip 1.5pt \hrule \vskip 5pt}

\def\L2{\ifmmode L_2\else $L_2$\fi}

\def\DeltaT{\ifmmode \Delta T\else $\Delta T$\fi}
\def\deltat{\ifmmode \Delta t\else $\Delta t$\fi}
\def\fknee{\ifmmode f_{\rm knee}\else $f_{\rm knee}$\fi}
\def\Fmax{\ifmmode F_{\rm max}\else $F_{\rm max}$\fi}
\def\solar{\ifmmode{\rm M}_{\mathord\odot}\else${\rm M}_{\mathord\odot}$\fi}
\def\Msolar{\ifmmode{\rm M}_{\mathord\odot}\else${\rm M}_{\mathord\odot}$\fi}
\def\Lsolar{\ifmmode{\rm L}_{\mathord\odot}\else${\rm L}_{\mathord\odot}$\fi}

\def\inv{\ifmmode^{-1}\else$^{-1}$\fi}
\def\mo{\ifmmode^{-1}\else$^{-1}$\fi}
\def\sup#1{\ifmmode ^{\rm #1}\else $^{\rm #1}$\fi}
\def\expo#1{\ifmmode \times 10^{#1}\else $\times 10^{#1}$\fi}
\def\,{\thinspace}
\def\lsim{\mathrel{\raise .4ex\hbox{\rlap{$<$}\lower 1.2ex\hbox{$\sim$}}}}
\def\gsim{\mathrel{\raise .4ex\hbox{\rlap{$>$}\lower 1.2ex\hbox{$\sim$}}}}

\def\simprop{\mathrel{\raise .4ex\hbox{\rlap{$\propto$}\lower 1.2ex\hbox{$\sim$}}}}
\def\deg{\ifmmode^\circ\else$^\circ$\fi}
\def\pdeg{\ifmmode $\setbox0=\hbox{$^{\circ}$}\rlap{\hskip.11\wd0 .}$^{\circ}
          \else \setbox0=\hbox{$^{\circ}$}\rlap{\hskip.11\wd0 .}$^{\circ}$\fi}
\def\arcs{\ifmmode {^{\scriptstyle\prime\prime}}
          \else $^{\scriptstyle\prime\prime}$\fi}
\def\arcm{\ifmmode {^{\scriptstyle\prime}}
          \else $^{\scriptstyle\prime}$\fi}
\newdimen\sa  \newdimen\sb
\def\parcs{\sa=.07em \sb=.03em
     \ifmmode \hbox{\rlap{.}}^{\scriptstyle\prime\kern -\sb\prime}\hbox{\kern -\sa}
     \else \rlap{.}$^{\scriptstyle\prime\kern -\sb\prime}$\kern -\sa\fi}
\def\parcm{\sa=.08em \sb=.03em
     \ifmmode \hbox{\rlap{.}\kern\sa}^{\scriptstyle\prime}\hbox{\kern-\sb}
     \else \rlap{.}\kern\sa$^{\scriptstyle\prime}$\kern-\sb\fi}
\def\ra[#1 #2 #3.#4]{#1\sup{h}#2\sup{m}#3\sup{s}\llap.#4}
\def\dec[#1 #2 #3.#4]{#1\deg#2\arcm#3\arcs\llap.#4}
\def\deco[#1 #2 #3]{#1\deg#2\arcm#3\arcs}
\def\rra[#1 #2]{#1\sup{h}#2\sup{m}}

\def\dots{\relax\ifmmode \ldots\else $\ldots$\fi}
\def\WHzsr{\ifmmode $W\,Hz\mo\,sr\mo$\else W\,Hz\mo\,sr\mo\fi}
\def\mHz{\ifmmode $\,mHz$\else \,mHz\fi}
\def\GHz{\ifmmode $\,GHz$\else \,GHz\fi}
\def\mKs{\ifmmode $\,mK\,s$^{1/2}\else \,mK\,s$^{1/2}$\fi}
\def\muKs{\ifmmode \,\mu$K\,s$^{1/2}\else \,$\mu$K\,s$^{1/2}$\fi}
\def\muKRJs{\ifmmode \,\mu$K$_{\rm RJ}$\,s$^{1/2}\else \,$\mu$K$_{\rm RJ}$\,s$^{1/2}$\fi}
\def\muKHz{\ifmmode \,\mu$K\,Hz$^{-1/2}\else \,$\mu$K\,Hz$^{-1/2}$\fi}
\def\MJysr{\ifmmode \,$MJy\,sr\mo$\else \,MJy\,sr\mo\fi}
\def\MJysrmK{\ifmmode \,$MJy\,sr\mo$\,mK$_{\rm CMB}\mo\else \,MJy\,sr\mo\,mK$_{\rm CMB}\mo$\fi}
\def\microns{\ifmmode \,\mu$m$\else \,$\mu$m\fi}
\def\micron{\microns}
\def\muK{\ifmmode \,\mu$K$\else \,$\mu$\hbox{K}\fi}
\def\microK{\ifmmode \,\mu$K$\else \,$\mu$\hbox{K}\fi}
\def\muW{\ifmmode \,\mu$W$\else \,$\mu$\hbox{W}\fi}
\def\kms{\ifmmode $\,km\,s$^{-1}\else \,km\,s$^{-1}$\fi}
\def\kmsMpc{\ifmmode $\,\kms\,Mpc\mo$\else \,\kms\,Mpc\mo\fi}
\setsymbol{LFI:center:frequency:70GHz:units}{70.3\,GHz}
\setsymbol{LFI:center:frequency:44GHz:units}{44.1\,GHz}
\setsymbol{LFI:center:frequency:30GHz:units}{28.5\,GHz}

\setsymbol{LFI:center:frequency:70GHz}{70.3}
\setsymbol{LFI:center:frequency:44GHz}{44.1}
\setsymbol{LFI:center:frequency:30GHz}{28.5}

\setsymbol{LFI:center:frequency:LFI18:Rad:M:units}{71.7\GHz}
\setsymbol{LFI:center:frequency:LFI19:Rad:M:units}{67.5\GHz}
\setsymbol{LFI:center:frequency:LFI20:Rad:M:units}{69.2\GHz}
\setsymbol{LFI:center:frequency:LFI21:Rad:M:units}{70.4\GHz}
\setsymbol{LFI:center:frequency:LFI22:Rad:M:units}{71.5\GHz}
\setsymbol{LFI:center:frequency:LFI23:Rad:M:units}{70.8\GHz}
\setsymbol{LFI:center:frequency:LFI24:Rad:M:units}{44.4\GHz}
\setsymbol{LFI:center:frequency:LFI25:Rad:M:units}{44.0\GHz}
\setsymbol{LFI:center:frequency:LFI26:Rad:M:units}{43.9\GHz}
\setsymbol{LFI:center:frequency:LFI27:Rad:M:units}{28.3\GHz}
\setsymbol{LFI:center:frequency:LFI28:Rad:M:units}{28.8\GHz}
\setsymbol{LFI:center:frequency:LFI18:Rad:S:units}{70.1\GHz}
\setsymbol{LFI:center:frequency:LFI19:Rad:S:units}{69.6\GHz}
\setsymbol{LFI:center:frequency:LFI20:Rad:S:units}{69.5\GHz}
\setsymbol{LFI:center:frequency:LFI21:Rad:S:units}{69.5\GHz}
\setsymbol{LFI:center:frequency:LFI22:Rad:S:units}{72.8\GHz}
\setsymbol{LFI:center:frequency:LFI23:Rad:S:units}{71.3\GHz}
\setsymbol{LFI:center:frequency:LFI24:Rad:S:units}{44.1\GHz}
\setsymbol{LFI:center:frequency:LFI25:Rad:S:units}{44.1\GHz}
\setsymbol{LFI:center:frequency:LFI26:Rad:S:units}{44.1\GHz}
\setsymbol{LFI:center:frequency:LFI27:Rad:S:units}{28.5\GHz}
\setsymbol{LFI:center:frequency:LFI28:Rad:S:units}{28.2\GHz}

\setsymbol{LFI:center:frequency:LFI18:Rad:M}{71.7}
\setsymbol{LFI:center:frequency:LFI19:Rad:M}{67.5}
\setsymbol{LFI:center:frequency:LFI20:Rad:M}{69.2}
\setsymbol{LFI:center:frequency:LFI21:Rad:M}{70.4}
\setsymbol{LFI:center:frequency:LFI22:Rad:M}{71.5}
\setsymbol{LFI:center:frequency:LFI23:Rad:M}{70.8}
\setsymbol{LFI:center:frequency:LFI24:Rad:M}{44.4}
\setsymbol{LFI:center:frequency:LFI25:Rad:M}{44.0}
\setsymbol{LFI:center:frequency:LFI26:Rad:M}{43.9}
\setsymbol{LFI:center:frequency:LFI27:Rad:M}{28.3}
\setsymbol{LFI:center:frequency:LFI28:Rad:M}{28.8}
\setsymbol{LFI:center:frequency:LFI18:Rad:S}{70.1}
\setsymbol{LFI:center:frequency:LFI19:Rad:S}{69.6}
\setsymbol{LFI:center:frequency:LFI20:Rad:S}{69.5}
\setsymbol{LFI:center:frequency:LFI21:Rad:S}{69.5}
\setsymbol{LFI:center:frequency:LFI22:Rad:S}{72.8}
\setsymbol{LFI:center:frequency:LFI23:Rad:S}{71.3}
\setsymbol{LFI:center:frequency:LFI24:Rad:S}{44.1}
\setsymbol{LFI:center:frequency:LFI25:Rad:S}{44.1}
\setsymbol{LFI:center:frequency:LFI26:Rad:S}{44.1}
\setsymbol{LFI:center:frequency:LFI27:Rad:S}{28.5}
\setsymbol{LFI:center:frequency:LFI28:Rad:S}{28.2}

\setsymbol{LFI:white:noise:sensitivity:70GHz:units}{134.7\muKs}
\setsymbol{LFI:white:noise:sensitivity:44GHz:units}{164.7\muKs}
\setsymbol{LFI:white:noise:sensitivity:30GHz:units}{143.4\muKs}

\setsymbol{LFI:white:noise:sensitivity:70GHz}{134.7}
\setsymbol{LFI:white:noise:sensitivity:44GHz}{164.7}
\setsymbol{LFI:white:noise:sensitivity:30GHz}{143.4}

\setsymbol{LFI:white:noise:sensitivity:LFI18:Rad:M:units}{512.0\muKs}
\setsymbol{LFI:white:noise:sensitivity:LFI19:Rad:M:units}{581.4\muKs}
\setsymbol{LFI:white:noise:sensitivity:LFI20:Rad:M:units}{590.8\muKs}
\setsymbol{LFI:white:noise:sensitivity:LFI21:Rad:M:units}{455.2\muKs}
\setsymbol{LFI:white:noise:sensitivity:LFI22:Rad:M:units}{492.0\muKs}
\setsymbol{LFI:white:noise:sensitivity:LFI23:Rad:M:units}{507.7\muKs}
\setsymbol{LFI:white:noise:sensitivity:LFI24:Rad:M:units}{462.2\muKs}
\setsymbol{LFI:white:noise:sensitivity:LFI25:Rad:M:units}{413.6\muKs}
\setsymbol{LFI:white:noise:sensitivity:LFI26:Rad:M:units}{478.6\muKs}
\setsymbol{LFI:white:noise:sensitivity:LFI27:Rad:M:units}{277.7\muKs}
\setsymbol{LFI:white:noise:sensitivity:LFI28:Rad:M:units}{312.3\muKs}
\setsymbol{LFI:white:noise:sensitivity:LFI18:Rad:S:units}{465.7\muKs}
\setsymbol{LFI:white:noise:sensitivity:LFI19:Rad:S:units}{555.6\muKs}
\setsymbol{LFI:white:noise:sensitivity:LFI20:Rad:S:units}{623.2\muKs}
\setsymbol{LFI:white:noise:sensitivity:LFI21:Rad:S:units}{564.1\muKs}
\setsymbol{LFI:white:noise:sensitivity:LFI22:Rad:S:units}{534.4\muKs}
\setsymbol{LFI:white:noise:sensitivity:LFI23:Rad:S:units}{542.4\muKs}
\setsymbol{LFI:white:noise:sensitivity:LFI24:Rad:S:units}{399.2\muKs}
\setsymbol{LFI:white:noise:sensitivity:LFI25:Rad:S:units}{392.6\muKs}
\setsymbol{LFI:white:noise:sensitivity:LFI26:Rad:S:units}{418.6\muKs}
\setsymbol{LFI:white:noise:sensitivity:LFI27:Rad:S:units}{302.9\muKs}
\setsymbol{LFI:white:noise:sensitivity:LFI28:Rad:S:units}{285.3\muKs}

\setsymbol{LFI:white:noise:sensitivity:LFI18:Rad:M}{512.0}
\setsymbol{LFI:white:noise:sensitivity:LFI19:Rad:M}{581.4}
\setsymbol{LFI:white:noise:sensitivity:LFI20:Rad:M}{590.8}
\setsymbol{LFI:white:noise:sensitivity:LFI21:Rad:M}{455.2}
\setsymbol{LFI:white:noise:sensitivity:LFI22:Rad:M}{492.0}
\setsymbol{LFI:white:noise:sensitivity:LFI23:Rad:M}{507.7}
\setsymbol{LFI:white:noise:sensitivity:LFI24:Rad:M}{462.2}
\setsymbol{LFI:white:noise:sensitivity:LFI25:Rad:M}{413.6}
\setsymbol{LFI:white:noise:sensitivity:LFI26:Rad:M}{478.6}
\setsymbol{LFI:white:noise:sensitivity:LFI27:Rad:M}{277.7}
\setsymbol{LFI:white:noise:sensitivity:LFI28:Rad:M}{312.3}
\setsymbol{LFI:white:noise:sensitivity:LFI18:Rad:S}{465.7}
\setsymbol{LFI:white:noise:sensitivity:LFI19:Rad:S}{555.6}
\setsymbol{LFI:white:noise:sensitivity:LFI20:Rad:S}{623.2}
\setsymbol{LFI:white:noise:sensitivity:LFI21:Rad:S}{564.1}
\setsymbol{LFI:white:noise:sensitivity:LFI22:Rad:S}{534.4}
\setsymbol{LFI:white:noise:sensitivity:LFI23:Rad:S}{542.4}
\setsymbol{LFI:white:noise:sensitivity:LFI24:Rad:S}{399.2}
\setsymbol{LFI:white:noise:sensitivity:LFI25:Rad:S}{392.6}
\setsymbol{LFI:white:noise:sensitivity:LFI26:Rad:S}{418.6}
\setsymbol{LFI:white:noise:sensitivity:LFI27:Rad:S}{302.9}
\setsymbol{LFI:white:noise:sensitivity:LFI28:Rad:S}{285.3}

\setsymbol{LFI:knee:frequency:70GHz:units}{29.5\mHz}
\setsymbol{LFI:knee:frequency:44GHz:units}{56.2\mHz}
\setsymbol{LFI:knee:frequency:30GHz:units}{113.7\mHz}

\setsymbol{LFI:knee:frequency:70GHz}{29.5}
\setsymbol{LFI:knee:frequency:44GHz}{56.2}
\setsymbol{LFI:knee:frequency:30GHz}{113.7}

\setsymbol{LFI:knee:frequency:LFI18:Rad:M:units}{16.3\mHz}
\setsymbol{LFI:knee:frequency:LFI19:Rad:M:units}{15.1\mHz}
\setsymbol{LFI:knee:frequency:LFI20:Rad:M:units}{18.7\mHz}
\setsymbol{LFI:knee:frequency:LFI21:Rad:M:units}{37.2\mHz}
\setsymbol{LFI:knee:frequency:LFI22:Rad:M:units}{12.7\mHz}
\setsymbol{LFI:knee:frequency:LFI23:Rad:M:units}{34.6\mHz}
\setsymbol{LFI:knee:frequency:LFI24:Rad:M:units}{46.2\mHz}
\setsymbol{LFI:knee:frequency:LFI25:Rad:M:units}{24.9\mHz}
\setsymbol{LFI:knee:frequency:LFI26:Rad:M:units}{67.6\mHz}
\setsymbol{LFI:knee:frequency:LFI27:Rad:M:units}{187.4\mHz}
\setsymbol{LFI:knee:frequency:LFI28:Rad:M:units}{122.2\mHz}
\setsymbol{LFI:knee:frequency:LFI18:Rad:S:units}{17.7\mHz}
\setsymbol{LFI:knee:frequency:LFI19:Rad:S:units}{22.0\mHz}
\setsymbol{LFI:knee:frequency:LFI20:Rad:S:units}{8.7\mHz}
\setsymbol{LFI:knee:frequency:LFI21:Rad:S:units}{25.9\mHz}
\setsymbol{LFI:knee:frequency:LFI22:Rad:S:units}{15.8\mHz}
\setsymbol{LFI:knee:frequency:LFI23:Rad:S:units}{129.8\mHz}
\setsymbol{LFI:knee:frequency:LFI24:Rad:S:units}{100.9\mHz}
\setsymbol{LFI:knee:frequency:LFI25:Rad:S:units}{38.9\mHz}
\setsymbol{LFI:knee:frequency:LFI26:Rad:S:units}{58.9\mHz}
\setsymbol{LFI:knee:frequency:LFI27:Rad:S:units}{104.4\mHz}
\setsymbol{LFI:knee:frequency:LFI28:Rad:S:units}{40.7\mHz}

\setsymbol{LFI:knee:frequency:LFI18:Rad:M}{16.3}
\setsymbol{LFI:knee:frequency:LFI19:Rad:M}{15.1}
\setsymbol{LFI:knee:frequency:LFI20:Rad:M}{18.7}
\setsymbol{LFI:knee:frequency:LFI21:Rad:M}{37.2}
\setsymbol{LFI:knee:frequency:LFI22:Rad:M}{12.7}
\setsymbol{LFI:knee:frequency:LFI23:Rad:M}{34.6}
\setsymbol{LFI:knee:frequency:LFI24:Rad:M}{46.2}
\setsymbol{LFI:knee:frequency:LFI25:Rad:M}{24.9}
\setsymbol{LFI:knee:frequency:LFI26:Rad:M}{67.6}
\setsymbol{LFI:knee:frequency:LFI27:Rad:M}{187.4}
\setsymbol{LFI:knee:frequency:LFI28:Rad:M}{122.2}
\setsymbol{LFI:knee:frequency:LFI18:Rad:S}{17.7}
\setsymbol{LFI:knee:frequency:LFI19:Rad:S}{22.0}
\setsymbol{LFI:knee:frequency:LFI20:Rad:S}{8.7}
\setsymbol{LFI:knee:frequency:LFI21:Rad:S}{25.9}
\setsymbol{LFI:knee:frequency:LFI22:Rad:S}{15.8}
\setsymbol{LFI:knee:frequency:LFI23:Rad:S}{129.8}
\setsymbol{LFI:knee:frequency:LFI24:Rad:S}{100.9}
\setsymbol{LFI:knee:frequency:LFI25:Rad:S}{38.9}
\setsymbol{LFI:knee:frequency:LFI26:Rad:S}{58.9}
\setsymbol{LFI:knee:frequency:LFI27:Rad:S}{104.4}
\setsymbol{LFI:knee:frequency:LFI28:Rad:S}{40.7}

\setsymbol{LFI:slope:70GHz:units}{$-1.03$\mHz}
\setsymbol{LFI:slope:44GHz:units}{$-0.89$\mHz}
\setsymbol{LFI:slope:30GHz:units}{$-0.87$\mHz}

\setsymbol{LFI:slope:70GHz}{$-1.03$}
\setsymbol{LFI:slope:44GHz}{$-0.89$}
\setsymbol{LFI:slope:30GHz}{$-0.87$}

\setsymbol{LFI:slope:LFI18:Rad:M:units}{$-1.04$\mHz}
\setsymbol{LFI:slope:LFI19:Rad:M:units}{$-1.09$\mHz}
\setsymbol{LFI:slope:LFI20:Rad:M:units}{$-0.69$\mHz}
\setsymbol{LFI:slope:LFI21:Rad:M:units}{$-1.56$\mHz}
\setsymbol{LFI:slope:LFI22:Rad:M:units}{$-1.01$\mHz}
\setsymbol{LFI:slope:LFI23:Rad:M:units}{$-0.96$\mHz}
\setsymbol{LFI:slope:LFI24:Rad:M:units}{$-0.83$\mHz}
\setsymbol{LFI:slope:LFI25:Rad:M:units}{$-0.91$\mHz}
\setsymbol{LFI:slope:LFI26:Rad:M:units}{$-0.95$\mHz}
\setsymbol{LFI:slope:LFI27:Rad:M:units}{$-0.87$\mHz}
\setsymbol{LFI:slope:LFI28:Rad:M:units}{$-0.88$\mHz}
\setsymbol{LFI:slope:LFI18:Rad:S:units}{$-1.15$\mHz}
\setsymbol{LFI:slope:LFI19:Rad:S:units}{$-1.00$\mHz}
\setsymbol{LFI:slope:LFI20:Rad:S:units}{$-0.95$\mHz}
\setsymbol{LFI:slope:LFI21:Rad:S:units}{$-0.92$\mHz}
\setsymbol{LFI:slope:LFI22:Rad:S:units}{$-1.01$\mHz}
\setsymbol{LFI:slope:LFI23:Rad:S:units}{$-0.95$\mHz}
\setsymbol{LFI:slope:LFI24:Rad:S:units}{$-0.73$\mHz}
\setsymbol{LFI:slope:LFI25:Rad:S:units}{$-1.16$\mHz}
\setsymbol{LFI:slope:LFI26:Rad:S:units}{$-0.79$\mHz}
\setsymbol{LFI:slope:LFI27:Rad:S:units}{$-0.82$\mHz}
\setsymbol{LFI:slope:LFI28:Rad:S:units}{$-0.91$\mHz}

\setsymbol{LFI:slope:LFI18:Rad:M}{$-1.04$}
\setsymbol{LFI:slope:LFI19:Rad:M}{$-1.09$}
\setsymbol{LFI:slope:LFI20:Rad:M}{$-0.69$}
\setsymbol{LFI:slope:LFI21:Rad:M}{$-1.56$}
\setsymbol{LFI:slope:LFI22:Rad:M}{$-1.01$}
\setsymbol{LFI:slope:LFI23:Rad:M}{$-0.96$}
\setsymbol{LFI:slope:LFI24:Rad:M}{$-0.83$}
\setsymbol{LFI:slope:LFI25:Rad:M}{$-0.91$}
\setsymbol{LFI:slope:LFI26:Rad:M}{$-0.95$}
\setsymbol{LFI:slope:LFI27:Rad:M}{$-0.87$}
\setsymbol{LFI:slope:LFI28:Rad:M}{$-0.88$}
\setsymbol{LFI:slope:LFI18:Rad:S}{$-1.15$}
\setsymbol{LFI:slope:LFI19:Rad:S}{$-1.00$}
\setsymbol{LFI:slope:LFI20:Rad:S}{$-0.95$}
\setsymbol{LFI:slope:LFI21:Rad:S}{$-0.92$}
\setsymbol{LFI:slope:LFI22:Rad:S}{$-1.01$}
\setsymbol{LFI:slope:LFI23:Rad:S}{$-0.95$}
\setsymbol{LFI:slope:LFI24:Rad:S}{$-0.73$}
\setsymbol{LFI:slope:LFI25:Rad:S}{$-1.16$}
\setsymbol{LFI:slope:LFI26:Rad:S}{$-0.79$}
\setsymbol{LFI:slope:LFI27:Rad:S}{$-0.82$}
\setsymbol{LFI:slope:LFI28:Rad:S}{$-0.91$}

\setsymbol{LFI:FWHM:70GHz:units}{13\parcm01}
\setsymbol{LFI:FWHM:44GHz:units}{27\parcm92}
\setsymbol{LFI:FWHM:30GHz:units}{32\parcm65}

\setsymbol{LFI:FWHM:70GHz}{13.01}
\setsymbol{LFI:FWHM:44GHz}{27.92}
\setsymbol{LFI:FWHM:30GHz}{32.65}

\setsymbol{LFI:FWHM:LFI18:units}{13\parcm39}
\setsymbol{LFI:FWHM:LFI19:units}{13\parcm01}
\setsymbol{LFI:FWHM:LFI20:units}{12\parcm75}
\setsymbol{LFI:FWHM:LFI21:units}{12\parcm74}
\setsymbol{LFI:FWHM:LFI22:units}{12\parcm87}
\setsymbol{LFI:FWHM:LFI23:units}{13\parcm27}
\setsymbol{LFI:FWHM:LFI24:units}{22\parcm98}
\setsymbol{LFI:FWHM:LFI25:units}{30\parcm46}
\setsymbol{LFI:FWHM:LFI26:units}{30\parcm31}
\setsymbol{LFI:FWHM:LFI27:units}{32\parcm65}
\setsymbol{LFI:FWHM:LFI28:units}{32\parcm66}

\setsymbol{LFI:FWHM:LFI18}{13.39}
\setsymbol{LFI:FWHM:LFI19}{13.01}
\setsymbol{LFI:FWHM:LFI20}{12.75}
\setsymbol{LFI:FWHM:LFI21}{12.74}
\setsymbol{LFI:FWHM:LFI22}{12.87}
\setsymbol{LFI:FWHM:LFI23}{13.27}
\setsymbol{LFI:FWHM:LFI24}{22.98}
\setsymbol{LFI:FWHM:LFI25}{30.46}
\setsymbol{LFI:FWHM:LFI26}{30.31}
\setsymbol{LFI:FWHM:LFI27}{32.65}
\setsymbol{LFI:FWHM:LFI28}{32.66}

\setsymbol{LFI:FWHM:uncertainty:LFI18:units}{0.170\arcm}
\setsymbol{LFI:FWHM:uncertainty:LFI19:units}{0.174\arcm}
\setsymbol{LFI:FWHM:uncertainty:LFI20:units}{0.170\arcm}
\setsymbol{LFI:FWHM:uncertainty:LFI21:units}{0.156\arcm}
\setsymbol{LFI:FWHM:uncertainty:LFI22:units}{0.164\arcm}
\setsymbol{LFI:FWHM:uncertainty:LFI23:units}{0.171\arcm}
\setsymbol{LFI:FWHM:uncertainty:LFI24:units}{0.652\arcm}
\setsymbol{LFI:FWHM:uncertainty:LFI25:units}{1.075\arcm}
\setsymbol{LFI:FWHM:uncertainty:LFI26:units}{1.131\arcm}
\setsymbol{LFI:FWHM:uncertainty:LFI27:units}{1.266\arcm}
\setsymbol{LFI:FWHM:uncertainty:LFI28:units}{1.287\arcm}

\setsymbol{LFI:FWHM:uncertainty:LFI18}{0.170}
\setsymbol{LFI:FWHM:uncertainty:LFI19}{0.174}
\setsymbol{LFI:FWHM:uncertainty:LFI20}{0.170}
\setsymbol{LFI:FWHM:uncertainty:LFI21}{0.156}
\setsymbol{LFI:FWHM:uncertainty:LFI22}{0.164}
\setsymbol{LFI:FWHM:uncertainty:LFI23}{0.171}
\setsymbol{LFI:FWHM:uncertainty:LFI24}{0.652}
\setsymbol{LFI:FWHM:uncertainty:LFI25}{1.075}
\setsymbol{LFI:FWHM:uncertainty:LFI26}{1.131}
\setsymbol{LFI:FWHM:uncertainty:LFI27}{1.266}
\setsymbol{LFI:FWHM:uncertainty:LFI28}{1.287}

\setsymbol{HFI:center:frequency:100GHz:units}{100\,GHz}
\setsymbol{HFI:center:frequency:143GHz:units}{143\,GHz}
\setsymbol{HFI:center:frequency:217GHz:units}{217\,GHz}
\setsymbol{HFI:center:frequency:353GHz:units}{353\,GHz}
\setsymbol{HFI:center:frequency:545GHz:units}{545\,GHz}
\setsymbol{HFI:center:frequency:857GHz:units}{857\,GHz}

\setsymbol{HFI:center:frequency:100GHz}{100}
\setsymbol{HFI:center:frequency:143GHz}{143}
\setsymbol{HFI:center:frequency:217GHz}{217}
\setsymbol{HFI:center:frequency:353GHz}{353}
\setsymbol{HFI:center:frequency:545GHz}{545}
\setsymbol{HFI:center:frequency:857GHz}{857}

\setsymbol{HFI:Ndetectors:100GHz}{8}
\setsymbol{HFI:Ndetectors:143GHz}{11}
\setsymbol{HFI:Ndetectors:217GHz}{12}
\setsymbol{HFI:Ndetectors:353GHz}{12}
\setsymbol{HFI:Ndetectors:545GHz}{3}
\setsymbol{HFI:Ndetectors:857GHz}{4}

\setsymbol{HFI:FWHM:Maps:100GHz:units}{9\parcm88}
\setsymbol{HFI:FWHM:Maps:143GHz:units}{7\parcm18}
\setsymbol{HFI:FWHM:Maps:217GHz:units}{4\parcm87}
\setsymbol{HFI:FWHM:Maps:353GHz:units}{4\parcm65}
\setsymbol{HFI:FWHM:Maps:545GHz:units}{4\parcm72}
\setsymbol{HFI:FWHM:Maps:857GHz:units}{4\parcm39}
\setsymbol{HFI:FWHM:Maps:100GHz}{9.88}
\setsymbol{HFI:FWHM:Maps:143GHz}{7.18}
\setsymbol{HFI:FWHM:Maps:217GHz}{4.87}
\setsymbol{HFI:FWHM:Maps:353GHz}{4.65}
\setsymbol{HFI:FWHM:Maps:545GHz}{4.72}
\setsymbol{HFI:FWHM:Maps:857GHz}{4.39}

\setsymbol{HFI:beam:ellipticity:Maps:100GHz}{1.15}
\setsymbol{HFI:beam:ellipticity:Maps:143GHz}{1.01}
\setsymbol{HFI:beam:ellipticity:Maps:217GHz}{1.06}
\setsymbol{HFI:beam:ellipticity:Maps:353GHz}{1.05}
\setsymbol{HFI:beam:ellipticity:Maps:545GHz}{1.14}
\setsymbol{HFI:beam:ellipticity:Maps:857GHz}{1.19}

\setsymbol{HFI:FWHM:Mars:100GHz:units}{9\parcm37}
\setsymbol{HFI:FWHM:Mars:143GHz:units}{7\parcm04}
\setsymbol{HFI:FWHM:Mars:217GHz:units}{4\parcm68}
\setsymbol{HFI:FWHM:Mars:353GHz:units}{4\parcm43}
\setsymbol{HFI:FWHM:Mars:545GHz:units}{3\parcm80}
\setsymbol{HFI:FWHM:Mars:857GHz:units}{3\parcm67}

\setsymbol{HFI:FWHM:Mars:100GHz}{9.37}
\setsymbol{HFI:FWHM:Mars:143GHz}{7.04}
\setsymbol{HFI:FWHM:Mars:217GHz}{4.68}
\setsymbol{HFI:FWHM:Mars:353GHz}{4.43}
\setsymbol{HFI:FWHM:Mars:545GHz}{3.80}
\setsymbol{HFI:FWHM:Mars:857GHz}{3.67}

\setsymbol{HFI:beam:ellipticity:Mars:100GHz}{1.18}
\setsymbol{HFI:beam:ellipticity:Mars:143GHz}{1.03}
\setsymbol{HFI:beam:ellipticity:Mars:217GHz}{1.14}
\setsymbol{HFI:beam:ellipticity:Mars:353GHz}{1.09}
\setsymbol{HFI:beam:ellipticity:Mars:545GHz}{1.25}
\setsymbol{HFI:beam:ellipticity:Mars:857GHz}{1.03}

\setsymbol{HFI:CMB:relative:calibration:100GHz}{$\lsim 1\%$}
\setsymbol{HFI:CMB:relative:calibration:143GHz}{$\lsim 1\%$}
\setsymbol{HFI:CMB:relative:calibration:217GHz}{$\lsim 1\%$}
\setsymbol{HFI:CMB:relative:calibration:353GHz}{$\lsim 1\%$}
\setsymbol{HFI:CMB:relative:calibration:545GHz}{}
\setsymbol{HFI:CMB:relative:calibration:857GHz}{}

\setsymbol{HFI:CMB:absolute:calibration:100GHz}{$\lsim 2\%$}
\setsymbol{HFI:CMB:absolute:calibration:143GHz}{$\lsim 2\%$}
\setsymbol{HFI:CMB:absolute:calibration:217GHz}{$\lsim 2\%$}
\setsymbol{HFI:CMB:absolute:calibration:353GHz}{$\lsim 2\%$}
\setsymbol{HFI:CMB:absolute:calibration:545GHz}{}
\setsymbol{HFI:CMB:absolute:calibration:857GHz}{}

\setsymbol{HFI:FIRAS:gain:calibration:accuracy:statistical:100GHz}{}
\setsymbol{HFI:FIRAS:gain:calibration:accuracy:statistical:143GHz}{}
\setsymbol{HFI:FIRAS:gain:calibration:accuracy:statistical:217GHz}{}
\setsymbol{HFI:FIRAS:gain:calibration:accuracy:statistical:353GHz}{2.5\%}
\setsymbol{HFI:FIRAS:gain:calibration:accuracy:statistical:545GHz}{1\%}
\setsymbol{HFI:FIRAS:gain:calibration:accuracy:statistical:857GHz}{0.5\%}

\setsymbol{HFI:FIRAS:gain:calibration:accuracy:systematic:100GHz}{}
\setsymbol{HFI:FIRAS:gain:calibration:accuracy:systematic:143GHz}{}
\setsymbol{HFI:FIRAS:gain:calibration:accuracy:systematic:217GHz}{}
\setsymbol{HFI:FIRAS:gain:calibration:accuracy:systematic:353GHz}{}
\setsymbol{HFI:FIRAS:gain:calibration:accuracy:systematic:545GHz}{7\%}
\setsymbol{HFI:FIRAS:gain:calibration:accuracy:systematic:857GHz}{7\%}

\setsymbol{HFI:FIRAS:zero:point:accuracy:100GHz:units}{0.8\MJysr}
\setsymbol{HFI:FIRAS:zero:point:accuracy:143GHz:units}{}
\setsymbol{HFI:FIRAS:zero:point:accuracy:217GHz:units}{}
\setsymbol{HFI:FIRAS:zero:point:accuracy:353GHz:units}{1.4\MJysr}
\setsymbol{HFI:FIRAS:zero:point:accuracy:545GHz:units}{2.2\MJysr}
\setsymbol{HFI:FIRAS:zero:point:accuracy:857GHz:units}{1.7\MJysr}

\setsymbol{HFI:FIRAS:zero:point:accuracy:100GHz}{0.8}
\setsymbol{HFI:FIRAS:zero:point:accuracy:143GHz}{}
\setsymbol{HFI:FIRAS:zero:point:accuracy:217GHz}{}
\setsymbol{HFI:FIRAS:zero:point:accuracy:353GHz}{1.4}
\setsymbol{HFI:FIRAS:zero:point:accuracy:545GHz}{2.2}
\setsymbol{HFI:FIRAS:zero:point:accuracy:857GHz}{1.7}

\setsymbol{HFI:unit:conversion:100GHz:units}{0.2415\MJysrmK}
\setsymbol{HFI:unit:conversion:143GHz:units}{0.3694\MJysrmK}
\setsymbol{HFI:unit:conversion:217GHz:units}{0.4811\MJysrmK}
\setsymbol{HFI:unit:conversion:353GHz:units}{0.2883\MJysrmK}
\setsymbol{HFI:unit:conversion:545GHz:units}{0.05826\MJysrmK}
\setsymbol{HFI:unit:conversion:857GHz:units}{0.002238\MJysrmK}

\setsymbol{HFI:unit:conversion:100GHz}{0.2415}
\setsymbol{HFI:unit:conversion:143GHz}{0.3694}
\setsymbol{HFI:unit:conversion:217GHz}{0.4811}
\setsymbol{HFI:unit:conversion:353GHz}{0.2883}
\setsymbol{HFI:unit:conversion:545GHz}{0.05826}
\setsymbol{HFI:unit:conversion:857GHz}{0.002238}

\setsymbol{HFI:colour:correction:alpha=-2:V1.01:100GHz}{0.9893}
\setsymbol{HFI:colour:correction:alpha=-2:V1.01:143GHz}{0.9759}
\setsymbol{HFI:colour:correction:alpha=-2:V1.01:217GHz}{1.0007}
\setsymbol{HFI:colour:correction:alpha=-2:V1.01:353GHz}{1.0028}
\setsymbol{HFI:colour:correction:alpha=-2:V1.01:545GHz}{1.0019}
\setsymbol{HFI:colour:correction:alpha=-2:V1.01:857GHz}{0.9889}

\setsymbol{HFI:colour:correction:alpha=0:V1.01:100GHz}{1.0008}
\setsymbol{HFI:colour:correction:alpha=0:V1.01:143GHz}{1.0148}
\setsymbol{HFI:colour:correction:alpha=0:V1.01:217GHz}{0.9909}
\setsymbol{HFI:colour:correction:alpha=0:V1.01:353GHz}{0.9888}
\setsymbol{HFI:colour:correction:alpha=0:V1.01:545GHz}{0.9878}
\setsymbol{HFI:colour:correction:alpha=0:V1.01:857GHz}{1.0014}

\newcommand{\snr}{S/N}
\newcommand{\Herschel}{\textit{Herschel}}
\newcommand{\xx}{\vec{x}}
\renewcommand{\micron}{$\mu$m}
\def\erf{\mathop{\rm erf}}
\begin{document}
\author{\small
Planck Collaboration:
P.~A.~R.~Ade\inst{91}
\and
N.~Aghanim\inst{64}
\and
F.~Arg\"{u}eso\inst{21}
\and
C.~Armitage-Caplan\inst{96}
\and
M.~Arnaud\inst{77}
\and
M.~Ashdown\inst{74, 6}
\and
F.~Atrio-Barandela\inst{19}
\and
J.~Aumont\inst{64}
\and
C.~Baccigalupi\inst{90}
\and
A.~J.~Banday\inst{99, 10}
% ...
}
\institute{\small
APC, AstroParticule et Cosmologie, Universit\'{e} Paris Diderot, CNRS/IN2P3, CEA/lrfu, Observatoire de Paris, Sorbonne Paris Cit\'{e}, 10, rue Alice Domon et L\'{e}onie Duquet, 75205 Paris Cedex 13, France\\
\and
Aalto University Mets\"{a}hovi Radio Observatory, Mets\"{a}hovintie 114, FIN-02540 Kylm\"{a}l\"{a}, Finland\\
\and
African Institute for Mathematical Sciences, 6-8 Melrose Road, Muizenberg, Cape Town, South Africa\\
\and
Agenzia Spaziale Italiana Science Data Center, Via del Politecnico snc, 00133, Roma, Italy\\
\and
Agenzia Spaziale Italiana, Viale Liegi 26, Roma, Italy\\
\and
Astrophysics Group, Cavendish Laboratory, University of Cambridge, J J Thomson Avenue, Cambridge CB3 0HE, U.K.\\
\and
Astrophysics \& Cosmology Research Unit, School of Mathematics, Statistics \& Computer Science, University of KwaZulu-Natal, Westville Campus, Private Bag X54001, Durban 4000, South Africa\\
\and
Atacama Large Millimeter/submillimeter Array, ALMA Santiago Central Offices, Alonso de Cordova 3107, Vitacura, Casilla 763 0355, Santiago, Chile\\
\and
CITA, University of Toronto, 60 St. George St., Toronto, ON M5S 3H8, Canada\\
\and
CNRS, IRAP, 9 Av. colonel Roche, BP 44346, F-31028 Toulouse cedex 4, France\\
\and
California Institute of Technology, Pasadena, California, U.S.A.\\
\and
Centre for Theoretical Cosmology, DAMTP, University of Cambridge, Wilberforce Road, Cambridge CB3 0WA, U.K.\\
\and
Centro de Estudios de F\'{i}sica del Cosmos de Arag\'{o}n (CEFCA), Plaza San Juan, 1, planta 2, E-44001, Teruel, Spain\\
\and
Computational Cosmology Center, Lawrence Berkeley National Laboratory, Berkeley, California, U.S.A.\\
\and
Consejo Superior de Investigaciones Cient\'{\i}ficas (CSIC), Madrid, Spain\\
\and
DSM/Irfu/SPP, CEA-Saclay, F-91191 Gif-sur-Yvette Cedex, France\\
\and
DTU Space, National Space Institute, Technical University of Denmark, Elektrovej 327, DK-2800 Kgs. Lyngby, Denmark\\
\and
D\'{e}partement de Physique Th\'{e}orique, Universit\'{e} de Gen\`{e}ve, 24, Quai E. Ansermet,1211 Gen\`{e}ve 4, Switzerland\\
\and
Departamento de F\'{\i}sica Fundamental, Facultad de Ciencias, Universidad de Salamanca, 37008 Salamanca, Spain\\
\and
Departamento de F\'{\i}sica, Universidad de Oviedo, Avda. Calvo Sotelo s/n, Oviedo, Spain\\
\and
Departamento de Matem\'{a}ticas, Universidad de Oviedo, Avda. Calvo Sotelo s/n, Oviedo, Spain\\
\and
Department of Astronomy and Astrophysics, University of Toronto, 50 Saint George Street, Toronto, Ontario, Canada\\
\and
Department of Astrophysics/IMAPP, Radboud University Nijmegen, P.O. Box 9010, 6500 GL Nijmegen, The Netherlands\\
\and
Department of Electrical Engineering and Computer Sciences, University of California, Berkeley, California, U.S.A.\\
\and
Department of Physics \& Astronomy, University of British Columbia, 6224 Agricultural Road, Vancouver, British Columbia, Canada\\
\and
Department of Physics and Astronomy, Dana and David Dornsife College of Letter, Arts and Sciences, University of Southern California, Los Angeles, CA 90089, U.S.A.\\
\and
Department of Physics and Astronomy, University College London, London WC1E 6BT, U.K.\\
\and
Department of Physics, Florida State University, Keen Physics Building, 77 Chieftan Way, Tallahassee, Florida, U.S.A.\\
\and
Department of Physics, Gustaf H\"{a}llstr\"{o}min katu 2a, University of Helsinki, Helsinki, Finland\\
\and
Department of Physics, Princeton University, Princeton, New Jersey, U.S.A.\\
\and
Department of Physics, University of California, One Shields Avenue, Davis, California, U.S.A.\\
\and
Department of Physics, University of California, Santa Barbara, California, U.S.A.\\
\and
Department of Physics, University of Illinois at Urbana-Champaign, 1110 West Green Street, Urbana, Illinois, U.S.A.\\
\and
Dipartimento di Fisica e Astronomia G. Galilei, Universit\`{a} degli Studi di Padova, via Marzolo 8, 35131 Padova, Italy\\
\and
Dipartimento di Fisica e Scienze della Terra, Universit\`{a} di Ferrara, Via Saragat 1, 44122 Ferrara, Italy\\
\and
Dipartimento di Fisica, Universit\`{a} La Sapienza, P. le A. Moro 2, Roma, Italy\\
\and
Dipartimento di Fisica, Universit\`{a} degli Studi di Milano, Via Celoria, 16, Milano, Italy\\
\and
Dipartimento di Fisica, Universit\`{a} degli Studi di Trieste, via A. Valerio 2, Trieste, Italy\\
\and
Dipartimento di Fisica, Universit\`{a} di Roma Tor Vergata, Via della Ricerca Scientifica, 1, Roma, Italy\\
\and
Discovery Center, Niels Bohr Institute, Blegdamsvej 17, Copenhagen, Denmark\\
\and
Dpto. Astrof\'{i}sica, Universidad de La Laguna (ULL), E-38206 La Laguna, Tenerife, Spain\\
\and
European Southern Observatory, ESO Vitacura, Alonso de Cordova 3107, Vitacura, Casilla 19001, Santiago, Chile\\
\and
European Space Agency, ESAC, Planck Science Office, Camino bajo del Castillo, s/n, Urbanizaci\'{o}n Villafranca del Castillo, Villanueva de la Ca\~{n}ada, Madrid, Spain\\
\and
European Space Agency, ESTEC, Keplerlaan 1, 2201 AZ Noordwijk, The Netherlands\\
\and
Finnish Centre for Astronomy with ESO (FINCA), University of Turku, V\"{a}is\"{a}l\"{a}ntie 20, FIN-21500, Piikki\"{o}, Finland\\
\and
Haverford College Astronomy Department, 370 Lancaster Avenue, Haverford, Pennsylvania, U.S.A.\\
\and
Helsinki Institute of Physics, Gustaf H\"{a}llstr\"{o}min katu 2, University of Helsinki, Helsinki, Finland\\
\and
INAF - Osservatorio Astrofisico di Catania, Via S. Sofia 78, Catania, Italy\\
\and
INAF - Osservatorio Astronomico di Padova, Vicolo dell'Osservatorio 5, Padova, Italy\\
\and
INAF - Osservatorio Astronomico di Roma, via di Frascati 33, Monte Porzio Catone, Italy\\
\and
INAF - Osservatorio Astronomico di Trieste, Via G.B. Tiepolo 11, Trieste, Italy\\
\and
INAF Istituto di Radioastronomia, Via P. Gobetti 101, 40129 Bologna, Italy\\
\and
INAF/IASF Bologna, Via Gobetti 101, Bologna, Italy\\
\and
INAF/IASF Milano, Via E. Bassini 15, Milano, Italy\\
\and
INFN, Sezione di Bologna, Via Irnerio 46, I-40126, Bologna, Italy\\
\and
INFN, Sezione di Roma 1, Universit\`{a} di Roma Sapienza, Piazzale Aldo Moro 2, 00185, Roma, Italy\\
\and
IPAG: Institut de Plan\'{e}tologie et d'Astrophysique de Grenoble, Universit\'{e} Joseph Fourier, Grenoble 1 / CNRS-INSU, UMR 5274, Grenoble, F-38041, France\\
\and
ISDC Data Centre for Astrophysics, University of Geneva, ch. d'Ecogia 16, Versoix, Switzerland\\
\and
IUCAA, Post Bag 4, Ganeshkhind, Pune University Campus, Pune 411 007, India\\
\and
Imperial College London, Astrophysics group, Blackett Laboratory, Prince Consort Road, London, SW7 2AZ, U.K.\\
\and
Infrared Processing and Analysis Center, California Institute of Technology, Pasadena, CA 91125, U.S.A.\\
\and
Institut N\'{e}el, CNRS, Universit\'{e} Joseph Fourier Grenoble I, 25 rue des Martyrs, Grenoble, France\\
\and
Institut Universitaire de France, 103, bd Saint-Michel, 75005, Paris, France\\
\and
Institut d'Astrophysique Spatiale, CNRS (UMR8617) Universit\'{e} Paris-Sud 11, B\^{a}timent 121, Orsay, France\\
\and
Institut d'Astrophysique de Paris, CNRS (UMR7095), 98 bis Boulevard Arago, F-75014, Paris, France\\
\and
Institute for Space Sciences, Bucharest-Magurale, Romania\\
\and
Institute of Astronomy and Astrophysics, Academia Sinica, Taipei, Taiwan\\
\and
Institute of Astronomy, University of Cambridge, Madingley Road, Cambridge CB3 0HA, U.K.\\
\and
Institute of Theoretical Astrophysics, University of Oslo, Blindern, Oslo, Norway\\
\and
Instituto de Astrof\'{\i}sica de Canarias, C/V\'{\i}a L\'{a}ctea s/n, La Laguna, Tenerife, Spain\\
\and
Instituto de F\'{\i}sica de Cantabria (CSIC-Universidad de Cantabria), Avda. de los Castros s/n, Santander, Spain\\
\and
Jet Propulsion Laboratory, California Institute of Technology, 4800 Oak Grove Drive, Pasadena, California, U.S.A.\\
\and
Jodrell Bank Centre for Astrophysics, Alan Turing Building, School of Physics and Astronomy, The University of Manchester, Oxford Road, Manchester, M13 9PL, U.K.\\
\and
Kavli Institute for Cosmology Cambridge, Madingley Road, Cambridge, CB3 0HA, U.K.\\
\and
LAL, Universit\'{e} Paris-Sud, CNRS/IN2P3, Orsay, France\\
\and
LERMA, CNRS, Observatoire de Paris, 61 Avenue de l'Observatoire, Paris, France\\
\and
Laboratoire AIM, IRFU/Service d'Astrophysique - CEA/DSM - CNRS - Universit\'{e} Paris Diderot, B\^{a}t. 709, CEA-Saclay, F-91191 Gif-sur-Yvette Cedex, France\\
\and
Laboratoire Traitement et Communication de l'Information, CNRS (UMR 5141) and T\'{e}l\'{e}com ParisTech, 46 rue Barrault F-75634 Paris Cedex 13, France\\
\and
Laboratoire de Physique Subatomique et de Cosmologie, Universit\'{e} Joseph Fourier Grenoble I, CNRS/IN2P3, Institut National Polytechnique de Grenoble, 53 rue des Martyrs, 38026 Grenoble cedex, France\\
\and
Laboratoire de Physique Th\'{e}orique, Universit\'{e} Paris-Sud 11 \& CNRS, B\^{a}timent 210, 91405 Orsay, France\\
\and
Lawrence Berkeley National Laboratory, Berkeley, California, U.S.A.\\
\and
Max-Planck-Institut f\"{u}r Astrophysik, Karl-Schwarzschild-Str. 1, 85741 Garching, Germany\\
\and
McGill Physics, Ernest Rutherford Physics Building, McGill University, 3600 rue University, Montr\'{e}al, QC, H3A 2T8, Canada\\
\and
MilliLab, VTT Technical Research Centre of Finland, Tietotie 3, Espoo, Finland\\
\and
National University of Ireland, Department of Experimental Physics, Maynooth, Co. Kildare, Ireland\\
\and
Niels Bohr Institute, Blegdamsvej 17, Copenhagen, Denmark\\
\and
Observational Cosmology, Mail Stop 367-17, California Institute of Technology, Pasadena, CA, 91125, U.S.A.\\
\and
Optical Science Laboratory, University College London, Gower Street, London, U.K.\\
\and
SB-ITP-LPPC, EPFL, CH-1015, Lausanne, Switzerland\\
\and
SISSA, Astrophysics Sector, via Bonomea 265, 34136, Trieste, Italy\\
\and
School of Physics and Astronomy, Cardiff University, Queens Buildings, The Parade, Cardiff, CF24 3AA, U.K.\\
\and
Space Research Institute (IKI), Russian Academy of Sciences, Profsoyuznaya Str, 84/32, Moscow, 117997, Russia\\
\and
Space Sciences Laboratory, University of California, Berkeley, California, U.S.A.\\
\and
Special Astrophysical Observatory, Russian Academy of Sciences, Nizhnij Arkhyz, Zelenchukskiy region, Karachai-Cherkessian Republic, 369167, Russia\\
\and
Stanford University, Dept of Physics, Varian Physics Bldg, 382 Via Pueblo Mall, Stanford, California, U.S.A.\\
\and
Sub-Department of Astrophysics, University of Oxford, Keble Road, Oxford OX1 3RH, U.K.\\
\and
Theory Division, PH-TH, CERN, CH-1211, Geneva 23, Switzerland\\
\and
UPMC Univ Paris 06, UMR7095, 98 bis Boulevard Arago, F-75014, Paris, France\\
\and
Universit\'{e} de Toulouse, UPS-OMP, IRAP, F-31028 Toulouse cedex 4, France\\
% ...
}

\title{\Planck\ 2013 results. XXVIII. The \Planck\ Catalogue of Compact Sources}
\authorrunning{Planck Collaboration}
\titlerunning{Planck Catalogue of Compact Sources}


\abstract{The \Planck\ Catalogue of Compact Sources (PCCS) is the catalogue of sources detected in the first 15 months of \Planck\ operations, the ``nominal'' mission. It consists of nine single-frequency catalogues of compact sources, both Galactic and extragalactic, detected over the entire sky. The PCCS covers the frequency range 30--857\,GHz with higher sensitivity (it is 90\,\% complete at 180\,mJy in the best channel) and better angular resolution (from $32.88$\arcm\ to $4.33$\arcm) than previous all-sky surveys in this frequency band. By construction its reliability is $>80$\,\% and more than $65$\,\% of the sources have been detected at least in two contiguous \Planck\ channels. 
In this paper we present the construction and validation of the PCCS, its contents and its statistical characterization.}


\maketitle

\section{Introduction}

This paper, one of a set associated with the 2013 release of data from the \Planck\footnote{\Planck\ (\url{http://www.esa.int/Planck}) is a project of the European Space Agency (ESA) with instruments provided by two scientific consortia funded by ESA member states (in particular the lead countries France and Italy), with contributions from NASA (USA) and telescope reflectors provided by a collaboration between ESA and a scientific consortium led and funded by Denmark.} mission~\citep{planck2013-p01}, describes the first release of the \Planck\ Catalogue of Compact Sources (PCCS).

The main goal of the \Planck\ mission is to measure anisotropies in the cosmic microwave background (CMB), the relic radiation of the big bang; this radiation is ``contaminated'' by foreground emission arising from cosmic structures of all scales located between the CMB and us -- galaxies, galaxy clusters, and gas and dust distributed on small as well as large scales within the Milky Way. In order to reveal the rich cosmological information concealed in the CMB such foreground emission must be characterized and separated \citep{planck2013-p06}.
As a by-product, the study of foregrounds delivers an extensive catalogue of discrete compact sources as well as a series of maps of the Galactic diffuse emission; both of these are valuable resources for a variety of studies in the fields of Galactic and extragalactic astrophysics \citep[e.g. ][]{planck2011-7.0, planck2013-XIV, planck2011-1.10, planck2011-7.7b, planck2013-p05a}.

In 1983 the IRAS FIR survey \citep{beichman88} revolutionized astronomy with the discovery of ULIRGS and debris disks, etc and is still relevant to active astrophysical research 30 years after its completion.
\Planck\ is at longer wavelengths but at comparable depths and promises to provide the community with an invaluable data set from the radio to sub-mm for many years to come.

The \Planck\ Early Release Compact Source Catalogue \citep[ERCSC;][]{planck2011-1.10} presented catalogues of discrete sources detected during \Planck's first 1.6 all-sky surveys. It has already been exploited for follow-up observations \citep[e.g.,][]{AMI2012,kurinsky13} and for astrophysical investigations including the first direct determination of the bright end of the extragalactic source counts at frequencies $\ge 100\,$GHz \citep{planck2012-VII}, the study of the spectral properties of radio sources \citep{planck2011-6.1, planck2011-6.2,bonavera11} and of their long-term variability \citep{Chen13}, accurate estimates of the luminosity function of dusty galaxies in the very local Universe (i.e., distances $\le 100$ Mpc) at several millimetre and submillimetre wavelengths \citep{negrello13} and of their dust mass and star-formation rate functions \citep{Clemens13}. Moreover, a $z=3.26$ strongly lensed submillimetre galaxy detected within the $\simeq 135\,\hbox{deg}^2$ of the phase 1 Herschel-ATLAS survey and possibly associated with a proto-cluster of dusty galaxies was found to be associated with an ERCSC source \citep{herranz13,fu12}, highlighting the potential of \Planck\ surveys for detecting extremely high-redshift sources.

This paper presents a new \Planck\ catalogue, the PCCS, which uses deeper observations (from the first 15 months of \Planck\ operations) and better calibration and analysis procedures \citep{planck2013-p02,planck2013-p02a,planck2013-p02b,planck2013-p03,planck2013-p03e,planck2013-p03f} to improve on the results from the ERCSC. The PCCS comprises nine single-frequency source lists, one for each \Planck\ frequency band. It contains high-reliability sources, both Galactic and extragalactic, detected over the entire sky.  The PCCS differs in philosophy from the ERCSC in that it puts more emphasis on the completeness of the catalogue, without greatly reducing the reliability of the detected sources ($>80$\,\% by construction). A comparison of the PCCS and ERCSC results is presented in Sect.~\ref{sec:ercsc}.

This paper describes the construction and content of the PCCS; scientific results from the catalogue will appear in later papers. In Sect.~\ref{sec:pccs} we describe the data, source detection pipelines, selection criteria, and photometry methods used in the production of the PCCS. In Sect.~\ref{sec:validation} we discuss the validation processes (both internal and external) performed to assess the quality of the catalogues. The main characteristics of the PCCS are summarized in Sect.~\ref{sec:characteristics}, and a  description of the content and use of the catalogue is presented in Sect.~\ref{sec:content}. Finally, in Sect.~\ref{sec:conclusions} we summarize our conclusions. Details of the different photometry estimators are described in Appendix \ref{sec:appendix}.

\section{The \Planck\ Catalogue of Compact Sources}\label{sec:pccs}

\begin{figure*}
\begin{center}
\includegraphics[width=\textwidth]{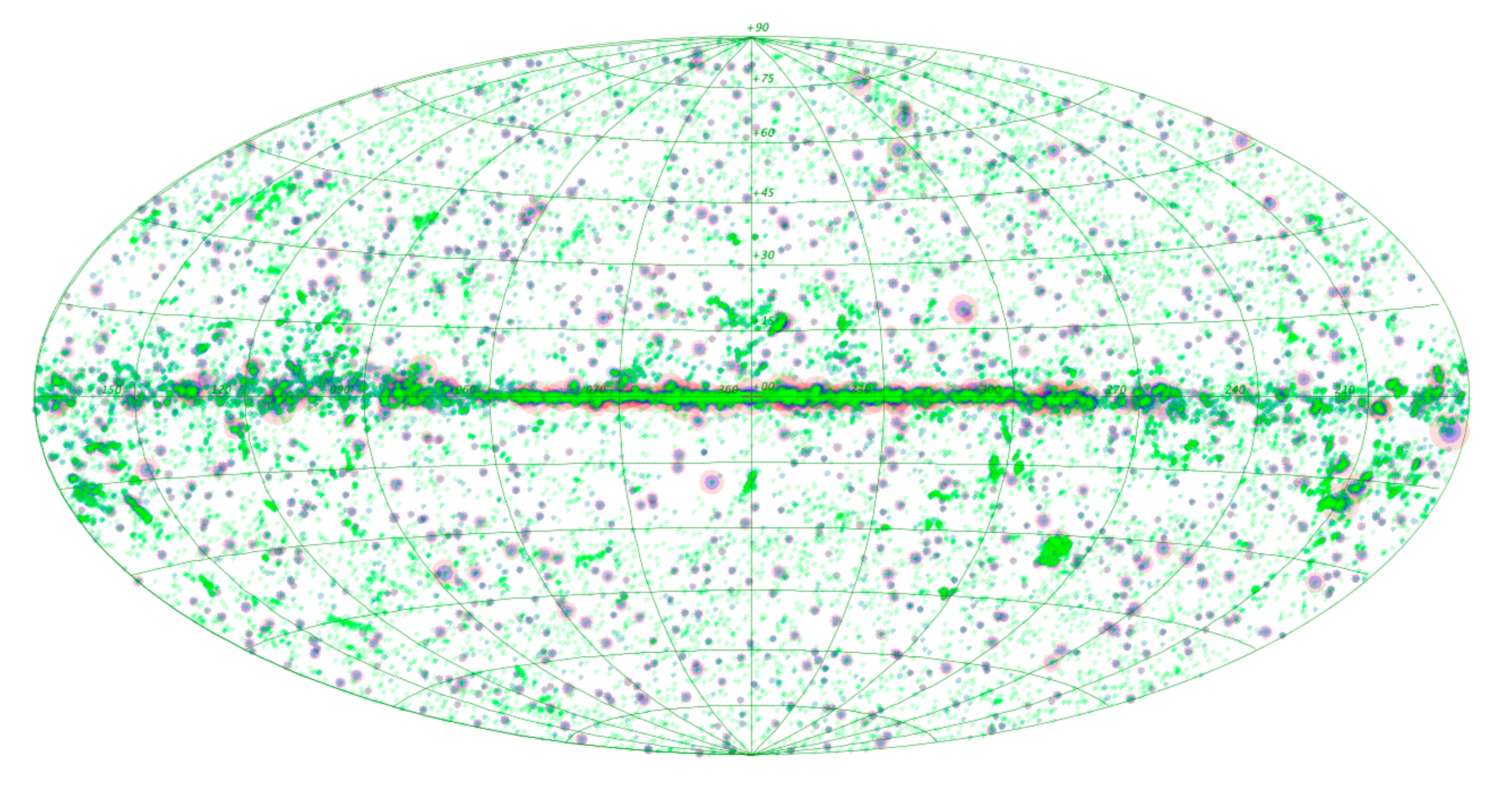}
\caption{Sky distribution of the PCCS sources at three different channels: 30\,GHz ({pink circles}); 143\,GHz ({magenta circles}); and 857\,GHz ({green circles}). The dimension of the circles is related to the brightness of the sources and the beam size of each channel. The figure is a full-sky Aitoff projection with the Galactic equator horizontal; longitude increases to the left with the Galactic centre in the centre of the map.
\label{fig:skydist}}
\end{center}
\end{figure*}


\begin{table*}
\begingroup
\newdimen\tblskip \tblskip=5pt
\caption{PCCS characteristics.}
\label{tab:pccs}
\nointerlineskip \vskip -3mm \footnotesize
\setbox\tablebox=\vbox{
\newdimen\digitwidth
\setbox0=\hbox{\rm 0}
\digitwidth=\wd0
\catcode`*=\active
\def*{\kern\digitwidth}
\newdimen\signwidth
\setbox0=\hbox{+}
\signwidth=\wd0
\catcode`!=\active
\def!{\kern\signwidth}
\halign to \hsize{\hbox to 1.9in{#\leaderfil}\tabskip 2em plus 0.2 em & \hfil#\hfil& \hfil#\hfil& \hfil#\hfil& \hfil#\hfil& \hfil#\hfil& \hfil#\hfil& \hfil#\hfil& \hfil#\hfil& \hfil#\hfil\tabskip=0pt\cr
\noalign{\doubleline}
\omit&\multispan9{\hfil Channel\hfil}\cr
\noalign{\vskip -3pt}
\omit&\multispan9\hrulefill\cr
\noalign{\vskip 2pt}
\omit& 30& 44& 70& 100& 143& 217& 353& 545& 857\cr
\noalign{\vskip 3pt\hrule\vskip 5pt}
Frequency [GHz]& 28.4& 44.1& 70.4& 100.0& 143.0& 217.0& 353.0& 545.0& 857.0\cr
Wavelength, $\lambda$ [$\mu$m]& 10561& 6807& 4260& 3000& 2098& 1382& 850& 550& 350\cr
Beam FWHM$^{\rm a}$ [arcmin]& 32.38& 27.10& 13.30& 9.65& 7.25& 4.99& 4.82& 4.68& 4.33\cr
Pixel size [arcmin]& 3.44& 3.44& 3.44& 1.72& 1.72& 1.72& 1.72& 1.72& 1.72\cr
\noalign{\vskip 2pt}
\noalign{{\snr\ thresholds:}}
\hskip 2em Full sky& 4.0& 4.0& 4.0& 4.6& 4.7& 4.8& \ldots& \ldots& \ldots\cr
\hskip 2em Extragalactic zone$^{\rm b}$& \ldots& \ldots& \ldots& \ldots& \ldots& \ldots& 4.9& 4.7& 4.9\cr
\hskip 2em Galactic zone$^{\rm b}$& \ldots& \ldots& \ldots& \ldots& \ldots& \ldots& 6.0& 7.0& 7.0\cr
\noalign{\vskip 2pt}
\noalign{{Number of sources:}}
\hskip 2em Full sky&           1256& 731& 939& 3850& 5675& 16070& 13613& 16933& 24381\cr
\hskip 2em $|b|>30\deg$& *572& 258& 332& *845& 1051& *1901& *1862& *3738& *7536\cr
\noalign{\vskip 2pt}
\noalign{{N($>$S)$^{\rm c}$:}}
\hskip 2em Full sky&           934& 535& 689& 3425& 5229& 15107& 13184& 15781& 23561\cr
\hskip 2em $|b|>30\deg$& 373& 151& 191& *629& *857& *1409& *1491& *2769& *6773\cr
\hskip 2em $|b|\leq30\deg$& 561& 384& 498& 2796& 4422& 13698& 11693& 13012& 16788\cr
\noalign{\vskip 2pt}
\noalign{{Flux densities:}}
\hskip 2em Minimum$^{\rm d}$ [mJy]&     461& *825& 566& 266& 169& 149& 289& 457& 658\cr
\hskip 2em 90\,\% completeness [mJy]& 575& 1047& 776& 300& 190& 180& 330& 570& 680\cr
\hskip 2em Uncertainty [mJy]&                109& *198& 149& *61& *38& *35& *69& 118& 166\cr
\noalign{\vskip 2pt}
Position uncertainty$^{\rm e}$ [arcmin]& 1.8& 2.1& 1.4& 1.0& 0.7& 0.7& 0.8& 0.5& 0.4\cr
\noalign{\vskip 5pt\hrule\vskip 3pt}}}
\endPlancktablewide
\tablenote {{\rm a}} {\tt FEBeCoP} band-averaged effective beam.  This table shows the exact values that were adopted for the PCCS. For HFI channels, these are the FWHM of the mean best-fit Gaussian.  For the LFI channels,  we use ${\rm FWHM}_{\rm eff} = \sqrt{(\Omega_{\rm eff}/{2\pi})8\log{2}}$, where $\Omega_{\rm eff }$ is the {\tt FEBeCoP} band-averaged effective solid angle (see \citealt{planck2013-p02d} and \citealt{planck2013-p03c} for a full description of the \Planck\ beams). When we constructed the PCCS for the LFI channels we used a value of the effective FWHM slightly different (by $\ll1$\%) of the final values specified in the \citet{planck2013-p02d} paper. This small correction will be made in later versions of the catalogue.\par
\tablenote {{\rm b}} The Galactic and extragalactic zones are defined in Sect.~\ref{sec:selection}.\par
\tablenote {{\rm c}} Number of sources above the 90\% completeness level.\par
\tablenote {{\rm d}} Minimum flux density of the catalogue at $|{\rm b}|>30\deg$ after excluding the faintest 10\,\% of sources.\par
\tablenote {{\rm e}} Positional uncertainty (median value) derived by comparison with PACO sample \citep{massardi11,bonavera11,bonaldi13} up to 353\,GHz and with \Herschel\ samples in the other channels (see Sect.~\ref{sec:highfreq} for more details).\par
\endgroup
\end{table*}

\subsection{Data}
\label{sec:data}

The data obtained from the \Planck\ nominal mission between 2009 August 12 and 2010 November 27 have been processed into full-sky maps by the Low Frequency Instrument (LFI; 30--70\,GHz) and High Frequency Instrument (HFI;100--857\,GHz) Data Processing Centres (DPCs)~\citep[see][]{planck2013-p02,planck2013-p03}. The data consist of two complete sky surveys and 60\,\% of the third survey. This implies that the flux densities of sources obtained from the nominal mission maps are the average of at least two observations separated by roughly six months.

The nine \Planck\ frequency channel maps were used as input to the source detection pipelines. The CMB dipole is removed during the map making stage. For the highest-frequency channels, 353, 545, and 857\,GHz, a model of the zodiacal emission~\citep{planck2013-pip88} was the only foreground emission subtracted from the maps before detecting the sources. The relevant properties of the frequency maps are summarized in Table~\ref{tab:pccs}.

\subsection{Source detection pipelines}
\label{sec:detection}

Compact sources were detected in each frequency map by looking for peaks after convolving with a linear filter that preserves the amplitude of the source while reducing both large scale structure (e.g., diffuse Galactic emission) and small scale fluctuations (e.g., instrumental noise) in the vicinity of the sources.

Although the matched filter is optimal for uniform Gaussian noise, the real data present additional challenges. For example, the power spectrum is needed to construct the matched filter and this quantity is not known a priori and has to be determined directly from the data. We have explored the performance of different filters using realistic \Planck\ simulations, among them our implementation of a matched filter and the first and second members of the Mexican Hat Wavelet family, MHW and MHW2 \citep{gnuevo06, caniego06}, and for these particular data we have chosen the last of these, MHW2, which performs better than the MHW and similarly to the matched filter.

The MHW2 has only one free parameter, the scale $R$, to be locally optimized, and is less sensitive to artefacts (e.g., missing pixels) or very bright structures in the image, like those found in the Galactic plane. These bright structures introduce instabilities in the determination of the power spectrum needed to construct the matched filter and reduce its performance. The MHW2 is robust and gives good performance at all Galactic latitudes. Besides, it has previously been used to detect compact sources in astronomical images, including  realistic simulations of \Planck\ \citep{caniego06,gnuevo06,leach08} and data from \textit{WMAP} \citep{caniego07,massardi09}.

The MHW2 filter in Fourier space is given by
\begin{equation}
\hat{\psi} \left(kR\right) \propto \left(kR\right)^{4} \tau(kR),
\label{eq:filtro_general}
\end{equation}
where $k$ is the wavenumber; $\tau$ is the beam profile or point spread function,  approximated by a Gaussian $\tau(\xx) =( 1 / 2\pi{\sigma_{\rm b}}^2 ) \exp[-\frac{1}{2}(\xx/\sigma_{\rm b})^2]$; and $\sigma_{\rm b}$ is the Gaussian beam dispersion.

Two independent implementations of the MHW2 algorithm have been used, one by the LFI DPC and another by the HFI DPC. The outputs of the two implementations have been compared and the results are compatible at the level of statistical uncertainty (see Sect.~\ref{sec:lfi_vs_hfi}). An additional algorithm,  the matrix filter \citep{herranz08, herranz09}, has been used to validate the catalogue; this is a multifrequency method that is also being used for the production of a multifrequency catalogue of non-thermal sources that will be published in a future paper.

The two MHW2 pipelines have a number of features in common. The full-sky {\tt HEALPix} maps \citep{gorski05} are divided into small, square patch maps using a gnomonic projection. The patches should be large enough to get a fair sample of the noise in each, but small enough that the noise and foreground characteristics are close to uniform across each patch. The number of patches is chosen to allow sufficient overlap to remove detections in the borders of the patches where edge effects become important. In both pipelines the scale $R$ of the filter is optimized by finding the maximum signal-to-noise ratio (\snr) of the sources in the filtered patch. The optimal scale is determined for each patch independently and, while it is always near to unity, it is usually smaller near the Galactic plane, e.g., $0.6<R<1.2$ at 30\,GHz, and bigger when the beam is smaller when compared to the pixel size, e.g., $1.1<R<1.5$ at 70\,GHz. Detections above a given \snr\ threshold are retained and the positions of the detected objects are translated from patch coordinates to spherical coordinates. The final stage is to remove multiple detections of the same object from different patches. No attempt has been made to remove detected sources from the maps.

\paragraph{LFI:} The compact source detection pipeline used in the LFI is part of the IFCAMEX software package\footnote{http://max.ifca.unican.es/IFCAMEX}. It can be used to detect sources with no prior information on their position (blind mode), or at the position of known objects (non-blind mode). For this analysis we blindly search for objects over the full sky with $\mathrm{\snr} \ge 2$ to produce a preliminary catalogue of potential sources with positions, flux densities. and uncertainties. In the second step, IFCAMEX is run in non-blind mode, using as input the coordinates of the objects detected in the first step and keeping those with $\mathrm{\snr} \ge 4$. In this case the patch is centred on the position of the source. The goal of this second iteration is to minimize the already small border and projection effects and to refine the estimation of the position and \snr\ of each detection, keeping only those objects that still have a \snr\ above the threshold, thus improving the reliability of the catalogue. In addition, and given the large size of the LFI beams, a fitting algorithm has been incorporated to find the centroids of the sources, achieving sub-pixel coordinate accuracy. Moreover, we have taken into account the effective non-Gaussian shape of the beams in the estimation of the flux density. This is done applying a correction factor to the IFCAMEX flux density estimation obtained by comparing the flux density of a simulated Gaussian test source of a given scale $R$ with that of source convolved with the effective non-Gaussian beam of the detector at each of the LFI frequencies. This correction factor is small, typically $<1\,\%$. Further details of this procedure can be found in \cite{massardi09}.
\paragraph{HFI:} The novel features of the HFI implementation were designed  to deal with the challenging environment for source detection in the high frequency channels. They aim to reduce the number of spurious detections with minimal impact on the number of real sources found. Each patch is filtered at the optimal scale, and also at four other scales bracketing the optimal scale. The dependence of the amplitude of the detection on the filter scale is compared with the predicted behaviour of a point source. The $\chi^2$ between the observed and predicted values is minimized to provide an alternative measurement of the amplitude. The values of the \snr, $\chi^2$, and the ratio of the two measurements of the amplitude determine whether a source is accepted or rejected. There is also an additional criterion for removing spurious detections that is based on the number of connected pixels, above a threshold, associated with a detection in the filtered patch at the optimal scale. The idea behind this is to reject artefacts that lie in narrow structures and may not be completely removed by the filtering. For the given scale of the wavelet, the number of expected connected pixels for a point source is evaluated and this is compared with the number of connections found for the detection. A combination of the \snr\ of the detection and the ratio of these numbers of connected pixels is used to determine whether rejection should occur. These additional criteria to reject artefacts help to improve the reliability of the catalogue without affecting its completeness and  its other statistical properties.

\subsection{Selection criteria}
\label{sec:selection}

\begin{figure}
\begin{center}
\includegraphics[width=\columnwidth]{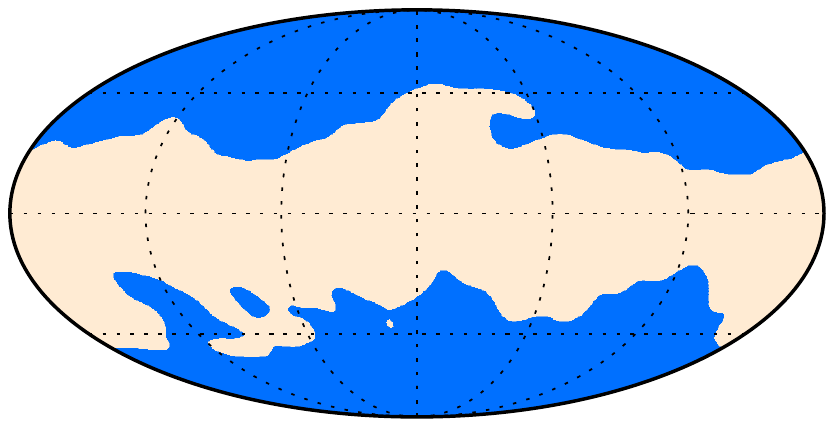}
\end{center}
\caption{The Galactic and extragalactic zones used to define the \snr\ thresholds to meet the reliability target. The figure is a full-sky Mollweide projection. See text for further details.}
\label{fig:zones}
\end{figure}
The source selection for the PCCS is made on the basis of the \snr. However, the background properties of the \Planck\ maps vary substantially depending on frequency and part of the sky. Up to 217\,GHz, the CMB is the dominant source of confusion at high Galactic latitudes. At high frequencies, confusion from Galactic foregrounds dominates the noise budget at low Galactic latitudes, and the cosmic infrared background at high Galactic latitudes. The SNR has therefore been adapted for each particular case.

The driving goal of the ERCSC was reliability greater than 90\,\%. In order to increase completeness and explore possibly interesting new sources at fainter flux densities, however, a lower reliability goal of 80\,\% was chosen for the PCCS. The \snr\ thresholds applied to each frequency channel have been determined, as far as possible, to meet this goal. The reliability of the catalogues has been assessed using the internal and external validation procedures described in Sect.~\ref{sec:validation}.

At 30, 44, and 70\,GHz, the reliability goal alone would permit \snr\ thresholds below 4. A secondary goal of minimizing the upward bias on fainter flux densities \citep[Eddington bias;][]{eddington40} led to the imposition of an \snr\ threshold of 4.

At higher frequencies, where the confusion caused by the Galactic emission starts to become an issue, the sky has been divided into two zones, one Galactic (52\,\% of the sky) and one extragalactic (48\,\% of the sky), using the G45 mask defined in \cite{planck2013-p08}. The zones are shown in Fig.~\ref{fig:zones}. At 100, 143, and 217\,GHz, the \snr\ threshold needed to achieve the target reliability is determined in the extragalactic zone, but applied uniformly across the sky. At 353, 545, and 857\,GHz, the need to control confusion from Galactic cirrus emission led to the adoption of different \snr\ thresholds in the two zones. This strategy ensures interesting depth and good reliability in the extragalactic zone, but also high reliability in the Galactic zone. The extragalactic zone has a lower threshold than the Galactic zone. The \snr\ thresholds are given in Table~\ref{tab:pccs}.

\subsection{Photometry}
\label{sec:photometry}

For each source in the PCCS we have obtained four different measures of the flux density. They are determined by the source detection algorithm: aperture photometry; point spread function (PSF) fitting; and Gaussian fitting. Only the first is obtained from the filtered maps, and the other measures are estimated from the full-sky maps at the positions of the sources. The source detection algorithm photometry, the aperture photometry and the PSF fitting use the \Planck\ band-average effective beams, calculated with {\tt FEBeCoP} (Fast Effective Beam Convolution in Pixel space) \citep{mitra2010,planck2013-p02d,planck2013-p03c}. Notice that only the PSF fitting uses a model of the PSF that depends on the position of the source and the scan pattern.

\paragraph{Detection pipeline photometry (\textbf{DETFLUX}).} As described in Sect.~\ref{sec:detection}, the detection pipelines assume that sources are point-like. The amplitude of the detected source is converted to flux density using the area of the beam and the conversion from map units into intensity units. If a source is resolved its flux density will be underestimated. In this case it may be better to use the GAUFLUX estimation.

\paragraph{Aperture photometry (\textbf{APERFLUX}).} The flux density is estimated by integrating the data in a circular aperture centred at the position of the source. An annulus around the aperture is used to evaluate the level of the background. The annulus is also used to make a local estimate of the noise to calculate the uncertainty in the estimate of the flux density. The flux density is corrected for the fraction of the beam solid angle falling outside the aperture and for the fraction of the beam solid angle falling in the annulus.
The aperture photometry was computed using an aperture with radius equal to the average FWHM of the effective beam, and an annulus with an inner radius of 1~FWHM and an outer radius of 2~FWHM. The effective beams were used to compute the beam solid angle corrections. For details see Appendix~\ref{sec:aperflux}.

\paragraph{PSF fit photometry (\textbf{PSFFLUX}).} The flux density is obtained by fitting a model of the PSF at the position of the source to the data. The model has two free parameters, the amplitude of the source and a background offset. The PSF is obtained from the effective beam. For details see Appendix~\ref{sec:psfflux}.

\paragraph{Gaussian fit photometry (\textbf{GAUFLUX}).} The flux density is obtained by fitting an elliptical Gaussian model to the source. The Gaussian is centred at the position of the source and its amplitude, size, and axial ratio are allowed to vary, as is the background offset. The flux density is calculated from the amplitude and the area of the Gaussian. For details see Appendix~\ref{sec:gauflux}.

\bigskip

\begin{figure}
\begin{center}
\includegraphics[width=\columnwidth]{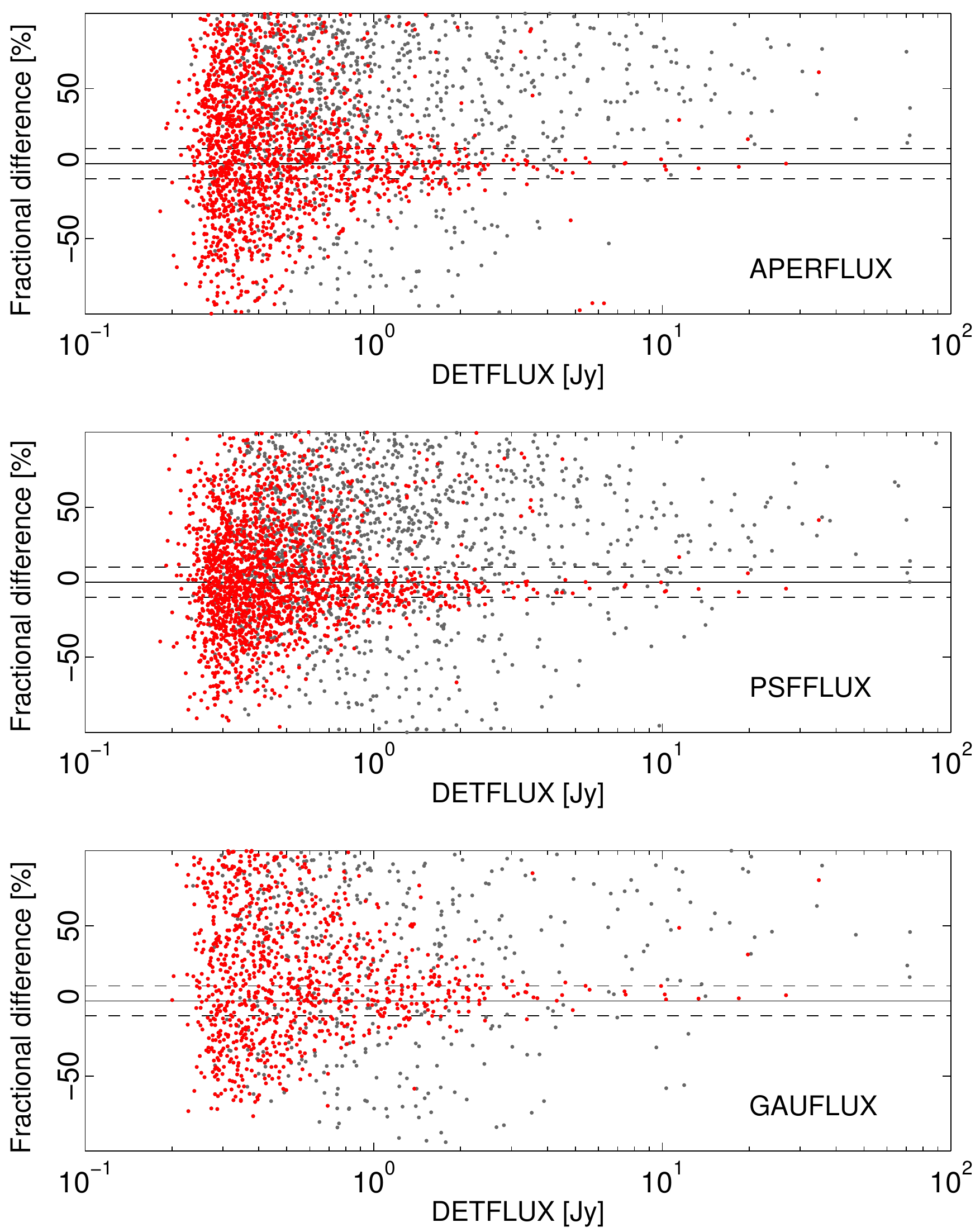}
\caption{Comparison of the APERFLUX, PSFFLUX and GAUFLUX flux density estimates with the DETFLUX ones for the 100\,GHz catalogue, $(S-S_{\rm DETFLUX})/S_{\rm DETFLUX}$. Grey points correspond to sources below $|{b}|<5^\circ$ while red ones show the ones for $|{b}|>5^\circ$. Dashed lines indicates the $\pm10\,\%$ uncertainty level.
\label{fig:photo_check}}
\end{center}
\end{figure}

\noindent Figure~\ref{fig:photo_check} shows a comparison between DETFLUX flux densities at 100\,GHz and the other three estimates. DETFLUX has been chosen as the reference photometry because it is the photometry used in the selection process and the only one estimated directly from the filtered patches (filtering attenuates the background fluctuations by a factor $\sim2$ and allows a much more robust estimation of the faintest flux densities). The dispersion increases at lower \snr\ and near the Galactic plane, where the different estimators behave differently in the presence of a strong background (indicated by grey points). At higher latitudes the agreement is much better for bright sources (the red points). This figure shows how  the four different flux density estimators can provide complementary information on the same object, for example if the object is extended or near the Galactic plane (see Sect. \ref{sec:cautionary}).

\subsection{Comparison between MHW2 pipelines}
\label{sec:lfi_vs_hfi}

\begin{figure}
\begin{center}
\includegraphics[width=\columnwidth]{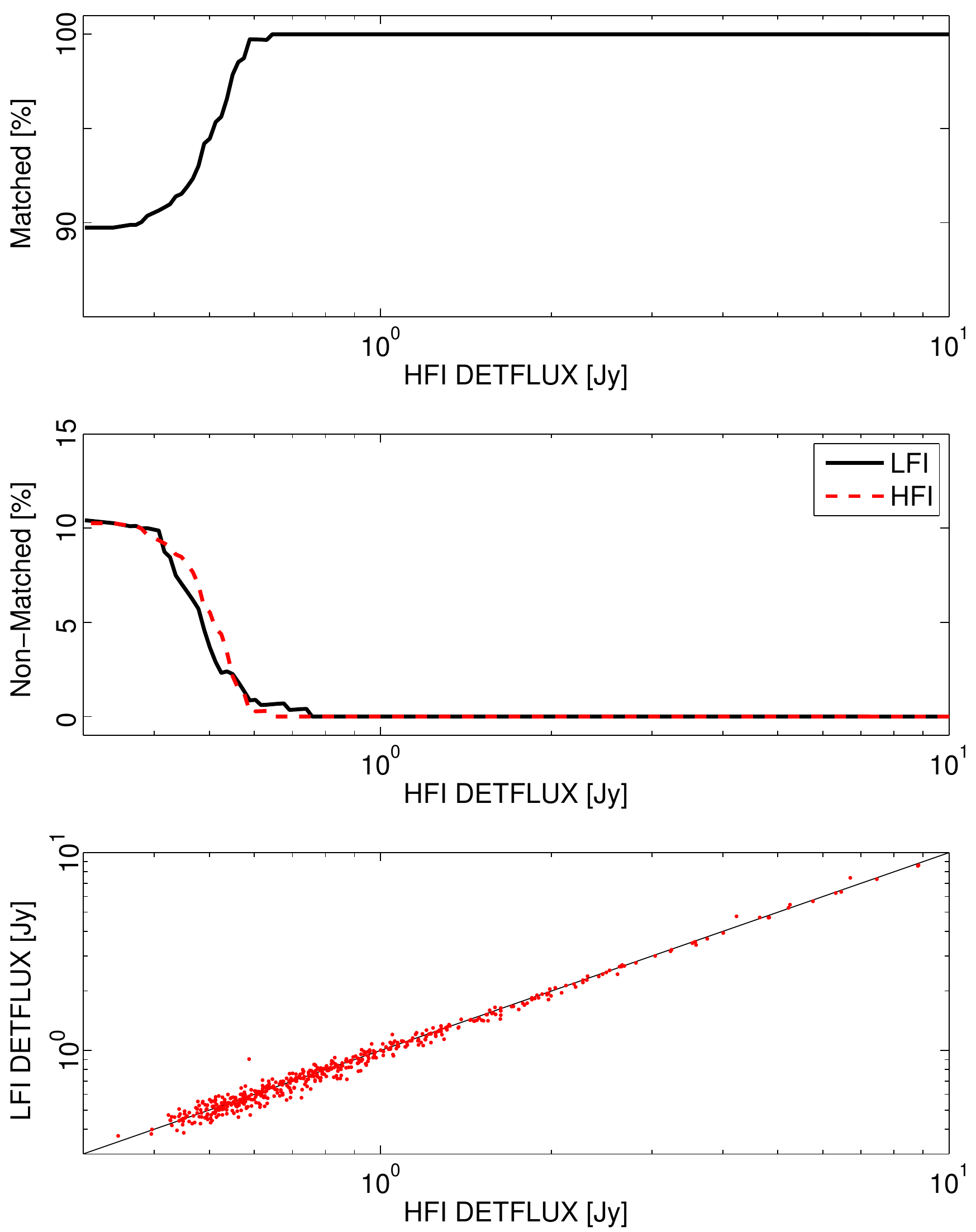}
\caption{Test of internal consistency between the two implementations of the MHW pipeline at 30\,GHz for $|{b}|>30^\circ$. \emph{Top panel:} Cumulative percentage of sources detected by both methods. \textit{Middle panel:} Cumulative percentage of sources detected by only one of the methods. \emph{Bottom panel:} Comparison of the recovered flux densities (DETFLUX).
\label{fig:hfi_lfi}}
\end{center}
\end{figure}

In order to ensure the internal consistency of the whole catalogue, we have checked that both implementations of the MHW2 algorithm are equivalent. Both were run on the LFI nominal maps producing two sets of catalogues and the outputs from both implementations have been compared (see Fig. \ref{fig:hfi_lfi} for an example). We have studied the number of sources detected by both implementations (``matched'') and the number of sources detected by only one  (``non-matched''). We have also compared the native (DETFLUX) photometry from both implementations. As shown in Fig.~\ref{fig:hfi_lfi} for the 30\,GHz channel, the only differences between the detections obtained by both implementations appear near the threshold where small changes in \snr\ values make the difference between a source being detected or not. These differences are always below 10\,\% in the faintest bin. More important is the good agreement between photometric results from the two pipelines.

\section{Validation of the PCCS}
\label{sec:validation}

The PCCS contents and the four different flux-density estimates have been validated by simulations (internal validation) and comparison with other observations (external validation). The validation of the non-thermal radio sources can be done with a large number of existing catalogues, whereas the validation of thermal sources is mostly done with simulations.  Detections identified with known sources have been marked in the catalogues.

\subsection{Internal validation}
\label{sec:int_val}

The catalogues for the HFI channels have been validated primarily through an internal Monte Carlo quality assessment (QA) process in which artificial sources are injected in both real maps and simulated maps. For each channel, we calculate statistical quantities describing the quality of detection, photometry, and astrometry for each detection code. The detection is described by the completeness and reliability of the catalogue: completeness is a function of intrinsic flux density, the selection threshold applied to detection (\snr), and location, while reliability is a function only of the detection \snr. The quality of photometry and astrometry is assessed by direct comparison of detected position and flux density with the known parameters of the artificial sources.  An input source is considered to be detected if a detection is made within one beam FWHM of the injected position.

The completeness is determined from the injection of unresolved point sources into the real maps. Bias due to the superimposition of sources is avoided by preventing injection within an exclusion radius of $\sigma_{\rm b}$ around both existing detections in the real map and previously injected sources.  The flux from real and injected point sources contributes to the noise estimation for each detection patch, reducing the \snr\ of all detections and biasing the completeness.  We prevent this effect by determining the noise properties on the maps before injecting sources, and have verified that remaining bias on detection and parameter estimates due to injected sources is negligible.  The injected sources are convolved with the effective beam \citep{planck2013-p02, planck2013-p03}.

We use two cumulative reliability estimates for the HFI catalogues.  The first, which we will call \emph{simulation reliability}, is determined from source injection into simulated maps and is defined as the fraction of detected sources that match the positions of injected sources.  If the simulations are accurate, such that the spurious and real detection number counts mirror the real catalogue, the reliability is exact.  To accept the simulations, we require that they pass the internal consistency tests outlined below.  Simulation reliability is used for the 100, 143, and 217\,GHz channels.

The simulations used to calculate simulation reliability consist of realizations of CMB, instrumental noise, and the diffuse Galactic emission component of the FFP6 simulations \citep[a set of realistic simulations based on the \Planck\ Sky Model; ][]{planck2013-p06,planck2013-p28,psm}. We require that the simulated catalogues pass two internal consistency tests: that they have the same injected source completeness as the real catalogues calculated as outlined above; and that they have total detected number counts, as a function of \snr, that are consistent with those in the real data. The intrinsic number counts are assumed to be power law functions of flux density and are fitted to the detection counts at higher flux densities, where the catalogues are reliable and complete, and extrapolated to lower flux densities.  Sources are injected with random positions.

The second reliability estimator is applied to the 353, 545, and 857\,GHz channels, where the simulations fail our internal consistency tests (due to deficiencies in the simulations of diffuse dust emission).  In the absence of accurate simulations capable of producing realistic realisations of spurious detections, we use an approximate reliability criterion that we will call \emph{injection reliability}.  Injection reliability makes use of source injection into the real maps to determine number counts of matched sources.  If the fiducial input source model is accurate, the matched counts are a good estimate of the real detection counts in the catalogue.  To form a reliability estimate, we take the ratio per \snr\ bin of the matched number counts over the number counts of the real catalogue (the latter of which is the sum of real and spurious number counts).

The input flux density model is assumed to be a power law and is fitted in the same way as for the simulation reliability. The extrapolation of the input source model to lower flux densities is the main source of uncertainty in the injection reliability estimate. However, it is also subject to bias due to the Poisson fluctuations of number counts in the real catalogue. The total numbers are large enough at low \snr\ in the higher frequency channels that the measurement of the increment of spurious sources is robust to these fluctuations. At higher \snr, however, we take as reliable any bin where the difference between expected real and measured total number counts is smaller than twice the Poisson noise of the total number counts. To minimise bias from fluctuations, we also assume the catalogues are completely reliable at \snr\ $> 10$. We have verified that the two reliability estimates are consistent with one another at $217$\,GHz, the only frequency where they can both be applied.

\subsubsection{Completeness}
\label{sec:qa_completeness}

\begin{figure}
\begin{center}
\includegraphics[width=\columnwidth]{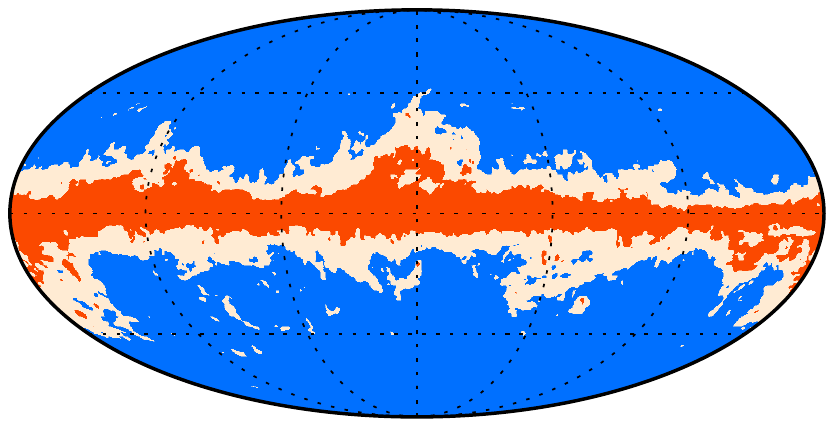}
\end{center}
\caption{The Galactic dust masks used to estimate completeness and reliability for some of the HFI channels. The unmasked zones correspond to sky fractions of 65\,\% and 85\,\%. The figure is a full-sky Mollweide projection. See text for further details.}
\label{fig:masks}
\end{figure}

We have estimated the completeness of each of the HFI catalogues in the 48\,\% extragalactic zone shown in Fig.~\ref{fig:zones}. We have also estimated completeness in the larger zones outside two Galactic dust masks shown in Fig.~\ref{fig:masks}: the less restrictive 85\,\% zone for $100$\,GHz and $143$\,GHz, and the 65\,\% zone for $217$\,GHz. These zones match those assumed for the reliability estimate at those channels. The completeness estimates are shown in Fig.~\ref{fig:qa_real}, along with full-sky maps of the sensitivity, defined as the flux density at 50\,\% differential completeness.

\subsubsection{Reliability}
\label{sec:qa_reliability}

\begin{figure*}
\begin{center}
\begin{tabular}{ccc}
\vspace{-0.56cm} \hspace{-0.1cm} \includegraphics[width=0.33\textwidth]{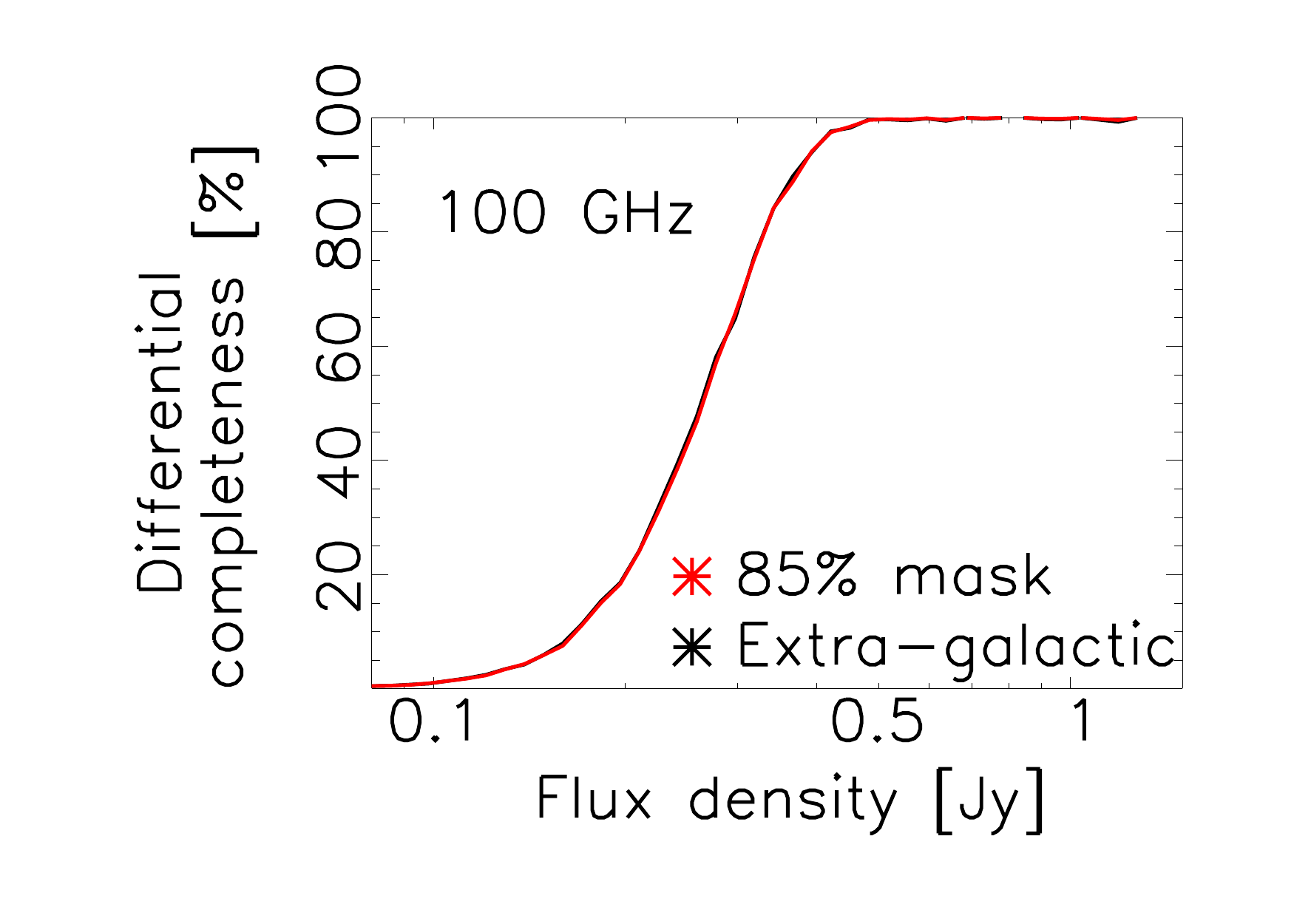} &
 \hspace{-0.7cm} \raisebox{0.3cm}{ \includegraphics[width=0.33\textwidth]{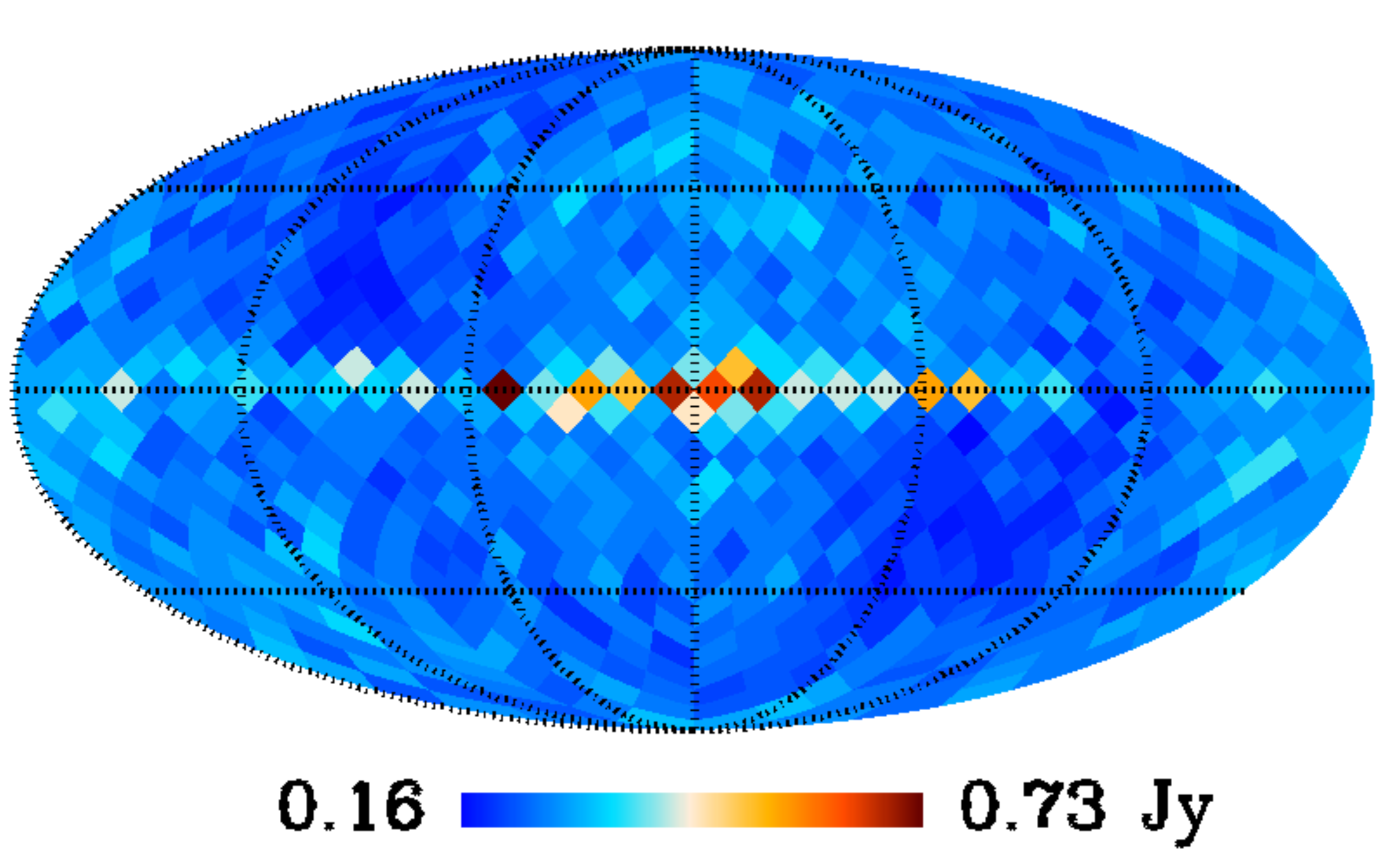} }&
\hspace{-1.5cm} \includegraphics[width=0.33\textwidth]{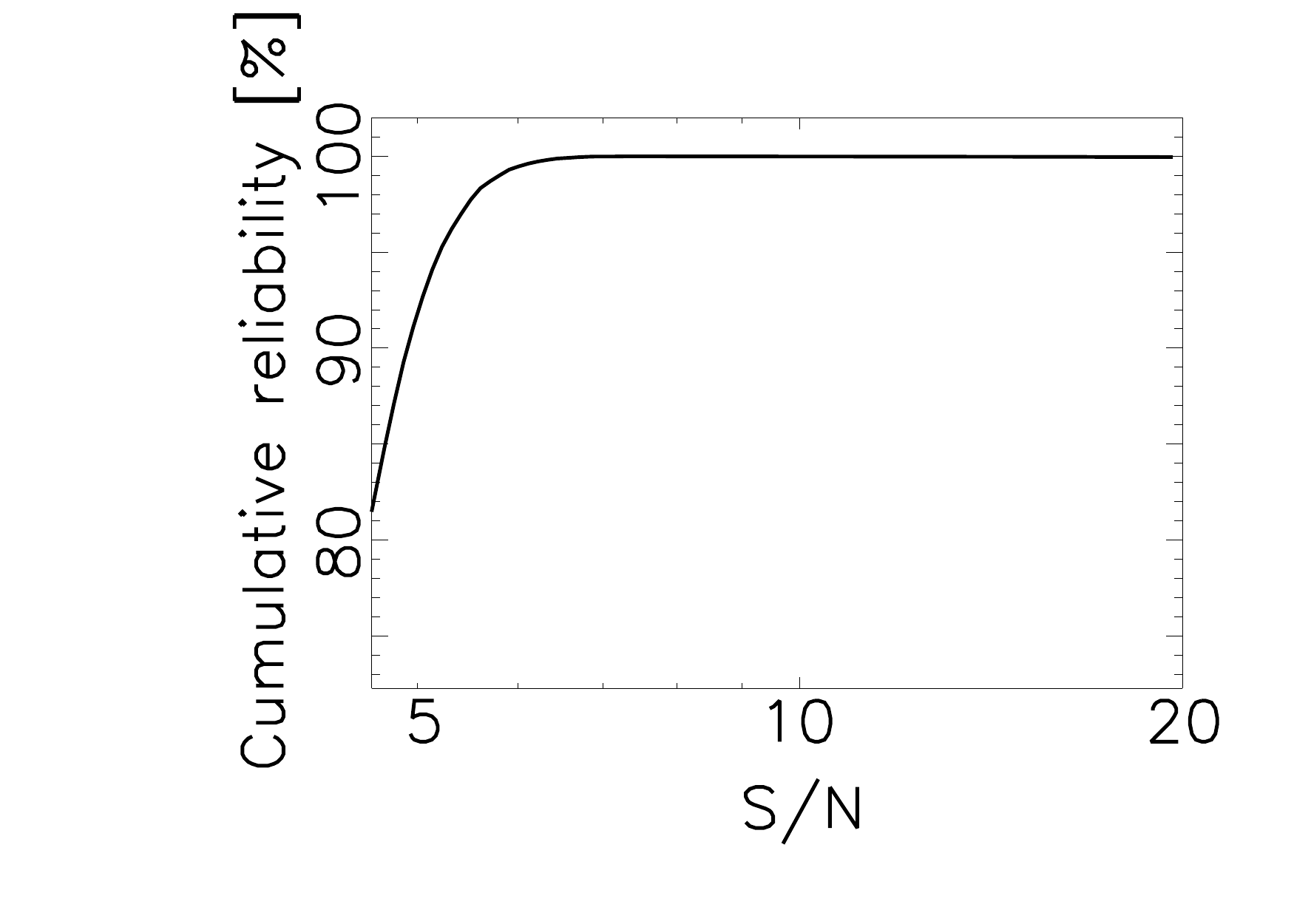} \\
\vspace{-0.6cm}  \hspace{-0.1cm} \includegraphics[width=0.33\textwidth]{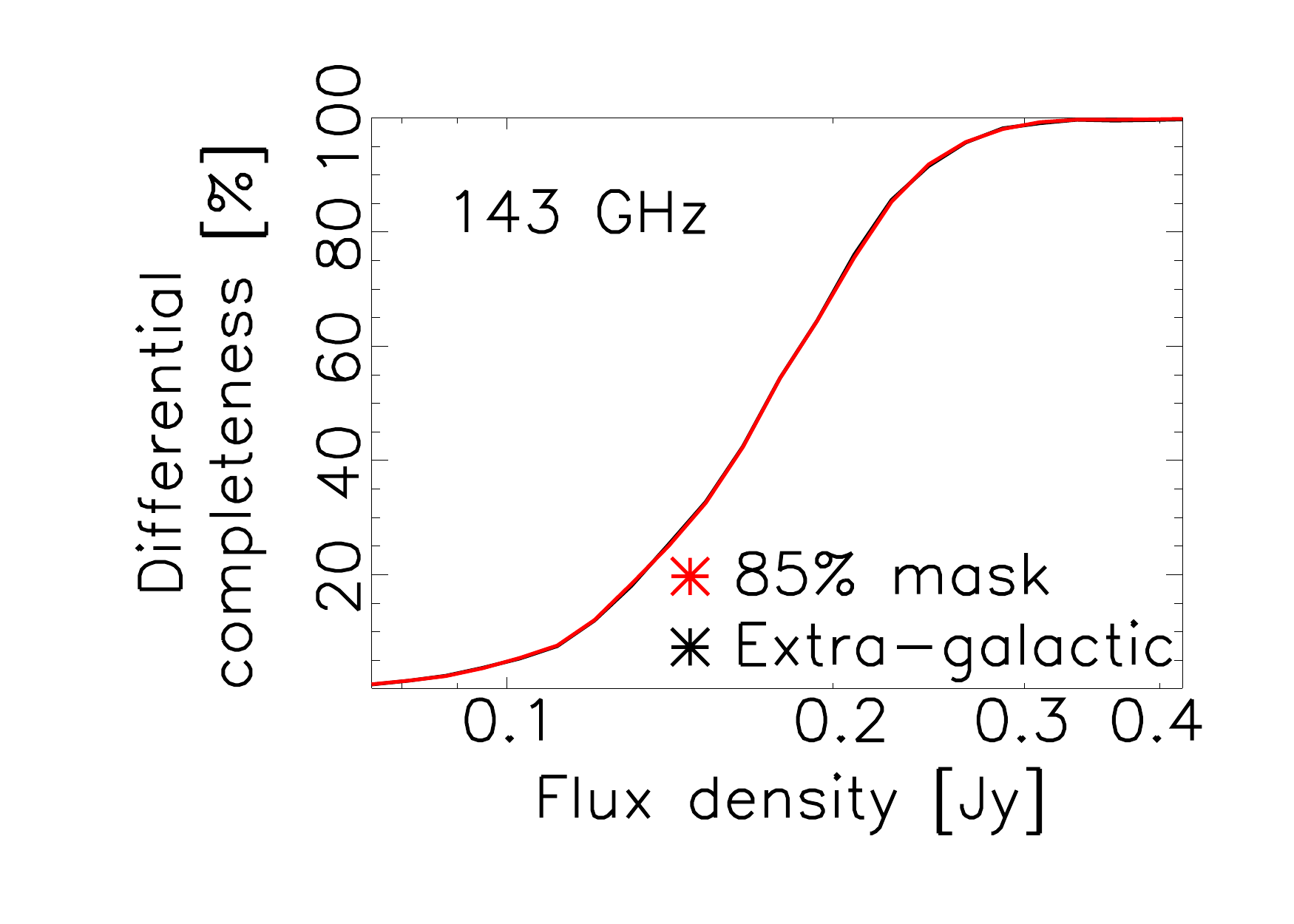} &
\hspace{-0.7cm} \raisebox{0.3cm}{\includegraphics[width=0.33\textwidth]{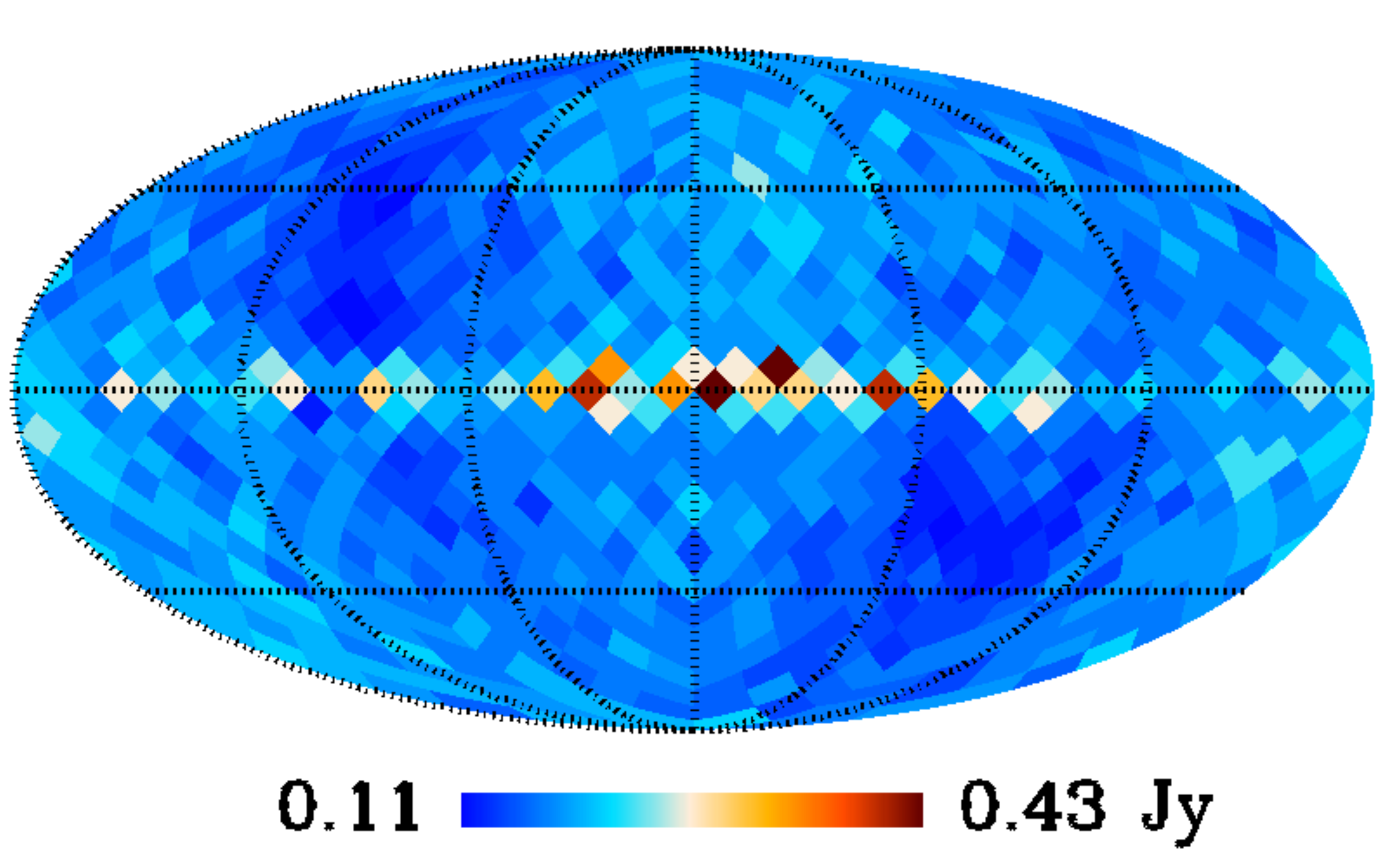} }&
\hspace{-1.5cm} \includegraphics[width=0.33\textwidth]{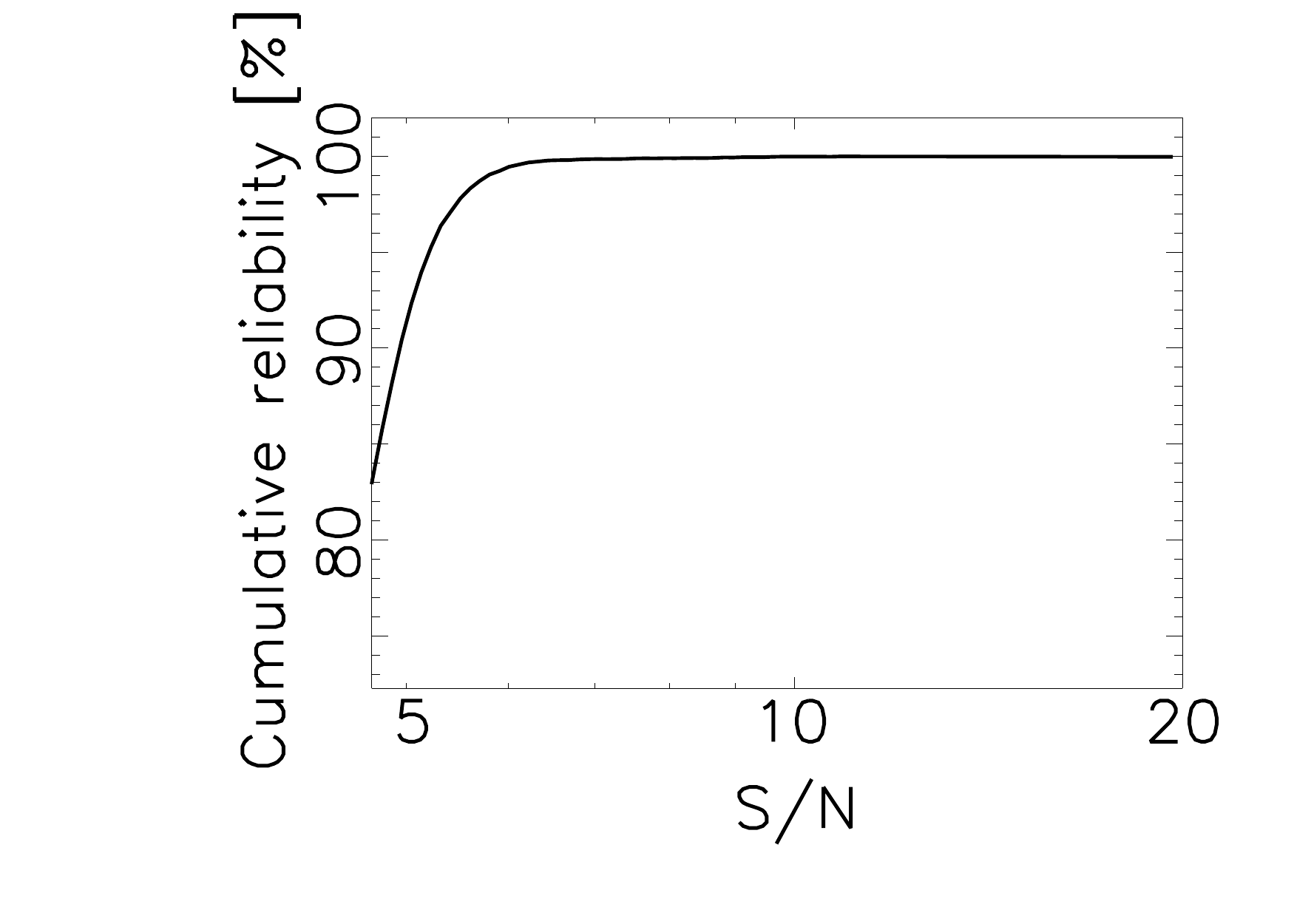} \\
\vspace{-0.6cm}  \hspace{-0.1cm} \includegraphics[width=0.33\textwidth]{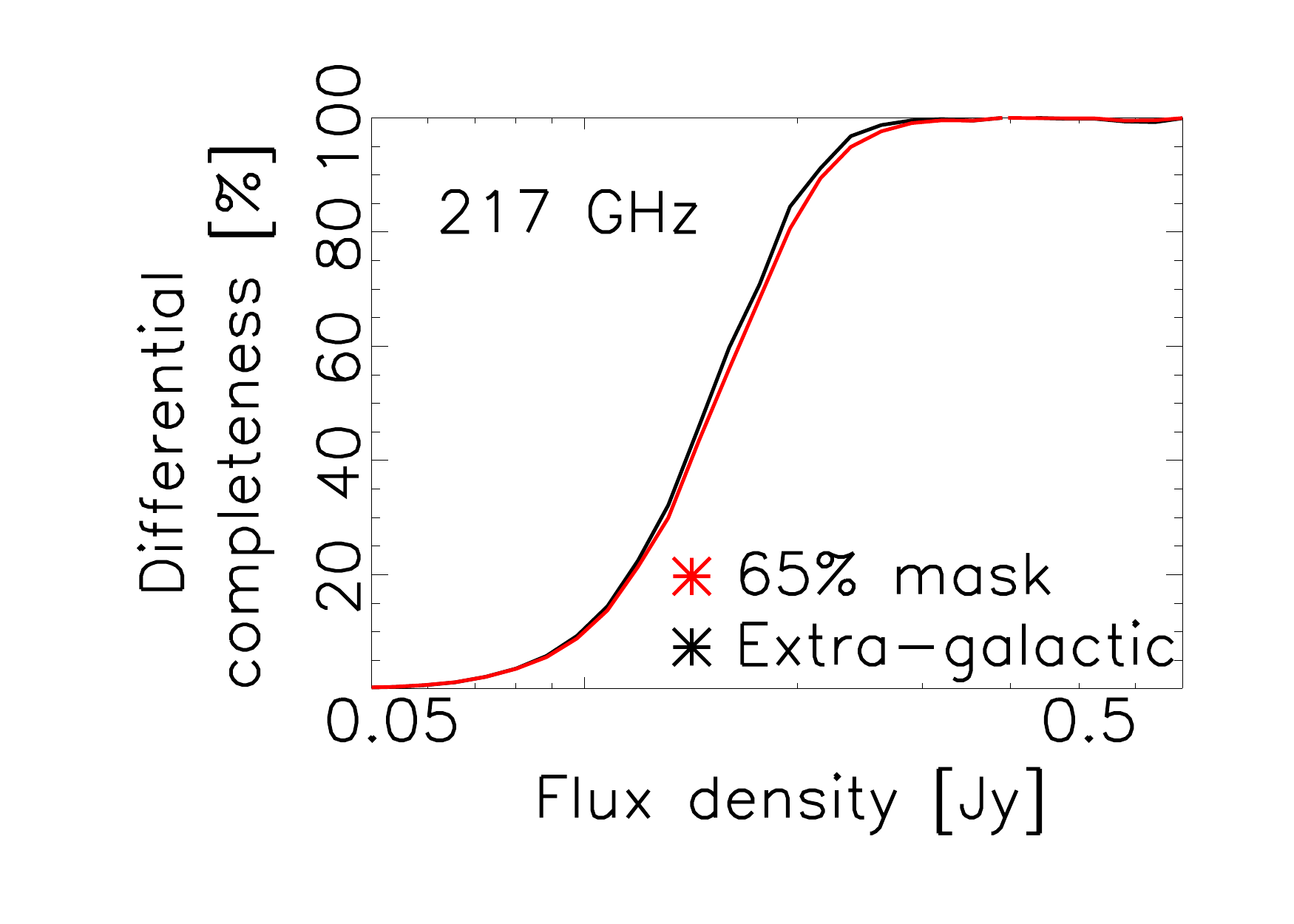} &
 \hspace{-0.7cm} \raisebox{0.3cm}{\includegraphics[width=0.33\textwidth]{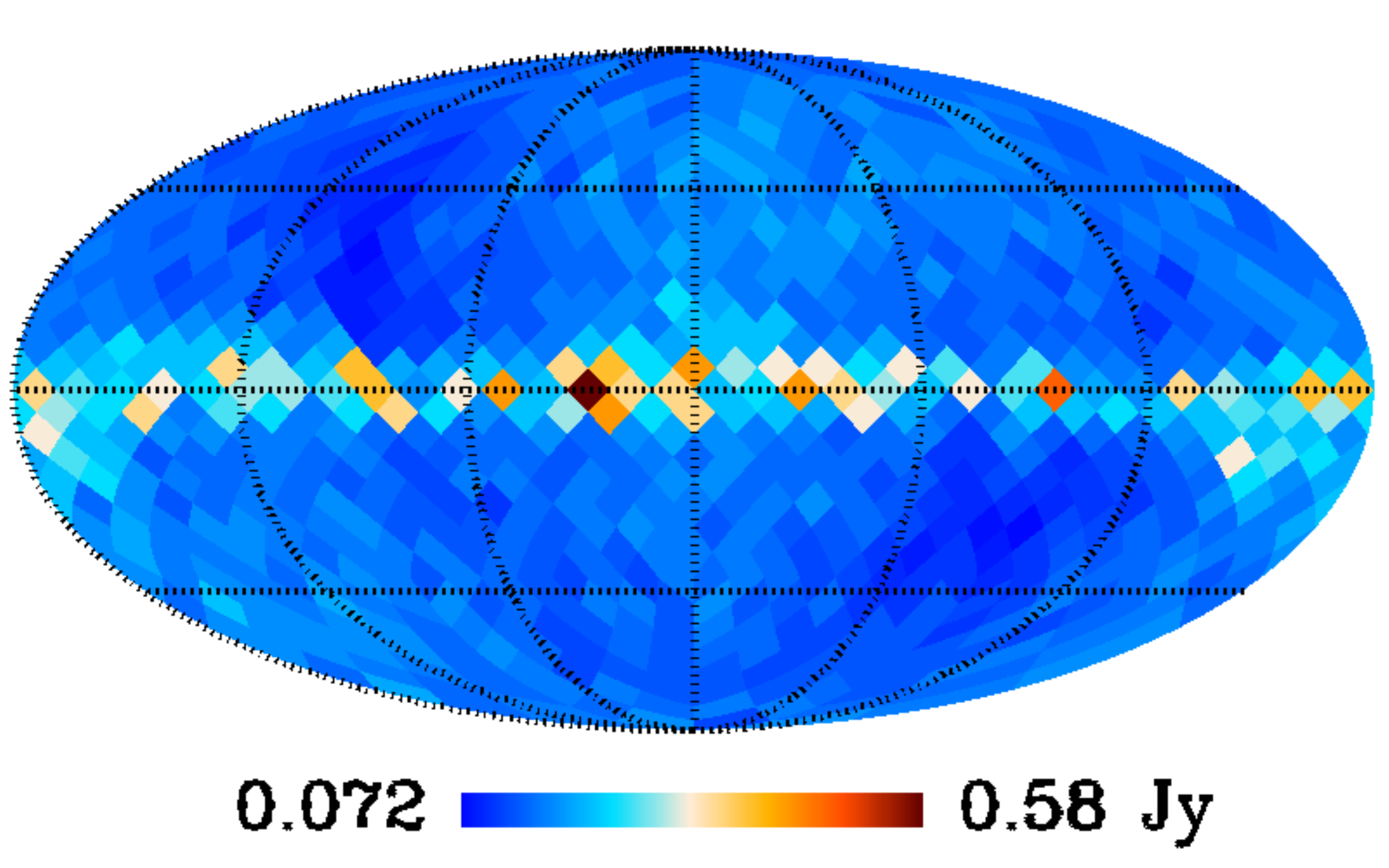} }&
\hspace{-1.5cm} \includegraphics[width=0.33\textwidth]{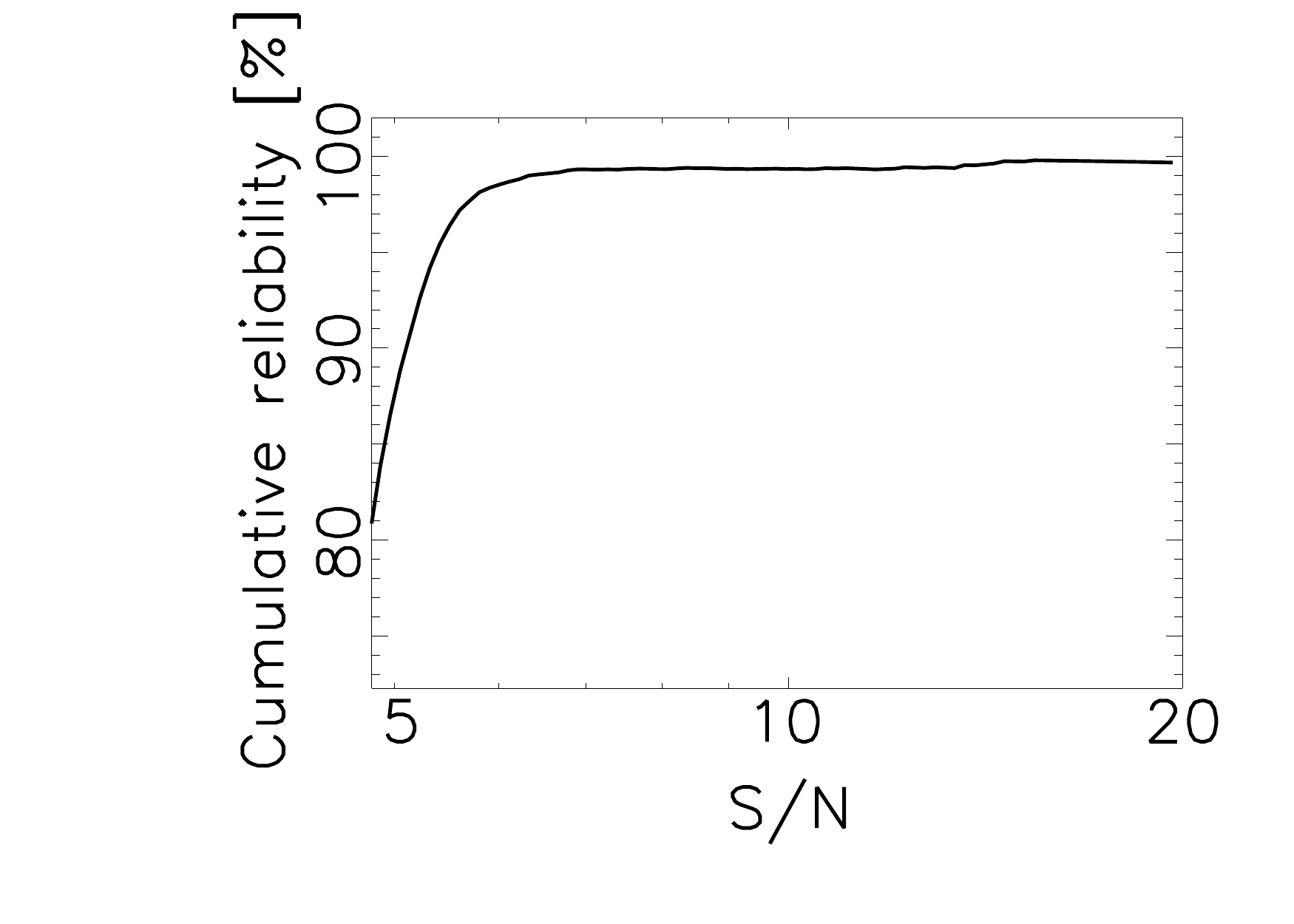}\\
\vspace{-0.6cm}  \hspace{-0.1cm} \includegraphics[width=0.33\textwidth]{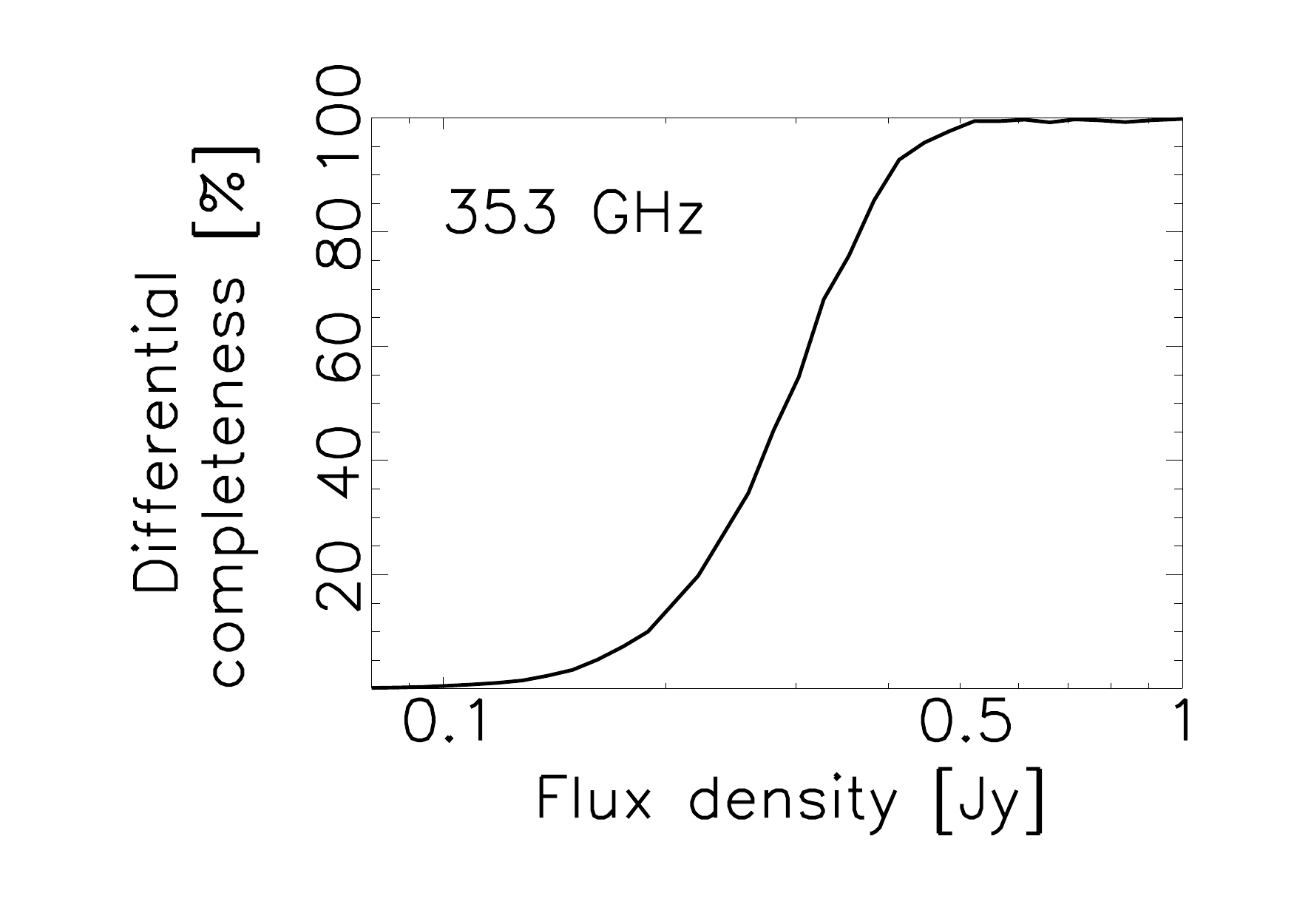} &
 \hspace{-0.7cm} \raisebox{0.3cm}{ \includegraphics[width=0.33\textwidth]{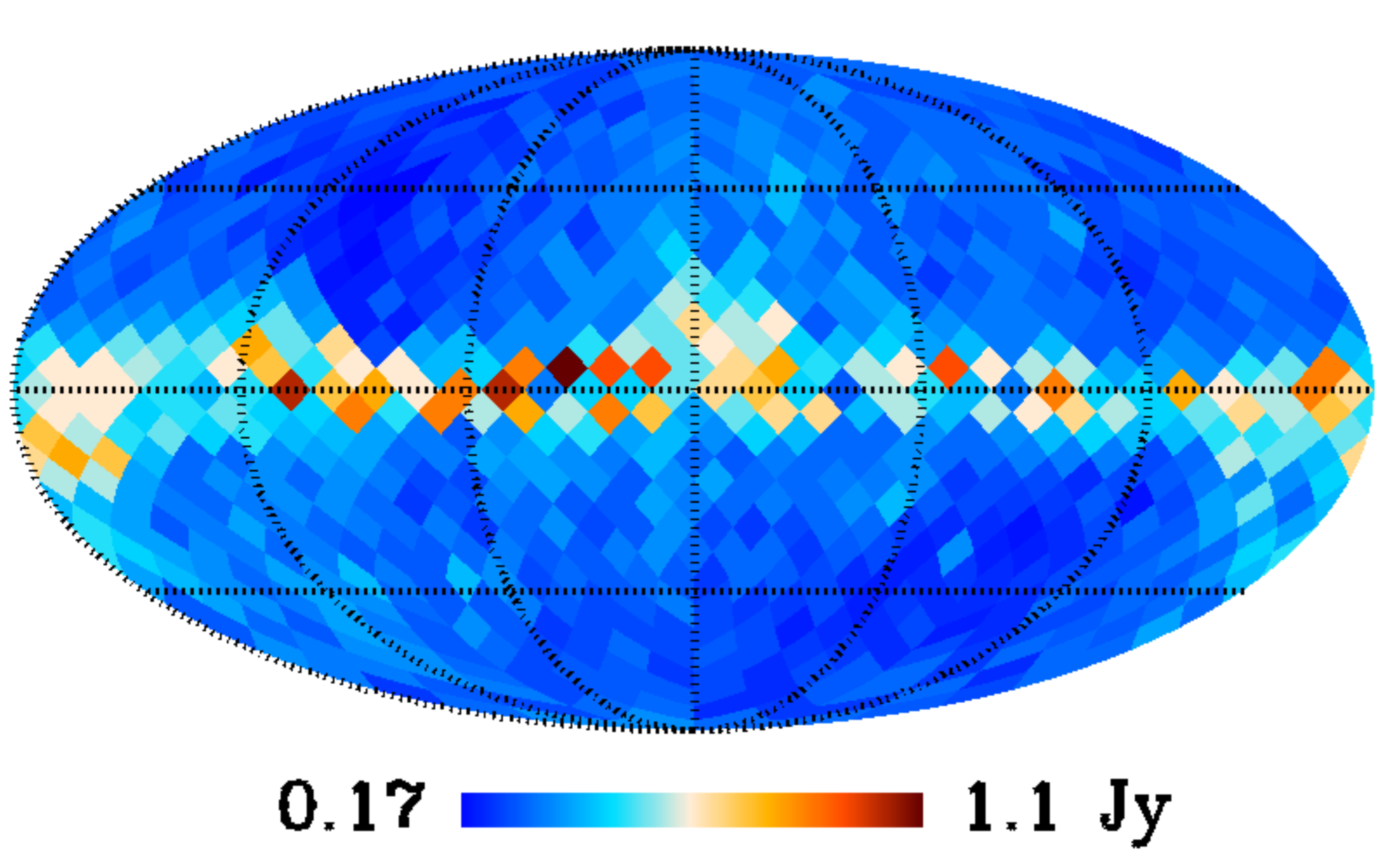} }&
\hspace{-1.5cm} \includegraphics[width=0.33\textwidth]{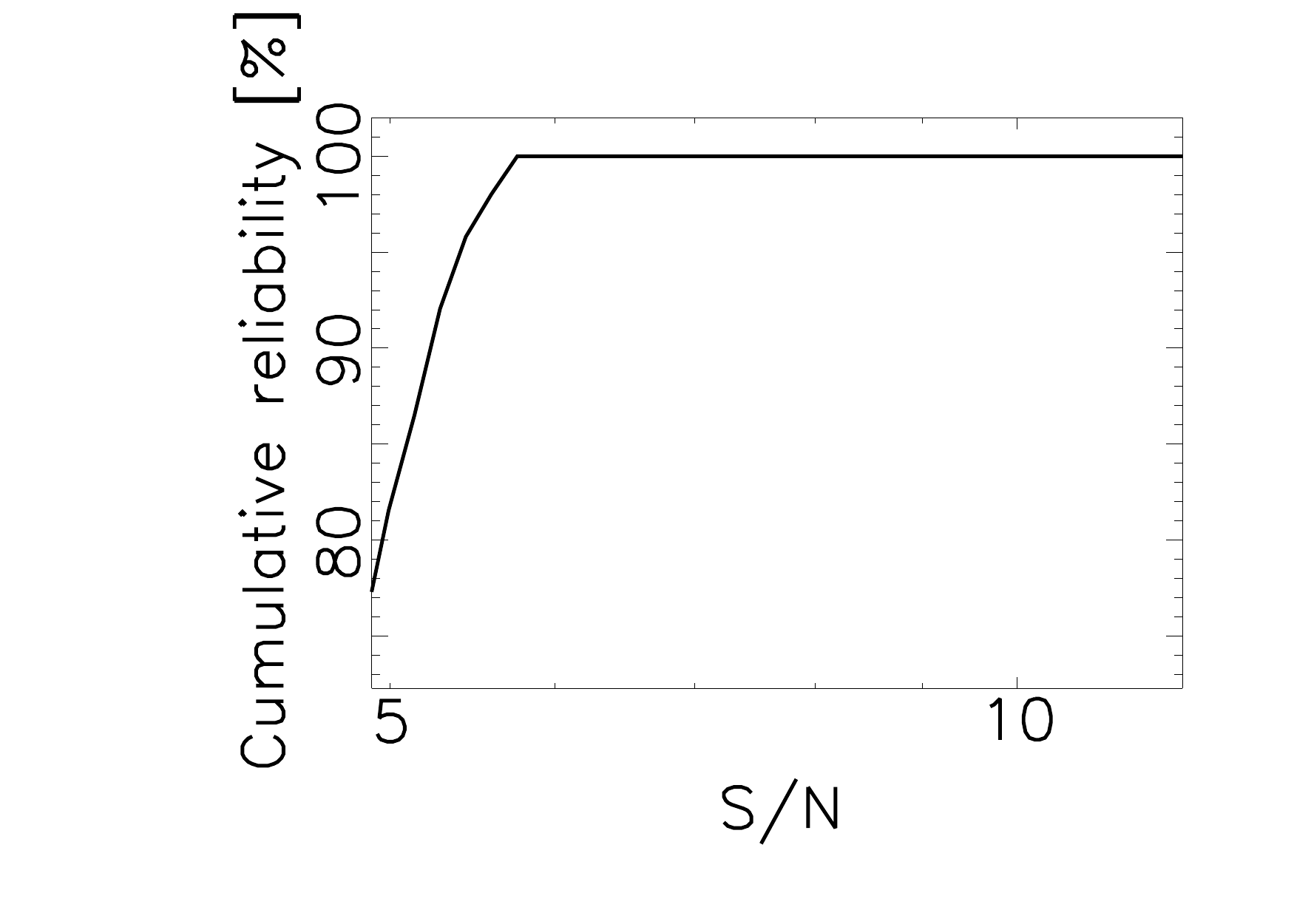} \\
\vspace{-0.6cm}  \hspace{-0.1cm} \includegraphics[width=0.33\textwidth]{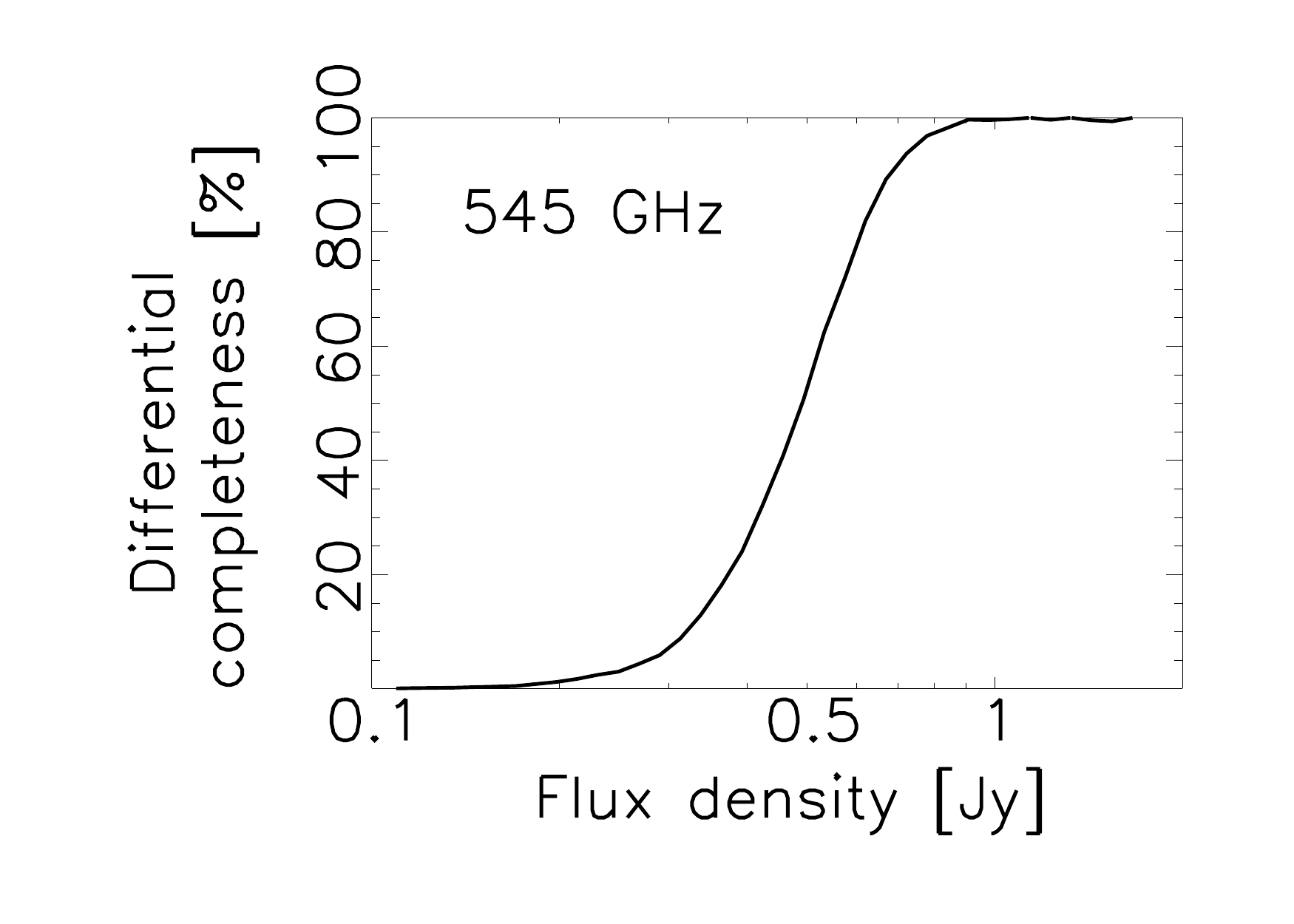} &
 \hspace{-0.7cm} \raisebox{0.3cm}{ \includegraphics[width=0.33\textwidth]{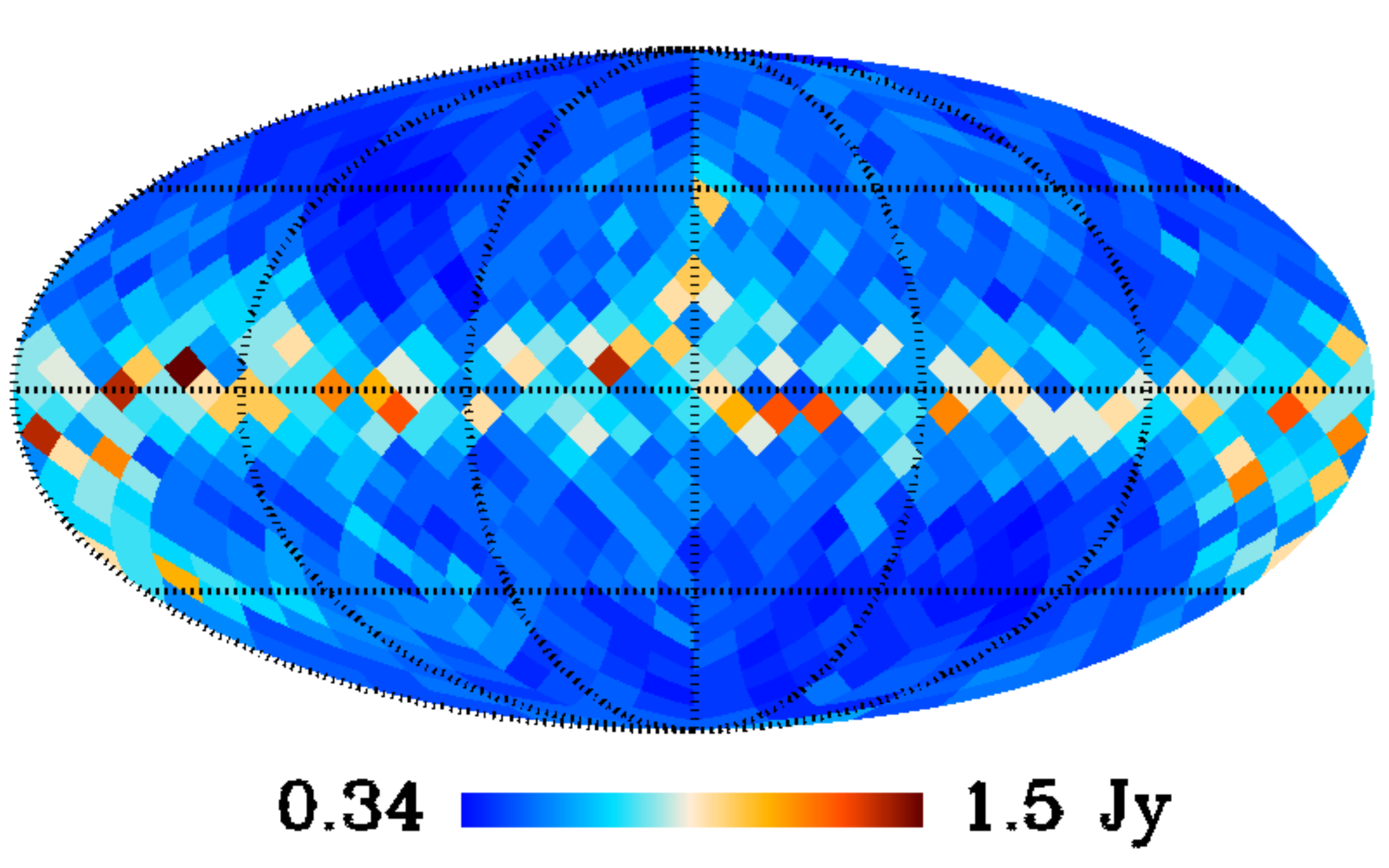} }&
\hspace{-1.5cm} \includegraphics[width=0.33\textwidth]{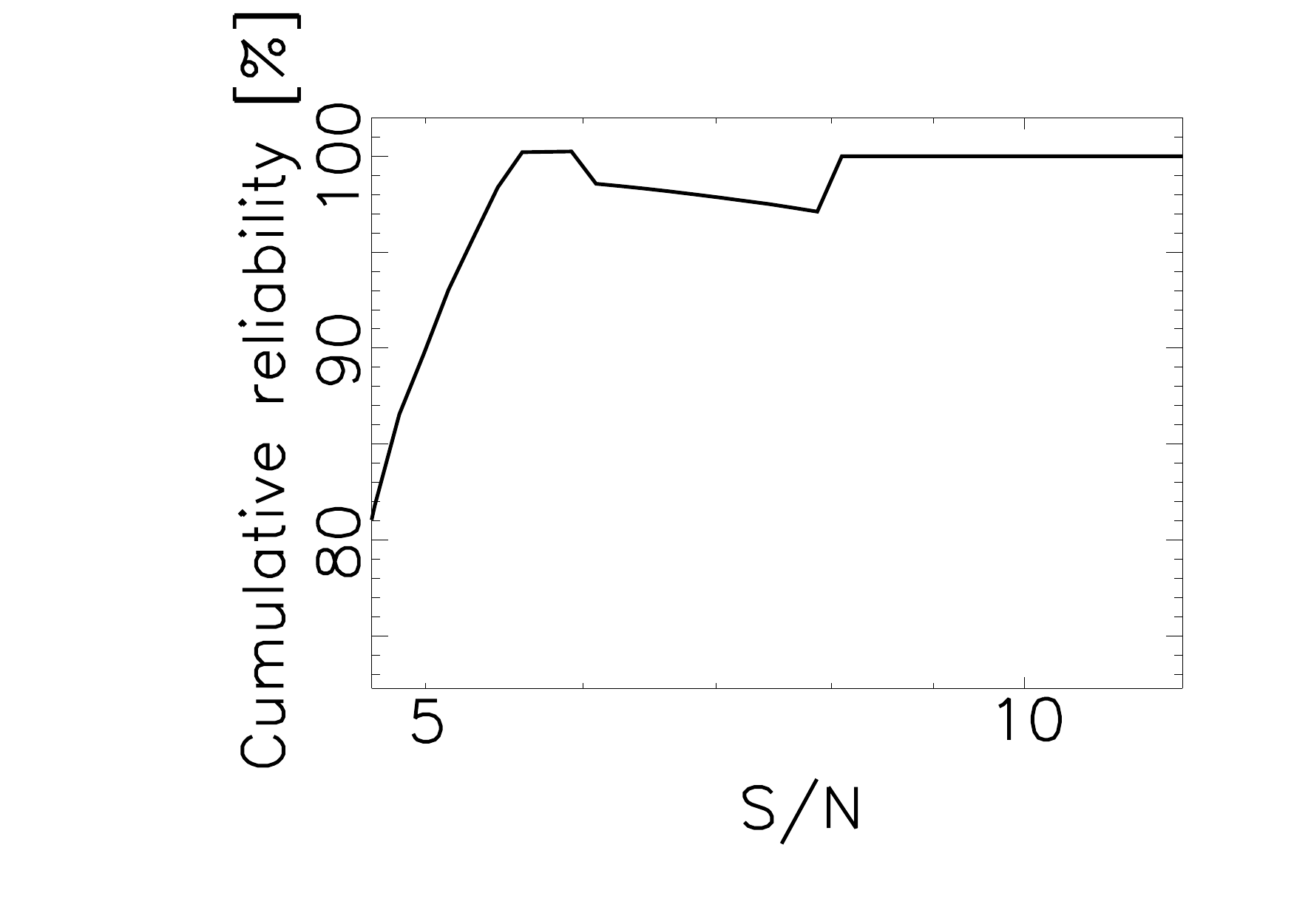} \\
\vspace{-0.6cm} \hspace{-0.1cm} \includegraphics[width=0.33\textwidth]{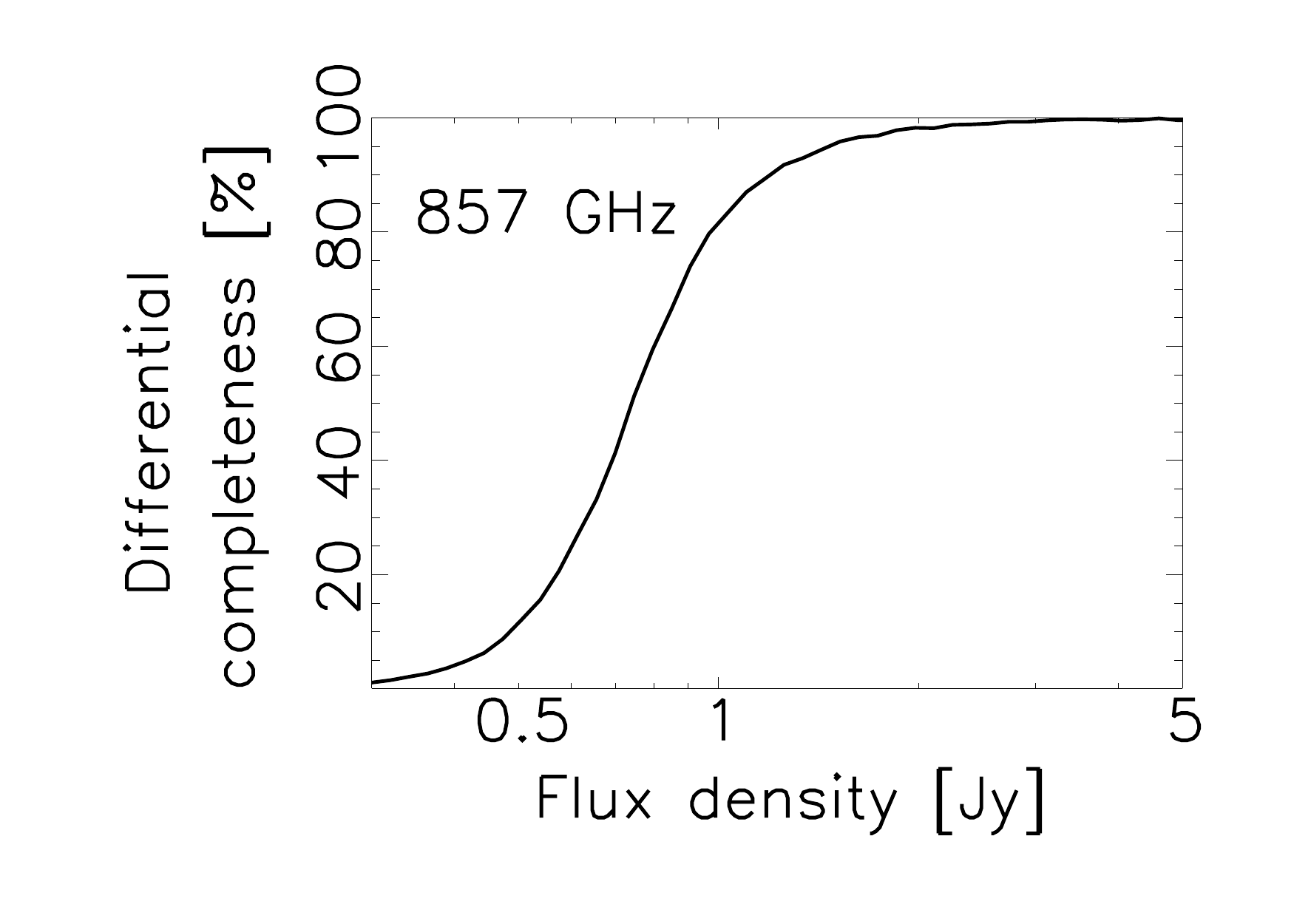} &
 \hspace{-0.7cm} \raisebox{0.3cm}{ \includegraphics[width=0.33\textwidth]{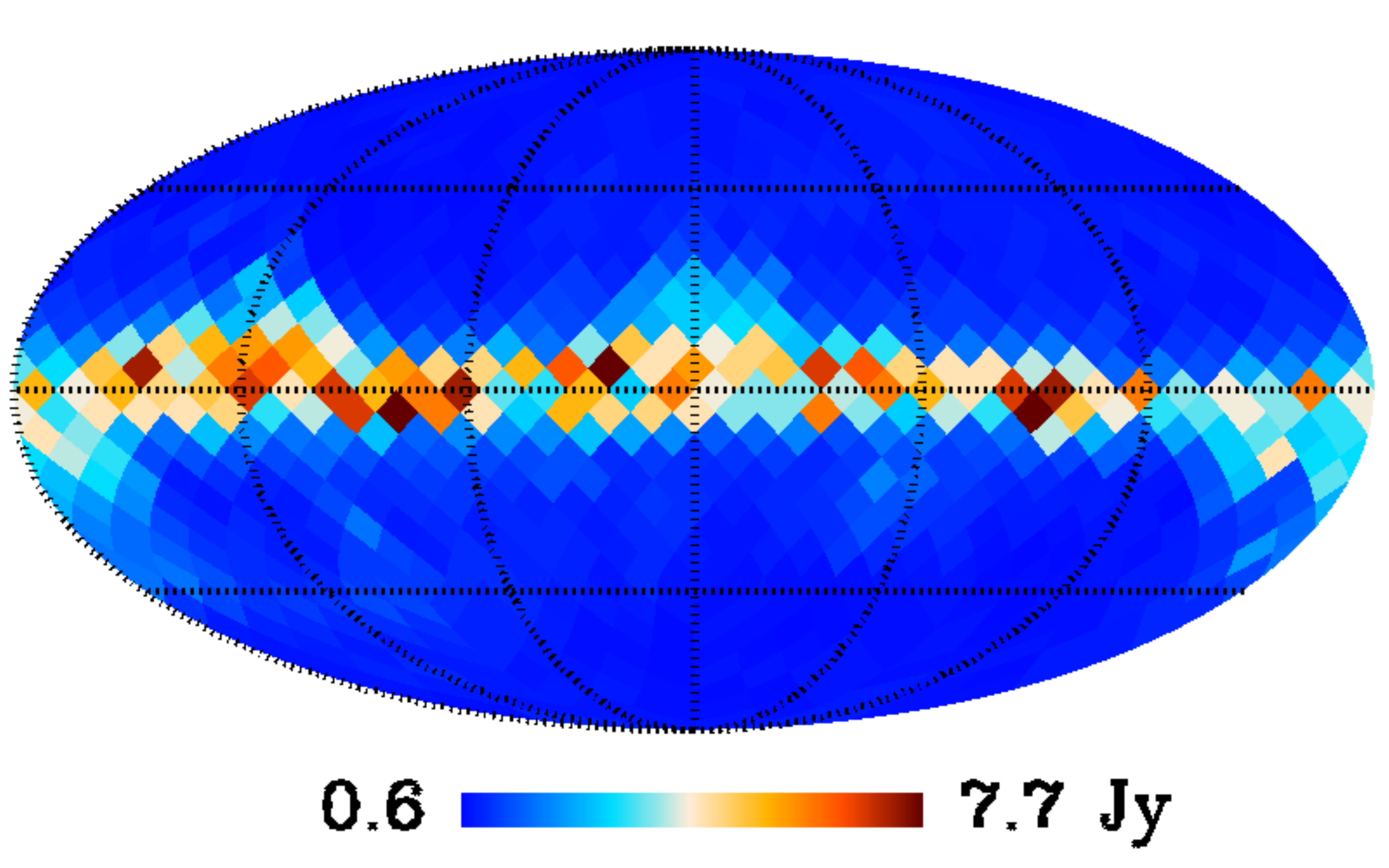} }&
\hspace{-1.5cm} \includegraphics[width=0.33\textwidth]{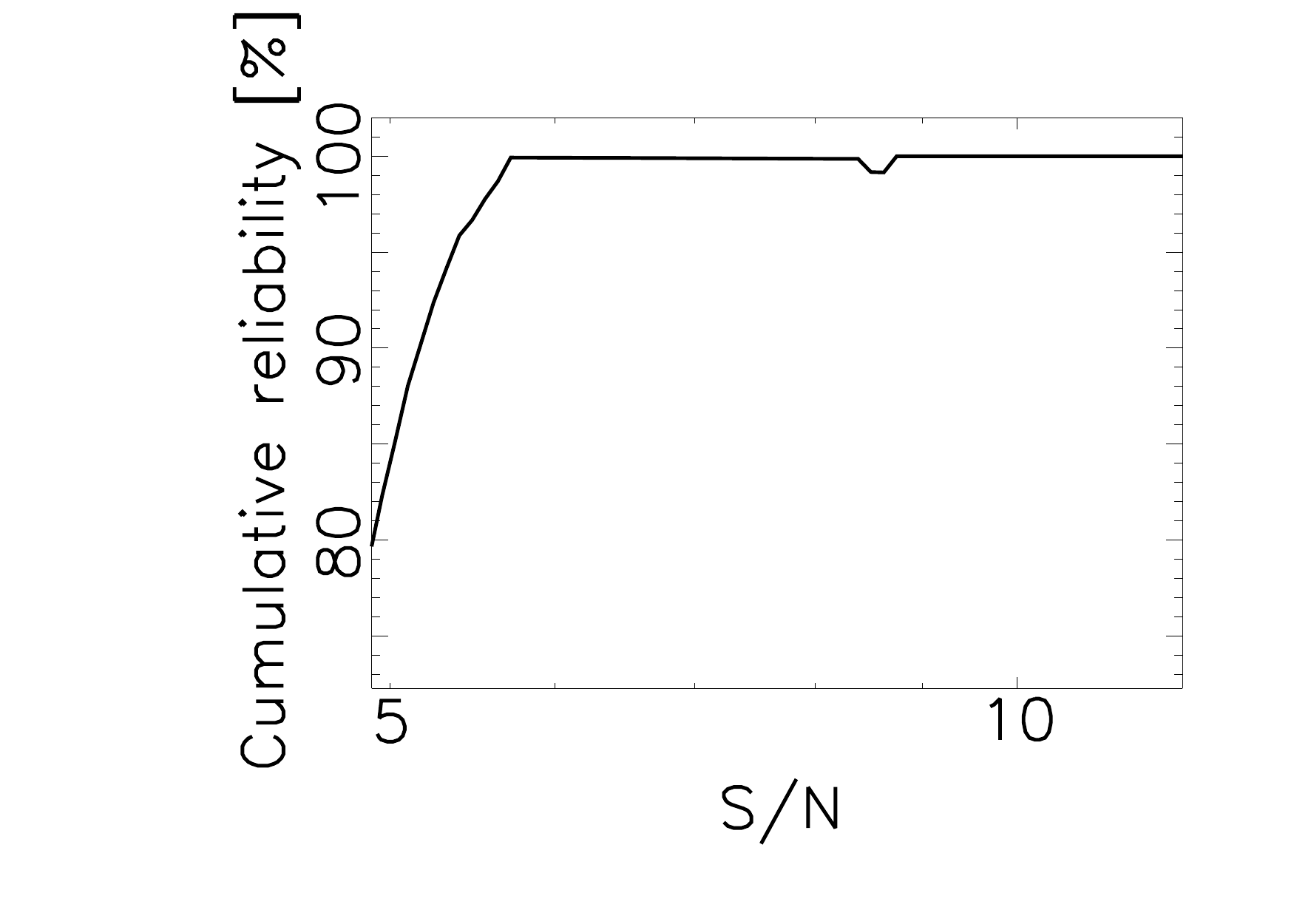}
\end{tabular}

\caption{Results of the internal validation for HFI channels.  The quantities plotted are (\emph{left}) completeness per bin, (\emph{middle}) a map of sensitivity (the 50\,\% completeness threshold in flux density), and (\emph{right}) cumulative reliability as a function of \snr.  The black curves in completeness are for the extragalactic zone described in Sect.~\ref{sec:selection}.  The red curves in completeness are for smaller masks used for the reliability estimation (if the extragalactic zone was not used).  See text for discussion of the limitations of the injection reliability estimate used for $353$\,GHz and above.}
\label{fig:qa_real}
\end{center}
\end{figure*}

The cumulative reliability, or fraction of detections above a given \snr\ that match a real source, is determined using the simulation reliability estimate for channels up to and including $217$\,GHz, and the injection reliability estimate at higher frequencies. These are shown in the right column of Fig.~\ref{fig:qa_real}.  For $100$\,GHz and $143$\,GHz, the reliability is calculated using the 85\,\% Galactic dust mask, for $217$\,GHz using the 65\,\% Galactic dust mask, and for the other channels using the 48\,\% extragalactic zone.  Injection reliability cannot accurately resolve the small departures from reliability at \snr\ $>5.8$, due to Poisson noise.  Some bins above this limit show departures from full reliability at greater than $2\,\sigma$ at $545$\,GHz and $857$\,GHz and these are responsible for the exaggerated stepping of the reliability.  These are likely purely statistical and are a limitation of the precision of injection reliability at higher \snr.

\subsubsection{Photometry and Astrometry}
\label{sec:qa_photometry}

\begin{figure*}
\begin{center}
\includegraphics[width=0.45\textwidth]{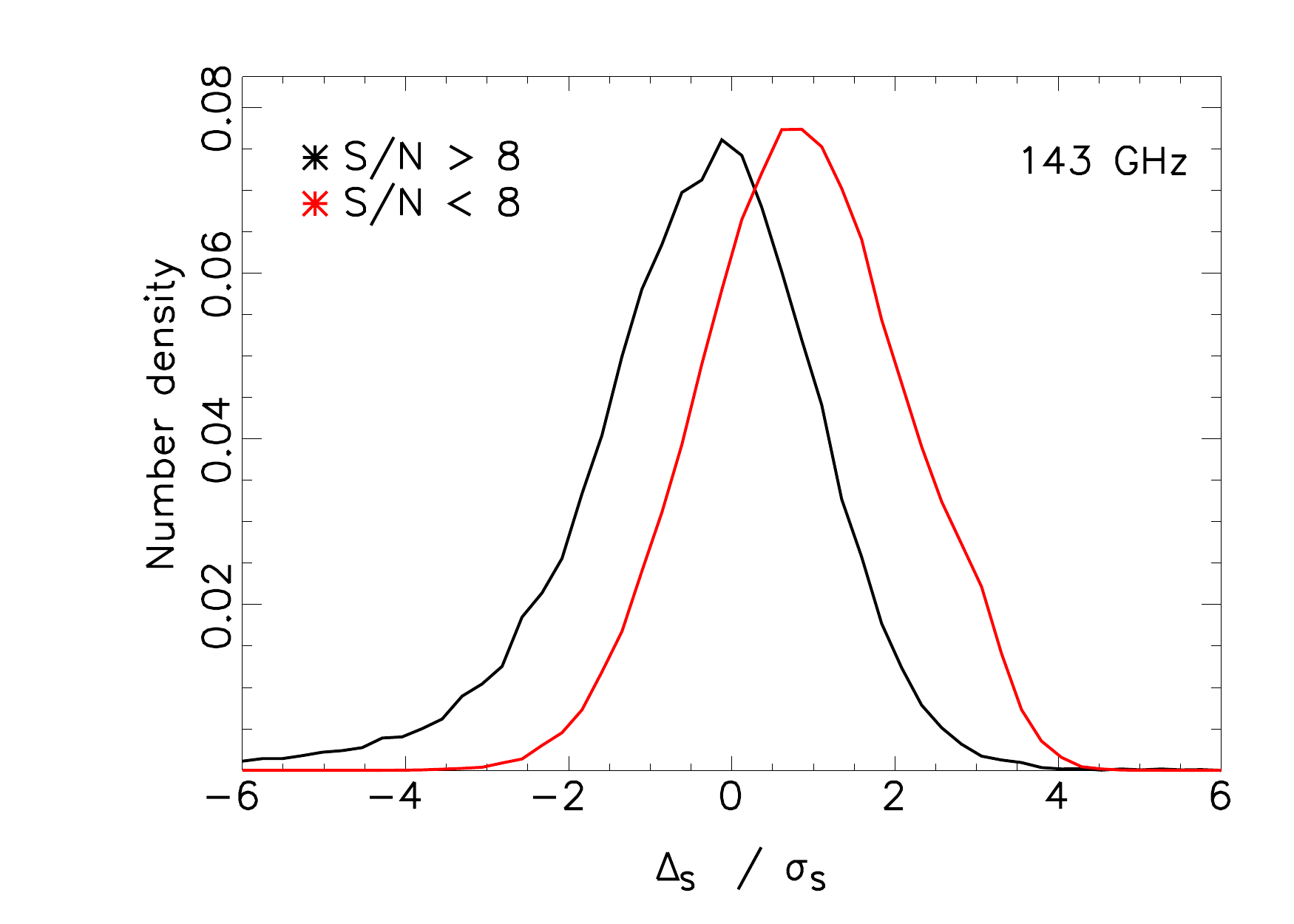}
\includegraphics[width=0.45\textwidth]{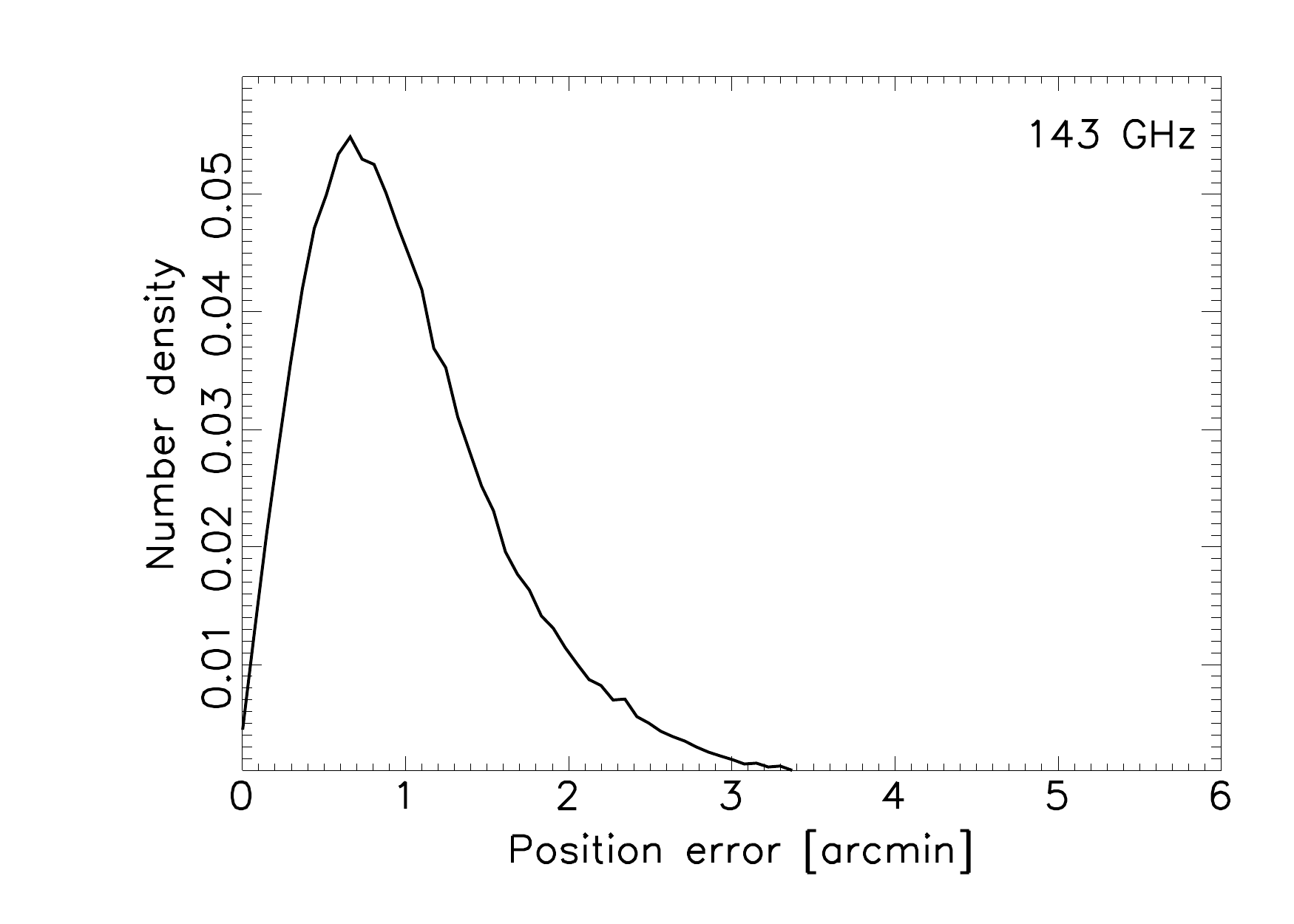} \\
\includegraphics[width=0.45\textwidth]{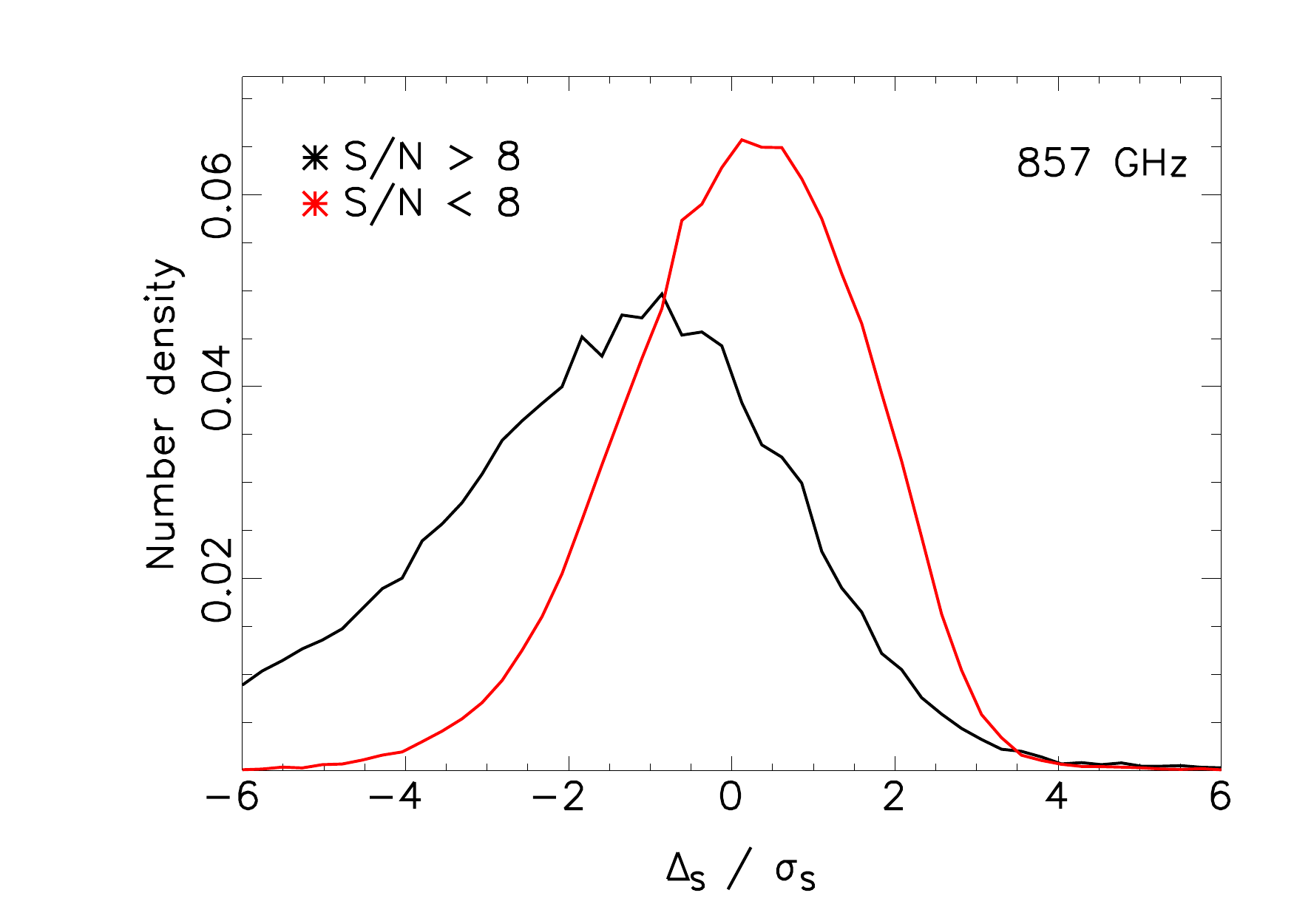}
\includegraphics[width=0.45\textwidth]{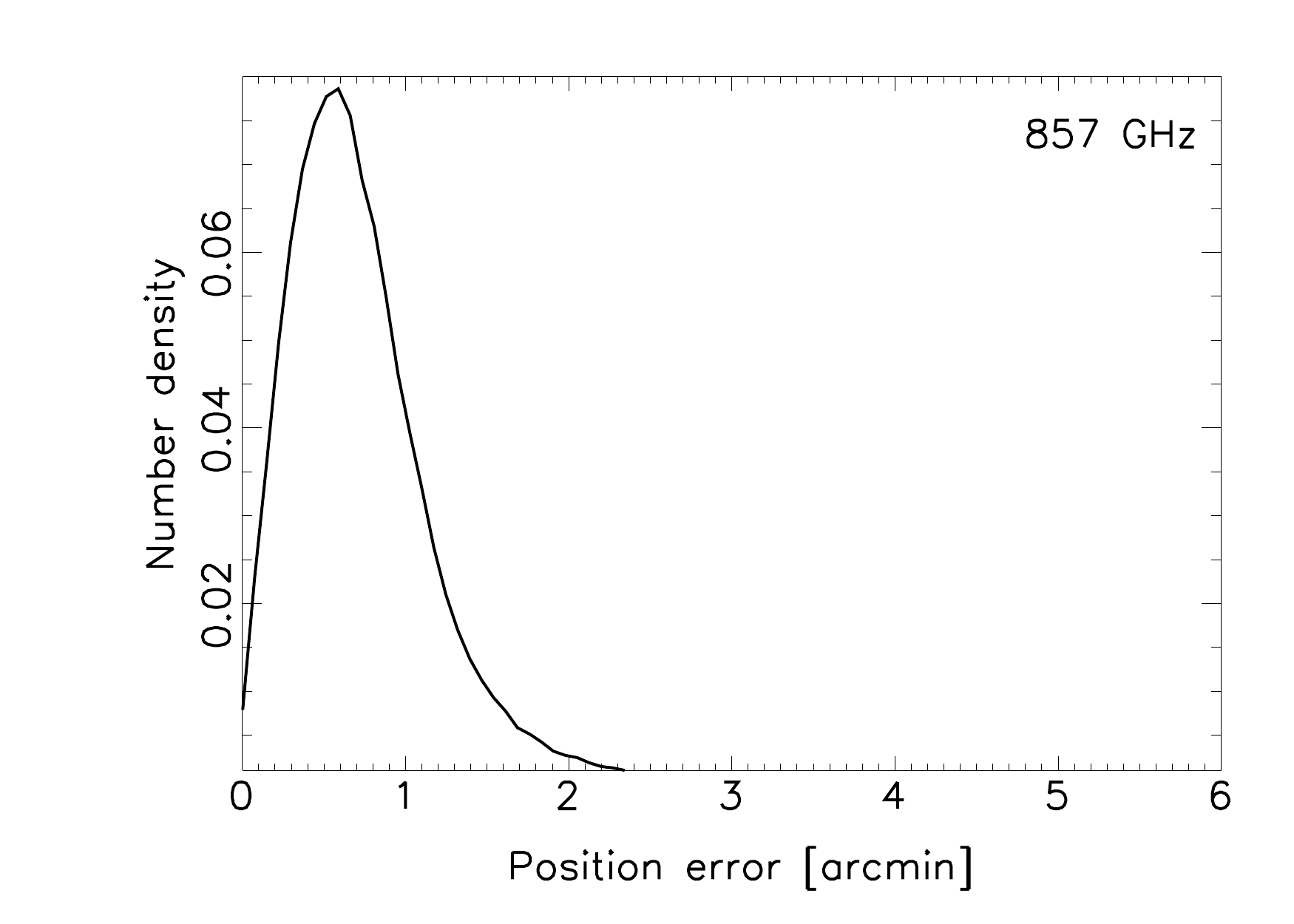} \\
\caption{Example distributions of photometric recovery (\emph{left}) and positional error (\emph{right}) for 143\,GHz (\emph{top row}) and 857\,GHz (\emph{bottom row}).}
\label{fig:phot_pos}
\end{center}
\end{figure*}

For the HFI channels we characterize the accuracy of source photometry by comparing the native flux density estimates (DETFLUX) of matched sources to the known flux densities of sources injected into the real maps. Examples of the distributions of photometric errors,  for $143$\,GHz and $857$\,GHz, are shown in Fig.~\ref{fig:phot_pos}, which  presents histograms  of the quantity $\Delta_S / \sigma_S$, where $\Delta_S$ is the difference between the estimated and the injected flux densities, and $\sigma_S$ is the flux density uncertainty.  The photometric accuracy is a function of \snr, with faint detections affected by upward bias due to noise fluctuations.  At lower HFI frequencies, the DETFLUX estimates are unbiased for bright sources.  At higher HFI frequencies, the DETFLUX estimates are biased low.  Table \ref{tab:qa} shows the DETFLUX bias per channel as well as the standard deviation of $\Delta_S / \sigma_S$ (which would be unity for Gaussian noise).

We characterize the accuracy of the astrometry by calculating the radial position offset between the positions of detected sources and the known positions of the sources injected into the real maps. The distribution of the radial offsets is shown in Fig.~\ref{fig:phot_pos} for $143$\,GHz and $857$\,GHz.

\begin{table}
\begingroup
\newdimen\tblskip \tblskip=5pt
\caption{Native photometry (DETFLUX) bias (mean multiplicative), photometric recovery uncertainty, and median radial position uncertainty from the internal validation, all calculated in the extragalactic zone.}
\label{tab:qa}
\nointerlineskip
\vskip -3mm
\footnotesize
\setbox\tablebox=\vbox{
   \newdimen\digitwidth
   \setbox0=\hbox{\rm 0}
   \digitwidth=\wd0
   \catcode`*=\active
   \def*{\kern\digitwidth}
   \newdimen\signwidth
   \setbox0=\hbox{+}
   \signwidth=\wd0
   \catcode`!=\active
   \def!{\kern\signwidth}
\halign to \hsize{\hbox to 0.8in{#\leaderfil}\tabskip=1em plus 0.2em&\hfil#\hfil&\hfil#\hfil&\hfil#\hfil\tabskip=0pt\cr
\noalign{\doubleline}
\omit \hfil Channel\hfil& DETFLUX bias$^{\rm a}$& stdev($\Delta_S/\sigma_S$)& Position error\cr
\omit&&& [arcmin]\cr
\noalign{\vskip 3pt\hrule\vskip 5pt}
100& 1.05& 1.33& 1.20\cr
143& 1.00& 0.95& 0.96\cr
217& 0.97& 2.79& 0.75\cr
353& 0.96& 2.49& 0.73\cr
545& 0.96& 1.97& 0.72\cr
857& 0.92& 7.06& 0.65\cr
\noalign{\vskip 3pt\hrule\vskip 3pt}}}
\endPlancktable
\tablenote {{\rm a}} For \snr\ $> 8$. \par
\endgroup
\end{table}

\subsection{External validation}
\label{sec:ext_val}

\subsubsection{Low frequencies: 30, 44, and 70\,GHz}

At the three lowest \Planck\ frequencies, it is possible to validate the PCCS source identifications, completeness, reliability, positional accuracy, and in some case even flux density accuracy using external data sets, particularly large-area radio surveys. This external validation was undertaken using the following catalogues and surveys: (1) the full-sky NEWPS catalogue, based on \textit{WMAP} maps \citep{caniego07,massardi09}; (2) in the southern hemisphere, the Australia Telescope 20\,GHz survey \citep[AT20G;][]{murphy10}; (3) in the northern hemisphere, where no large-area survey at similar frequencies to AT20G is available, the 8.4\,GHz {\it Combined Radio All-sky Targeted Eight GHz Survey }\citep[CRATES;][]{healey07}. These catalogues have comparable frequency coverage and source density to the PCCS. We also compared the PCCS with the \Planck\ ERCSC: this provides a useful check on the PCCS pipelines, although the ERCSC  is based on a subset of the data used for the PCCS and is not an independent catalogue. As discussed in \cite{planck2011-1.10}, more than 95\,\% of the ERCSC sources had a clear counterpart in external catalogues.
\begin{figure*}
\begin{center}
\includegraphics[width=\columnwidth]{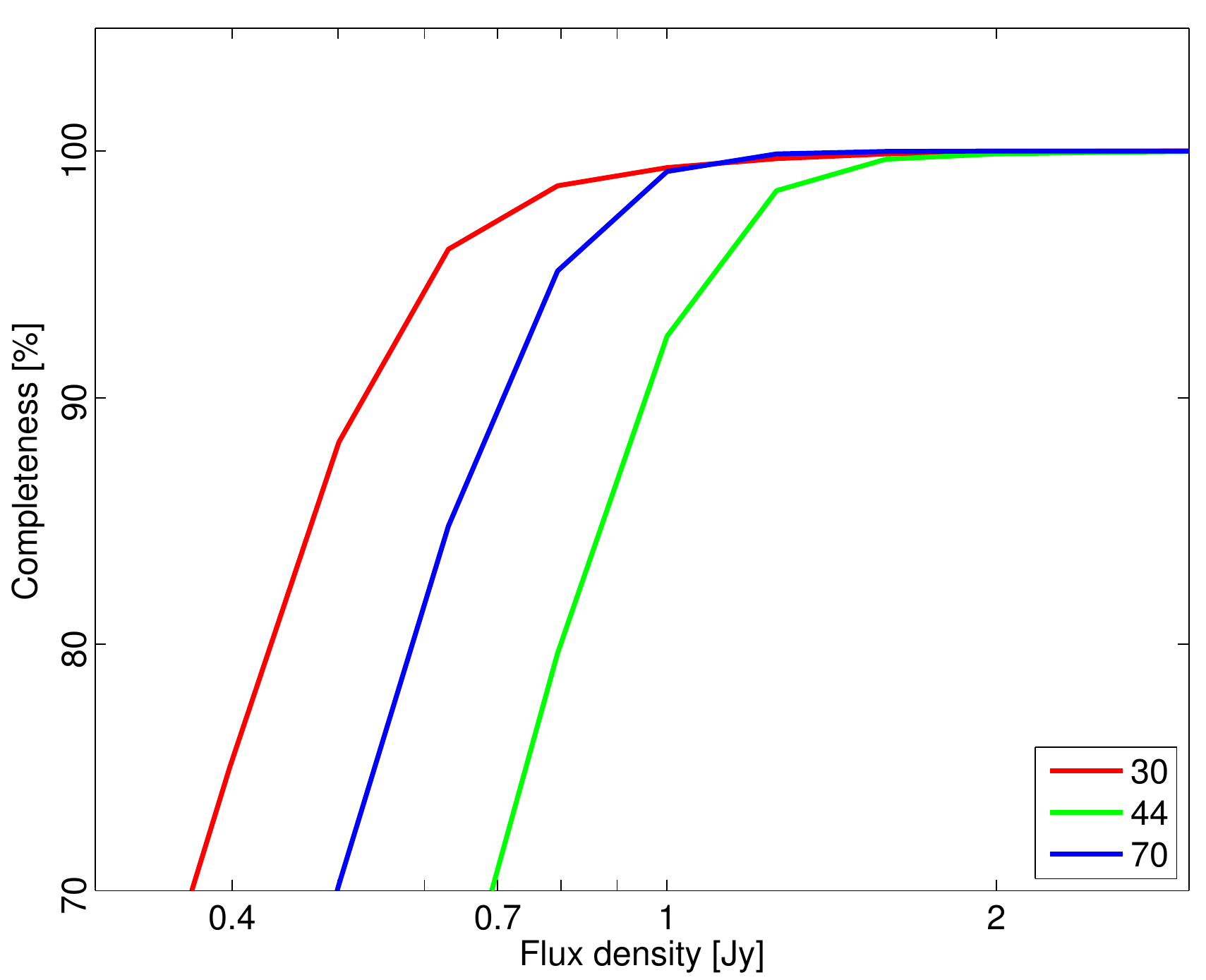}
\includegraphics[width=\columnwidth]{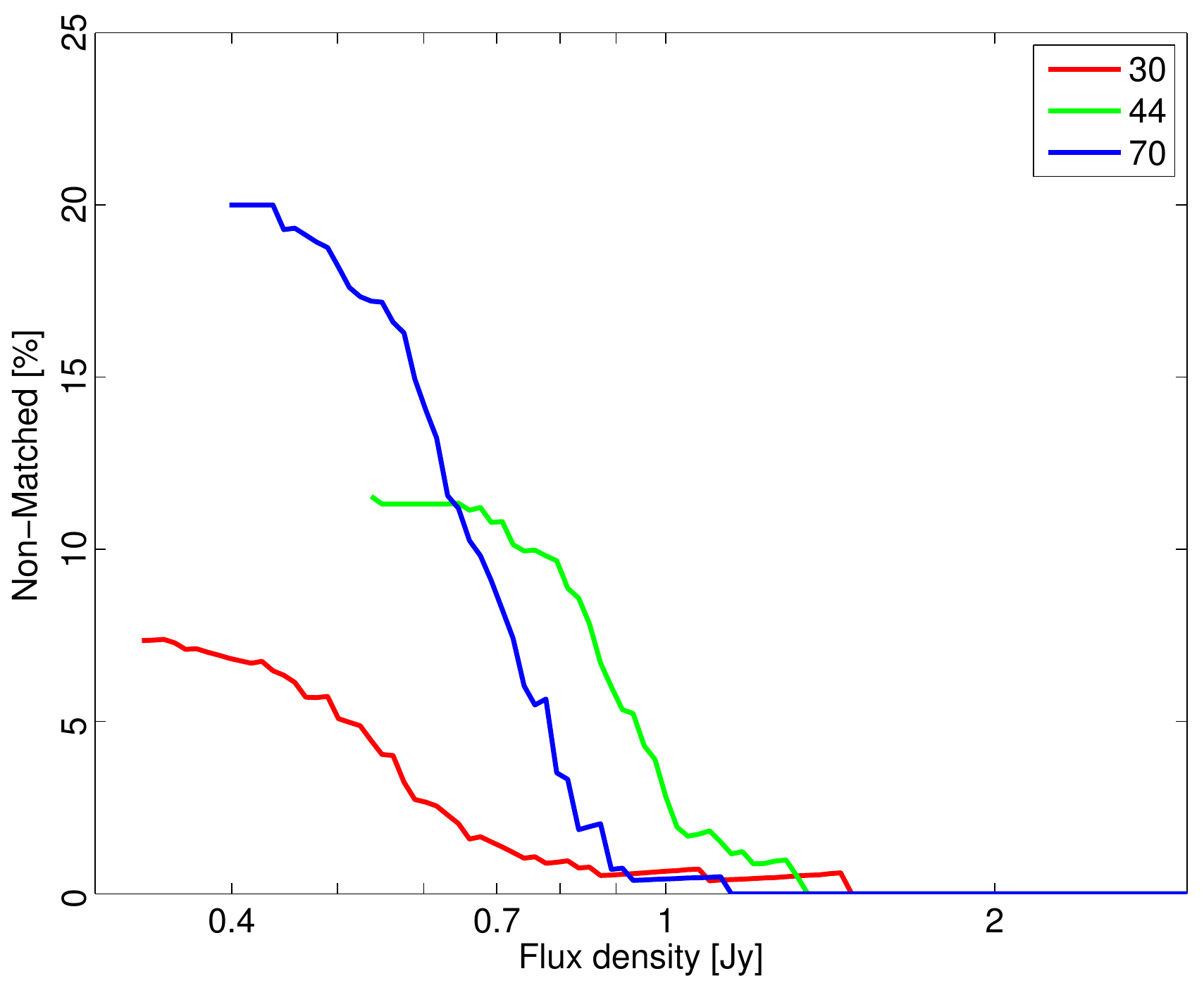}
\caption{External validation summary (completeness and number of non-matched sources) of the 30, 44, and 70\,GHz channels.
\label{fig:qa_lfi}}
\end{center}
\end{figure*}

For this comparison, a PCCS source is considered reliably identified if it falls within a circle of radius 0.65 times the \Planck\ effective beam FWHM (see Table \ref{tab:pccs}) that is centred on a source at the corresponding frequency in one of the above catalogues. Of the four reference catalogues, only the ERCSC covers the Galactic plane and therefore for $|{b}|<2\deg$ (the AT20G Galactic cut) the external validation relies on the previous identification effort performed for the ERCSC \citep{planck2011-6.2}.

Owing to its better sensitivity, the PCCS detects almost all the sources previously found by  \textit{WMAP} \citep{Bennett2013}. Therefore, for studying completeness, deeper samples like the AT20G or CRATES are needed. The problem is that those samples are at lower frequencies (20 and 8.4\,GHz, respectively) than the LFI, so spectral effects or variability could in some cases put the sources below the PCCS detection limit. The completeness values estimated by comparison with these catalogues should thus be considered as lower limits. For this reason we used an alternative completeness estimate that can be derived from knowledge of the noise in the maps.
If the native flux density estimates are subject to Gaussian errors with amplitude given by the noise of the filtered patches, the completeness per patch should be
\begin{equation}
C(S) = \frac{1}{2} + \frac{1}{2} \erf\left(\frac{S - q\sigma(\theta,\phi)}{\sigma_S(\theta,\phi)}\right),
\end{equation}
where $\sigma_{S}^2(\theta,\phi)$ is the variance of the filtered patch located at $(\theta,\phi)$, $q$ is the \snr\ threshold and $\erf(x) = \frac{2}{\sqrt{\pi}}\int_0^x e^{-t^2} dt$ is the standard error function. The true completeness will depart from this limit when the simplifying assumptions of non-Gaussian noise and uniform Gaussian beams are broken. The cumulative completeness is derived by making use of a model of the source counts $N(S)$ \citep{dezotti05}.

Figure~\ref{fig:qa_lfi} shows the estimated completeness and a summary of the external validation results at the LFI channels (30, 44, and 70\,GHz), after combining the information from the four reference radio catalogues. The completeness at 90\,\% level is also given in Table~\ref{tab:pccs}.

The unmatched sources are those detected by \textit{Planck} and appearing in the PCCS, but not present in any of the reference catalogues.  Many of them, however, are internally confirmed, either by a multifrequency detection method for the LFI frequencies or in adjacent HFI channels.  Sources in this category are robust detections, and therefore are probably sources previously undetected at frequencies between 10 and 20\,GHz or by IRAS at higher frequencies. The few remaining ones are likely to be peaks or structures in the distribution of Galactic emission, that may include supernova remnants, reflection nebulae, or planetary nebulae \citep{AMI2012}. They could also be faint thermal sources, or strongly variable ones. They may therefore constitute an interesting sample for follow-up observations. The status of the cross-matching is indicated in the EXT\_VAL column in the PCCS (see Sect.~\ref{sec:usage}).

\begin{figure*}
\begin{center}
\includegraphics[width=\columnwidth]{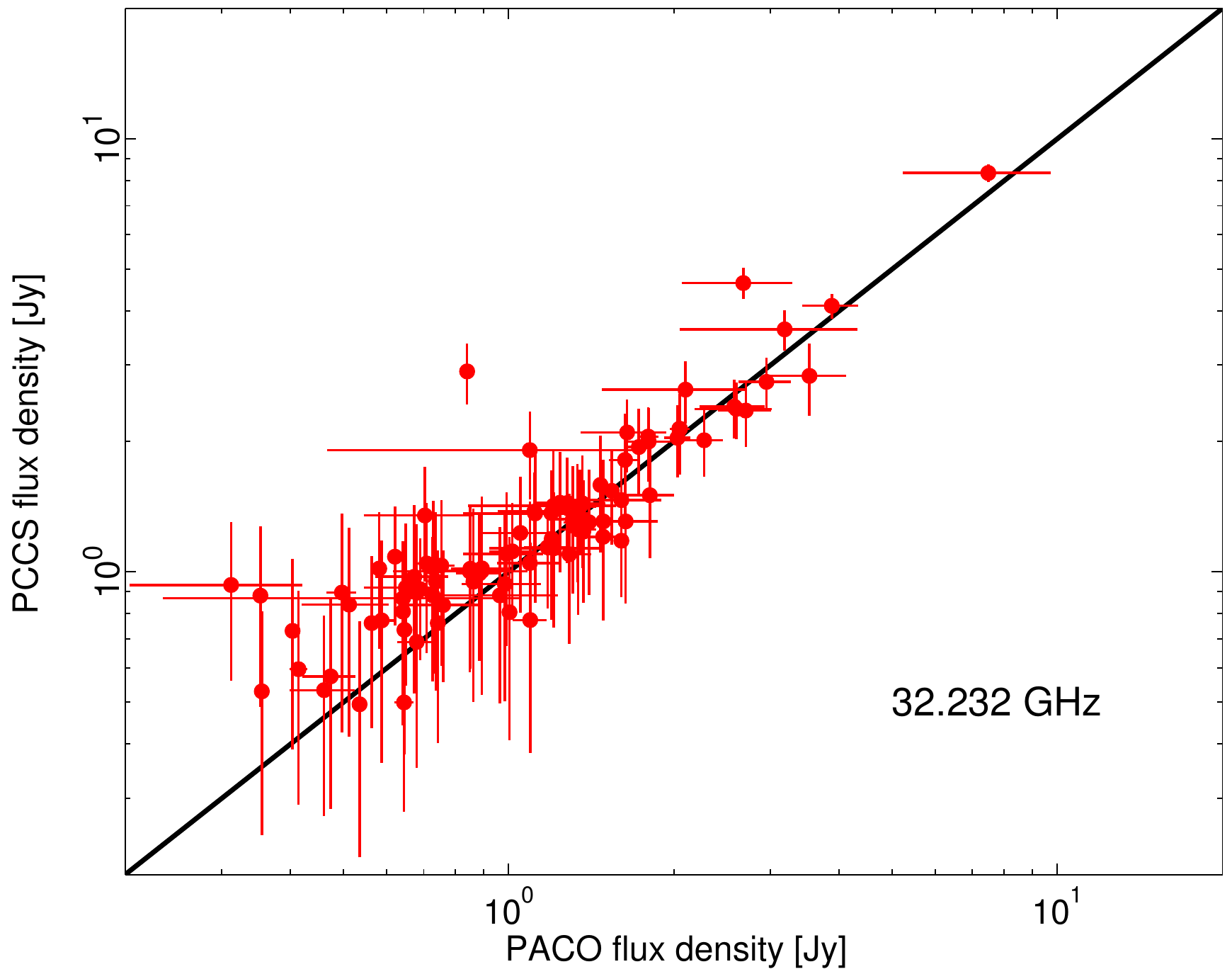}
\includegraphics[width=\columnwidth]{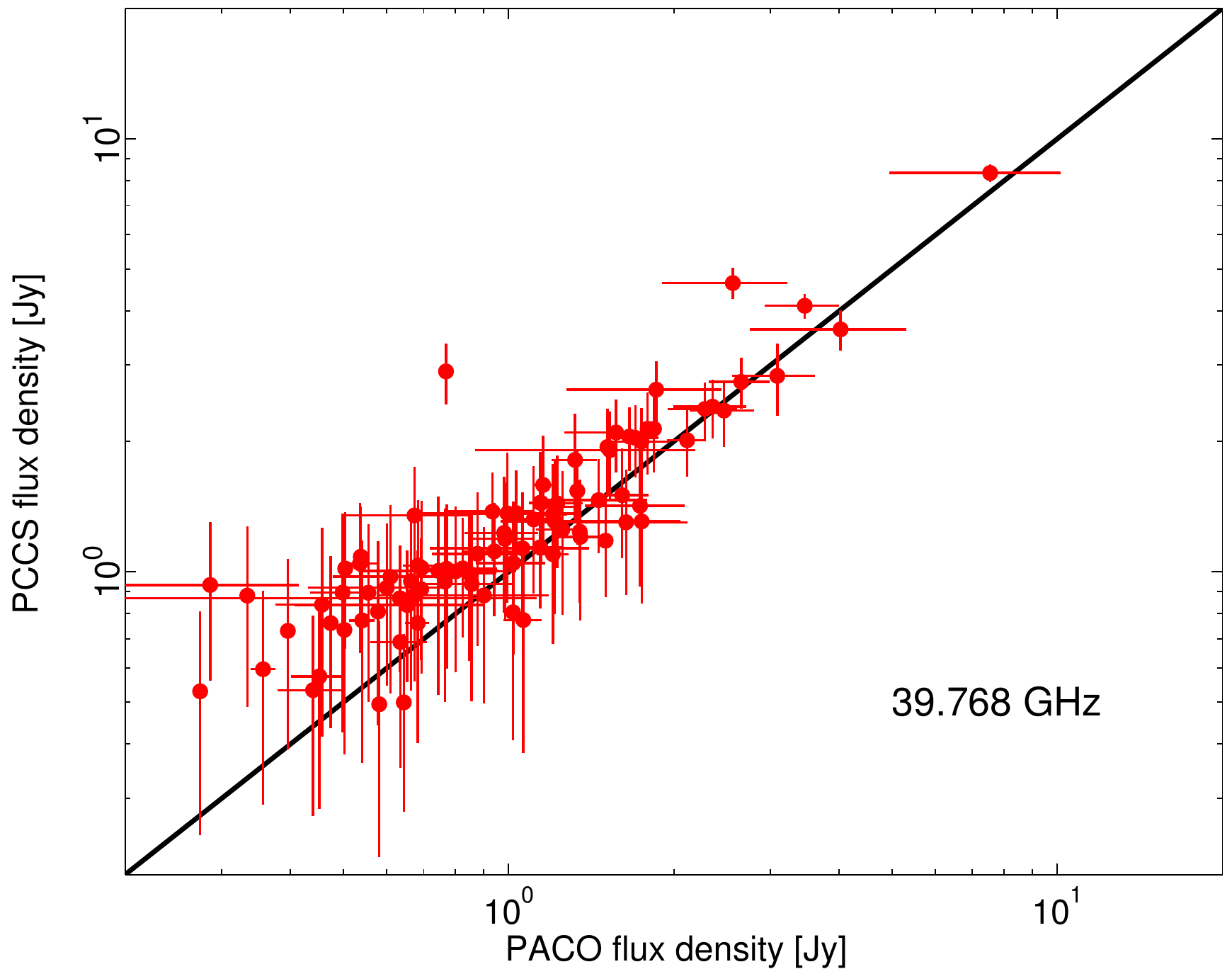}
\caption{Comparison between the PACO sample \citep{massardi11,bonavera11,bonaldi13} and the extrapolated, color-corrected PCCS flux densities (DETFLUX) at $32$ (\emph{left}) and $40$\,GHz (\emph{right}). The multiple PACO observations of each source have been averaged to a single flux density and therefore the uncertainties reflect the variability of the sources instead of the actual flux density accuracy of the measurements (a few mJy).
\label{fig:photo_paco}}
\end{center}
\end{figure*}

An absolute validation of the extracted photometry can be obtained by comparing the PCCS measurements with external data sets. For instance, the \Planck\ Australia Telescope Compact Array (ATCA) Co-eval Observations \citep[PACO;][]{massardi11,bonavera11,bonaldi13} have provided flux density measurements of well-defined samples of AT20G sources at frequencies below and overlapping with \Planck\ frequency bands, obtained almost simultaneously with \Planck\ observations. A total of 482 sources have been observed in the frequency range between 4.5 and 40\,GHz in the period between 2009 July and 2010 August. The multiple PACO observations have been averaged to a single flux density and therefore the uncertainties reflect the variability of the sources instead of the actual flux density accuracy of the measurements (typically, a few millijanskys). The comparison was performed at the PACO frequencies by extrapolating the PCCS flux densities using the spectral indices estimated between 30 and 44\,GHz and taking into account the corresponding colour correction \citep{planck2013-p02,planck2013-p03}. At both 30 and 44\,GHz the two flux density scales appear to be in good overall agreement ($-4\,\%$ $\pm8\,\%$ and 5\,\% $\pm10\,\%$, respectively) with any difference attributable partly to the background effect in the \Planck\ measurements and partly to variability in the radio sources, since the PCCS and PACO measurements were not exactly simultaneous (see Fig.~\ref{fig:photo_paco}). In particular, the PCCS flux densities of the faintest sources are, on average, overestimated due to faint sources that  exceed the detection threshold because they lie on top of positive intensity fluctuations. This effect, known as Eddington bias, can be statistically corrected (\citealt{caniego07}, Appendix B2; \citealt{Crawford10}). Note however that it mostly affects sources below the 90\% completeness limit.

\begin{figure*}
\begin{center}
\includegraphics[width=\columnwidth]{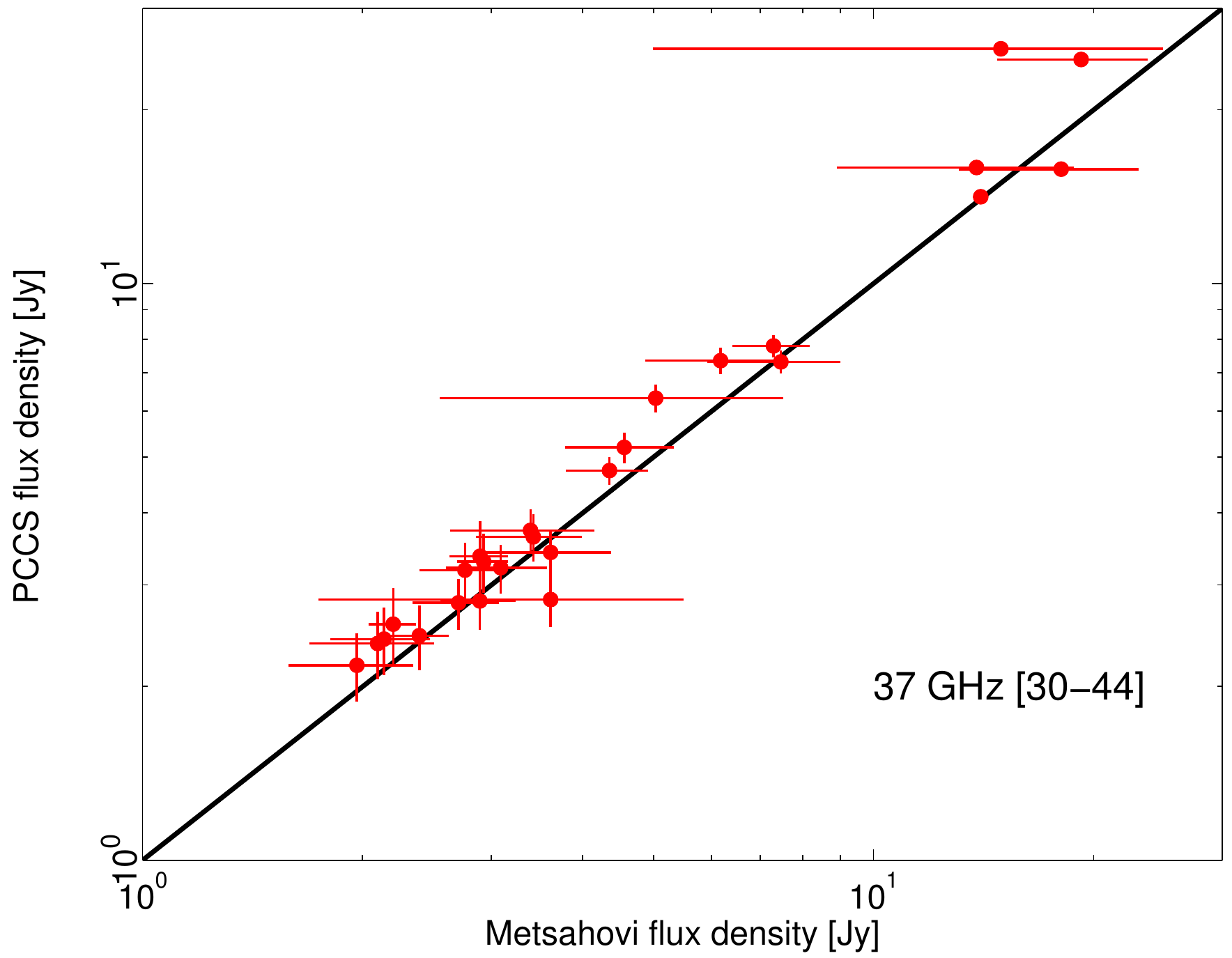}
\includegraphics[width=\columnwidth]{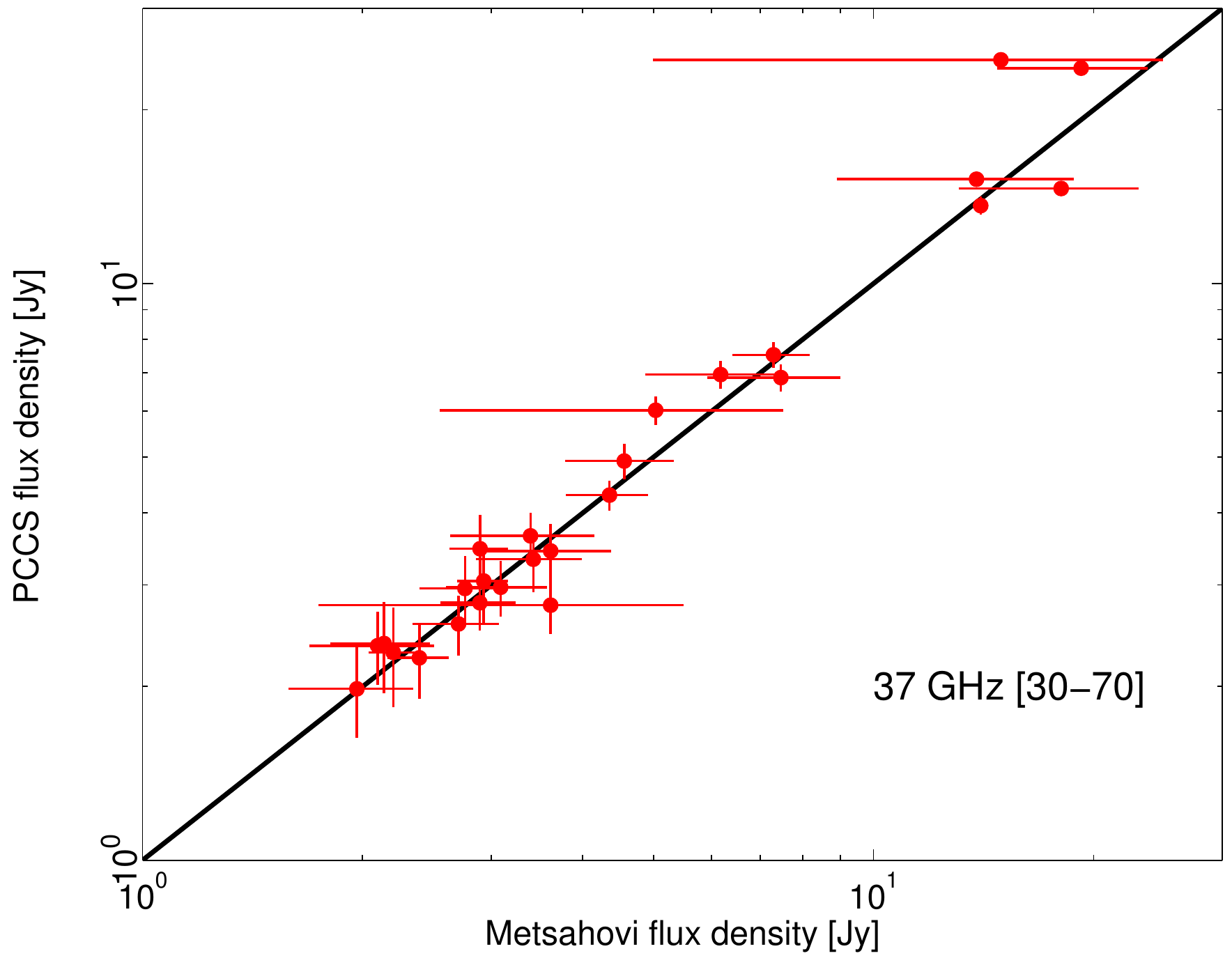}
\caption{Comparison between the Mets{\"a}hovi and the color-corrected PCCS flux densities (DETFLUX) interpolated to 37\,GHz using 30 and 44\,GHz \Planck\ data (\emph{left}) and 30 and 70\,GHz data (\emph{right}). The multiple observations of each source have been averaged to a single flux density and therefore the uncertainties reflect the variability of the sources instead of the actual flux density accuracy of the measurements (a few mJy).
\label{fig:photo_met}}
\end{center}
\end{figure*}

The Mets{\"a}hovi observatory is continuously monitoring bright radio sources in the northern sky \citep{planck2011-6.3a} at 37\,GHz. From their sample, sources brighter than 2\,Jy were selected and their flux densities averaged over the period of \Planck\ observations used for the PCCS. As in the PACO case, the uncertainties in the plot reflect the variability of the sources during the \Planck\ nominal mission period. The \Planck\ measurements were colour-corrected and extrapolated to the Mets{\"a}hovi frequency before the comparison (see Fig.~\ref{fig:photo_met}). The \Planck\ and Mets{\"a}hovi flux densities agree at the 0.2\,\% level with an uncertainty of $\pm4\,\%$.

\begin{figure*}
\begin{center}
\includegraphics[width=\columnwidth]{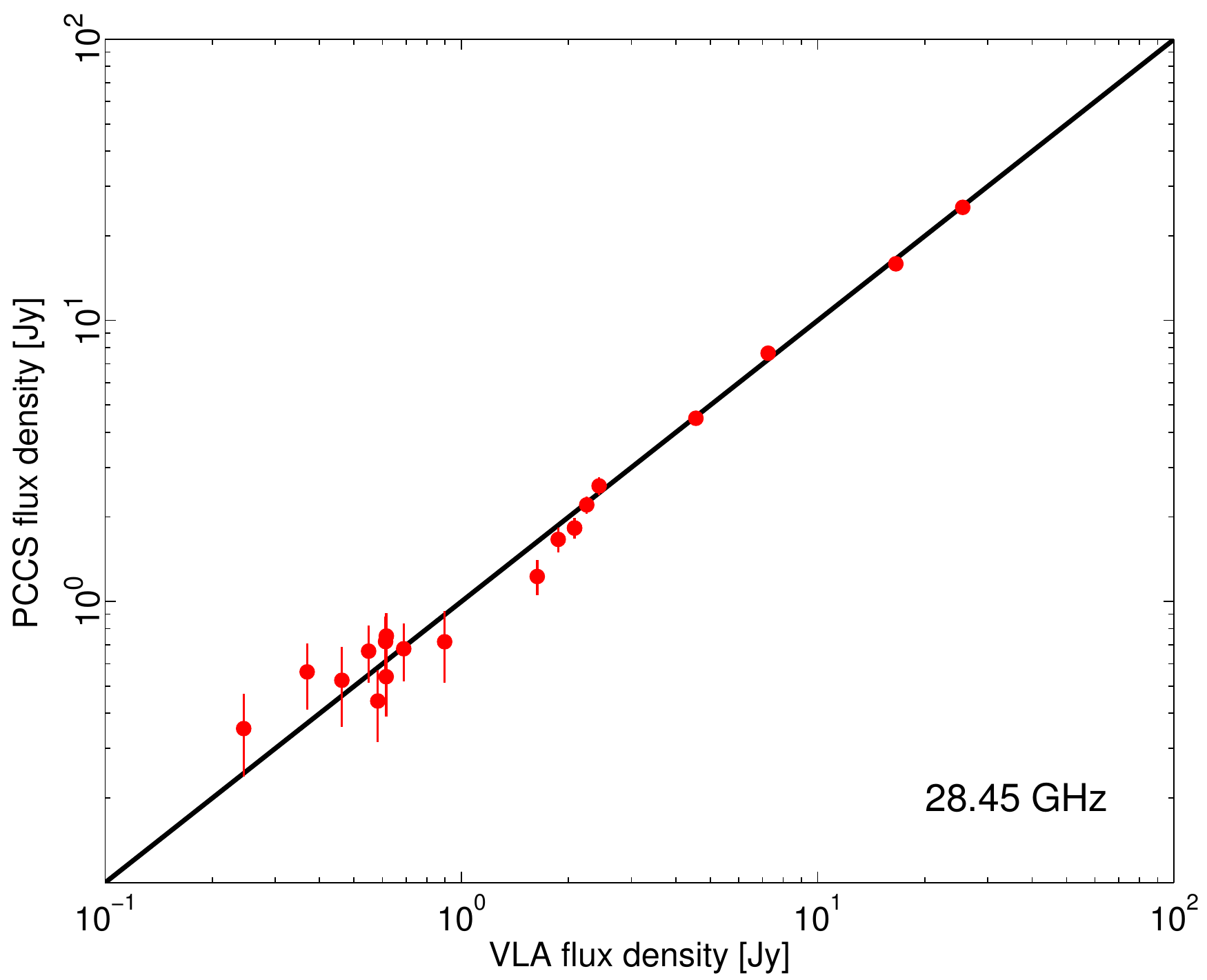}
\includegraphics[width=\columnwidth]{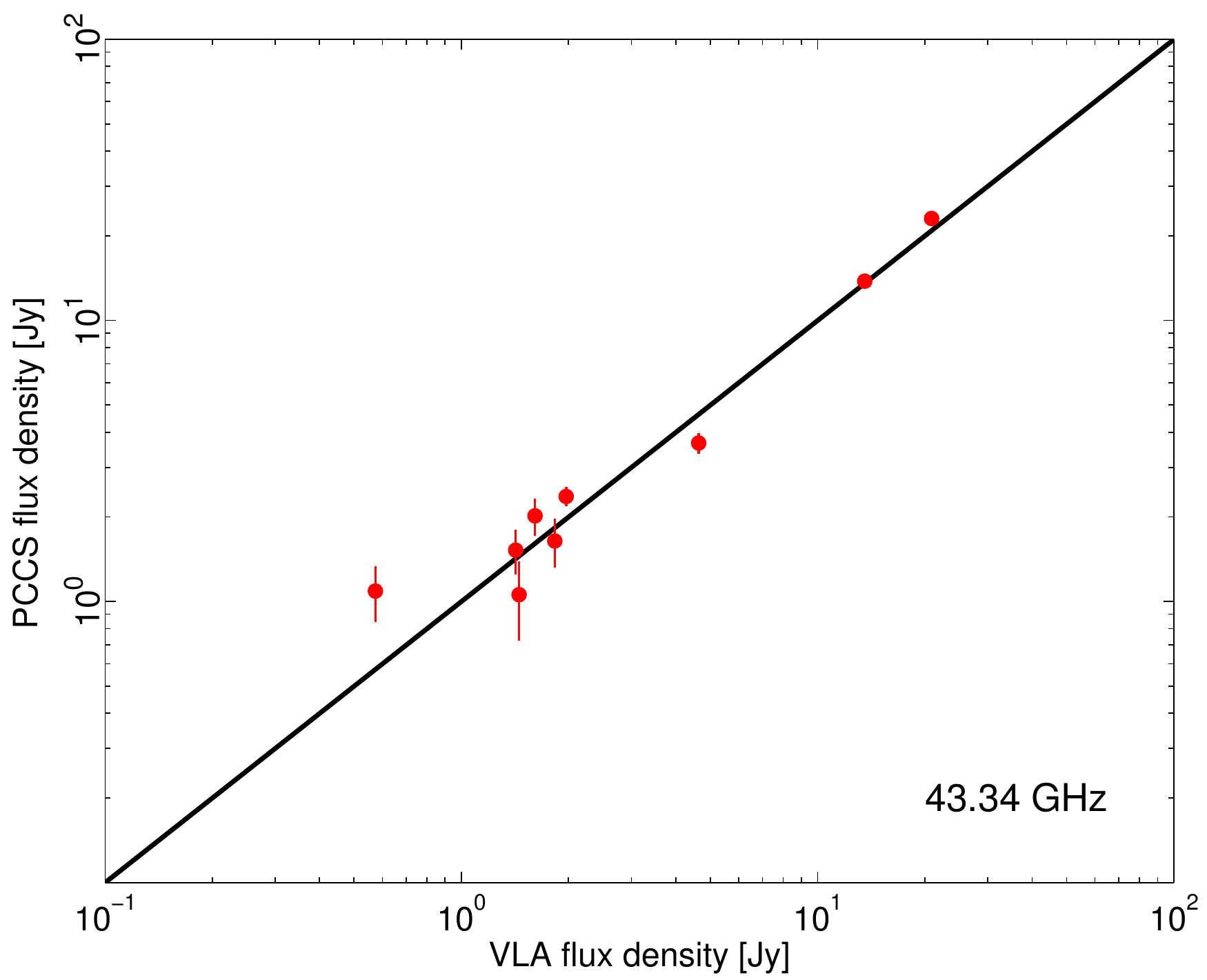}
\caption{\Planck\ flux densities for bright sources observed within a month of VLA observations at that frequency. \Planck\ values (APERFLUX) were colour-corrected and interpolated to $\sim$\,28\,GHz (\emph{left}) and $\sim$\,43\,GHz (\emph{right}).
\label{fig:photo_JVLA}}
\end{center}
\end{figure*}

On 2012 January 19--20, the Karl G. Jansky Very Large Array (VLA) was employed by Rick Perley of the NRAO staff to make observations of a number of bright, extragalactic radio sources also detected by \Planck\ within a month of that date.  The aim of these coordinated observations was to minimize scatter caused by the variability of bright radio sources, most of them blazars.  The VLA observations were made at a number of frequencies, spanning the two lowest LFI frequencies. \Planck\ data (APERFLUX), colour-corrected and interpolated to the VLA frequencies of 28.45 and 43.34\,GHz, were compared with nearly simultaneous \Planck\ observations (see Fig.~\ref{fig:photo_JVLA} for the 43\,GHz case). To lessen the effect of Eddington-like bias in the \Planck\ data, the fit was forced to pass through $(0,0)$. The slopes of the fitted lines show that the VLA and \Planck\ flux densities agree to about $2\pm1.6$\,\% at 28\,GHz, with \Planck\ slightly low.  At 43\,GHz the agreement is not as good, with \Planck\ PCCS flux densities running $\sim6$\,\% high on average. This value, however, is driven by one source, 3C\,84, known to be variable.  If it is excluded, \Planck\ and VLA flux densities at 44\,GHz agree to $\sim0.5\pm2.5$\,\%. The VLA flux density scale used in this comparison is the new one proposed by \citet{Per13}, based on observations of Mars.


\subsubsection{Intermediate frequencies: 143 and 217\,GHz}

\begin{figure*}
\begin{center}
\includegraphics[width=\columnwidth]{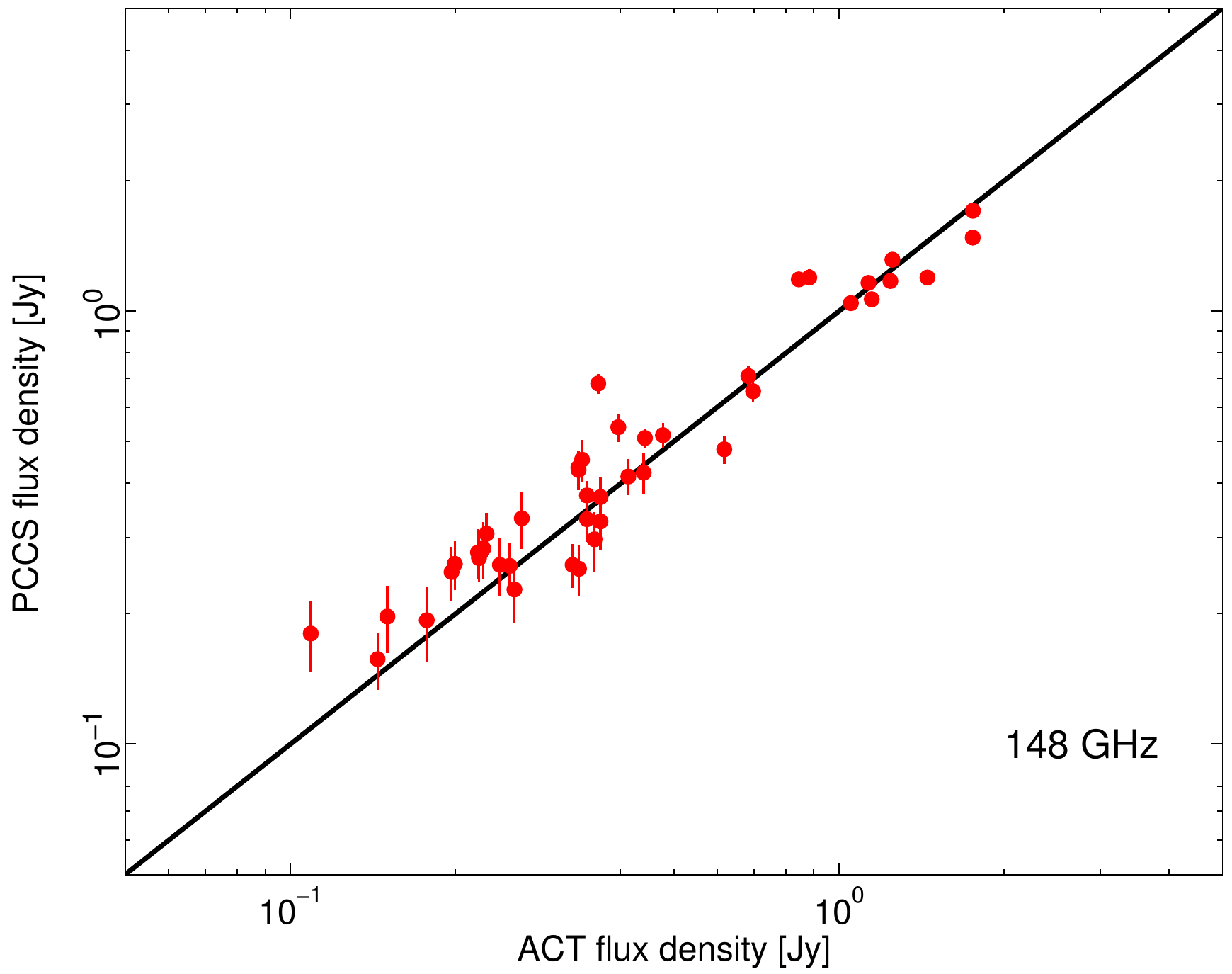}
\includegraphics[width=\columnwidth]{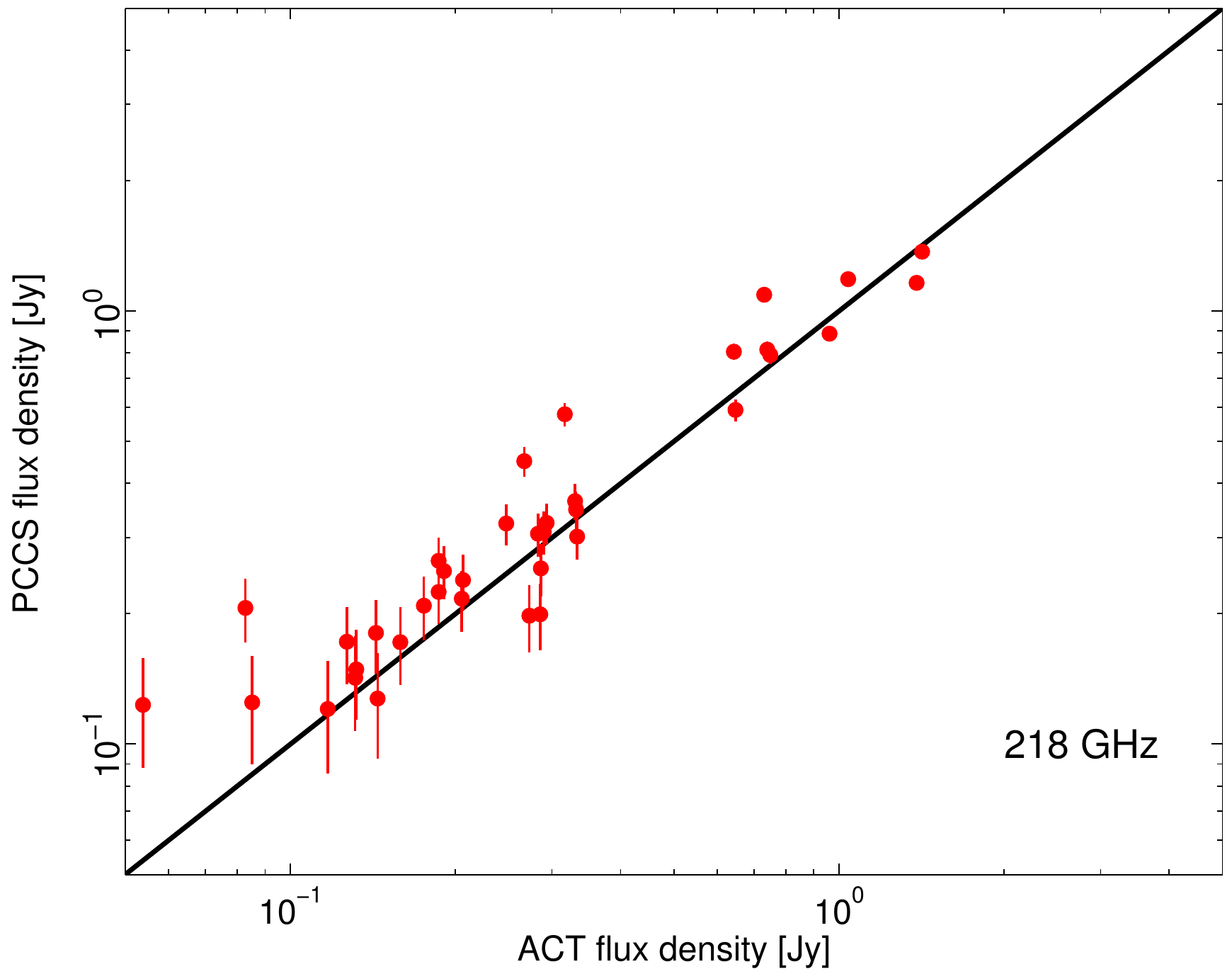}
\caption{Comparison between ACT and \Planck\ measurements (DETFLUX; colour-corrected). \emph{Left panel}: \Planck\ measurements were extrapolated to 148\,GHz. \Planck\ flux densities are on average 1\,\% fainter (or ACT's brighter). The uncertainty in the slope is 0.025 = 2.5\,\%. \emph{Right panel}: \Planck\ measurements were extrapolated to 218\,GHz.The slope is 1.033: \Planck\ flux densities are high (or ACT's low) by $3.3 \pm 3.4$\,\% on average.
\label{fig:photo_ACTa}}
\end{center}
\end{figure*}

A similar comparison was made between \Planck\ flux density measurements of around 40 sources catalogued by the Atacama Cosmology Telescope team (M. Gralla et al., in preparation). \Planck\ 143 and 217\,GHz measurements were colour-corrected and interpolated to match  the band centres of the ACT 148 and 218\,GHz channels. Since the ACT measurements were made over a wider span of time than the \Planck\ ones, source variability introduces a scatter (see Fig.~\ref{fig:photo_ACTa}).  Nevertheless, on average, \Planck\ and ACT observations agree to $1.0 \pm 2.5$\,\% at 148\,GHz, and $\sim 3.0 \pm 3.5$\,\%  at 218\,GHz. If we exclude 2--4 variable sources, the agreement at 218\,GHz improves to $\sim 1.0 \pm 3.5$\,\%.

\subsubsection{High frequencies: 353, 545, and 857\,GHz}
\label{sec:highfreq}

\begin{figure*}
\begin{center}
\includegraphics[angle=90,width=\textwidth]{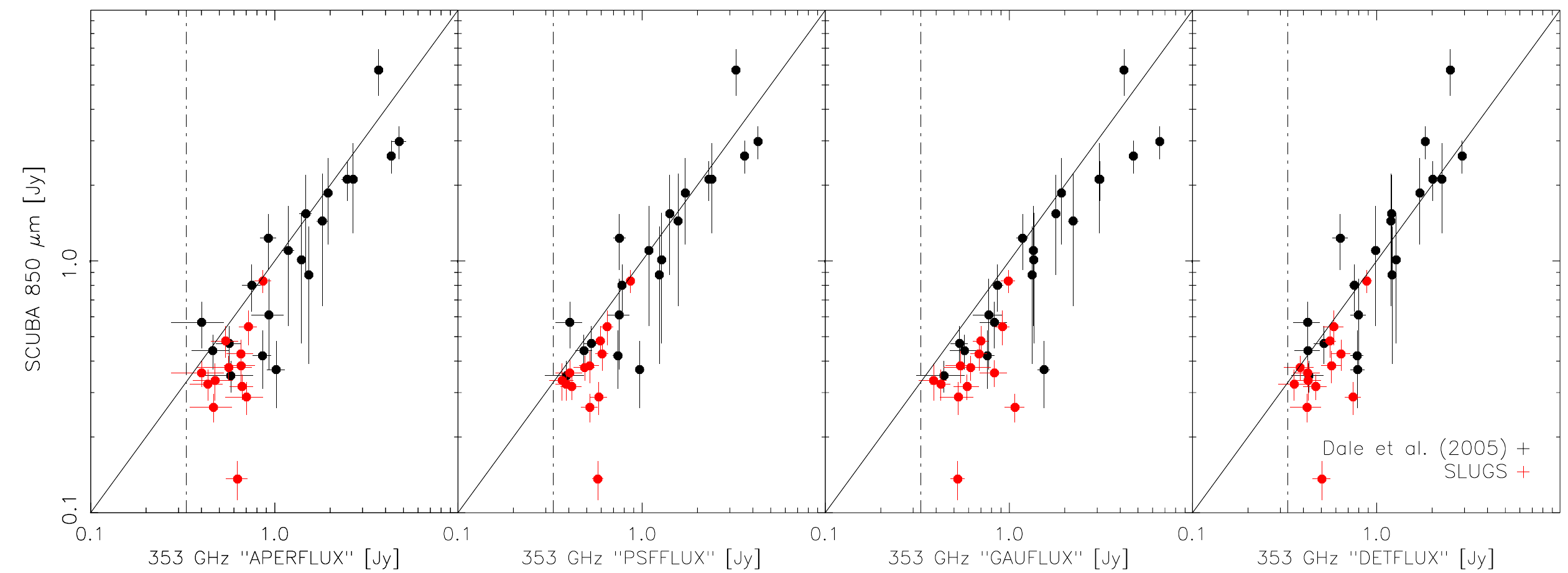}
\caption{Comparison between SCUBA and \Planck\ flux densities at 353\,GHz. All four PCCS flux densities estimates are shown, from left to right, APERFLUX, PSFFLUX, GAUFLUX, and DETFLUX. A colour correction of 0.898 has been applied to the \Planck\ flux densities. The vertical dashed line shows the 90\,\% completeness level of the PCCS. The diagonals show the line of equality between the flux densities.}
\label{fig:photo_scuba}
\end{center}
\end{figure*}

Figure~\ref{fig:photo_scuba} shows a comparison between \Planck\ flux densities at 353\,GHz and those from two SCUBA catalogues (\citealt{dale05}; \citealt{dunne00} [SLUGS]) at 850\,\micron. A colour correction of 0.898 has been applied to the \Planck\ flux densities \citep{planck2013-p03}. The flux densities are in broad agreement between the two catalogues. The uncertainties in SCUBA measurements for extended sources make it difficult to draw strong conclusions about the suitability of the four PCCS flux density estimates.

The \Herschel/SPIRE instrument~\citep{Griffin10} is performing many science programs, among which the wide surveys (extragalactic and Galactic) can be used to cross-check the flux densities of SPIRE and HFI at the common channels: 857\,GHz with 350\,\micron, and 545\,GHz with 500\,\micron. The H-ATLAS survey~\citep{eales10} is of particular interest since many common bright sources (typically with flux densities above a few hundred millijanskys) can be compared.

\begin{figure*}
\begin{center}
\includegraphics[angle=90,width=\textwidth]{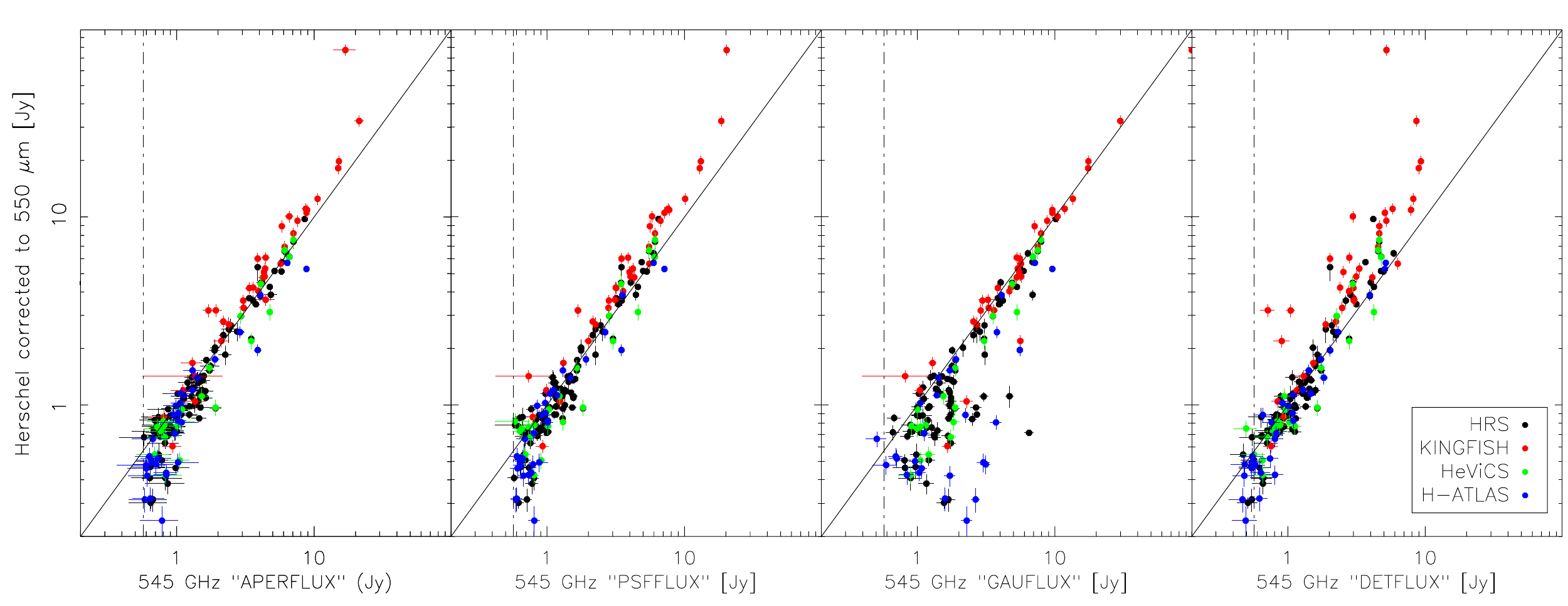}\\
\includegraphics[angle=90,width=\textwidth]{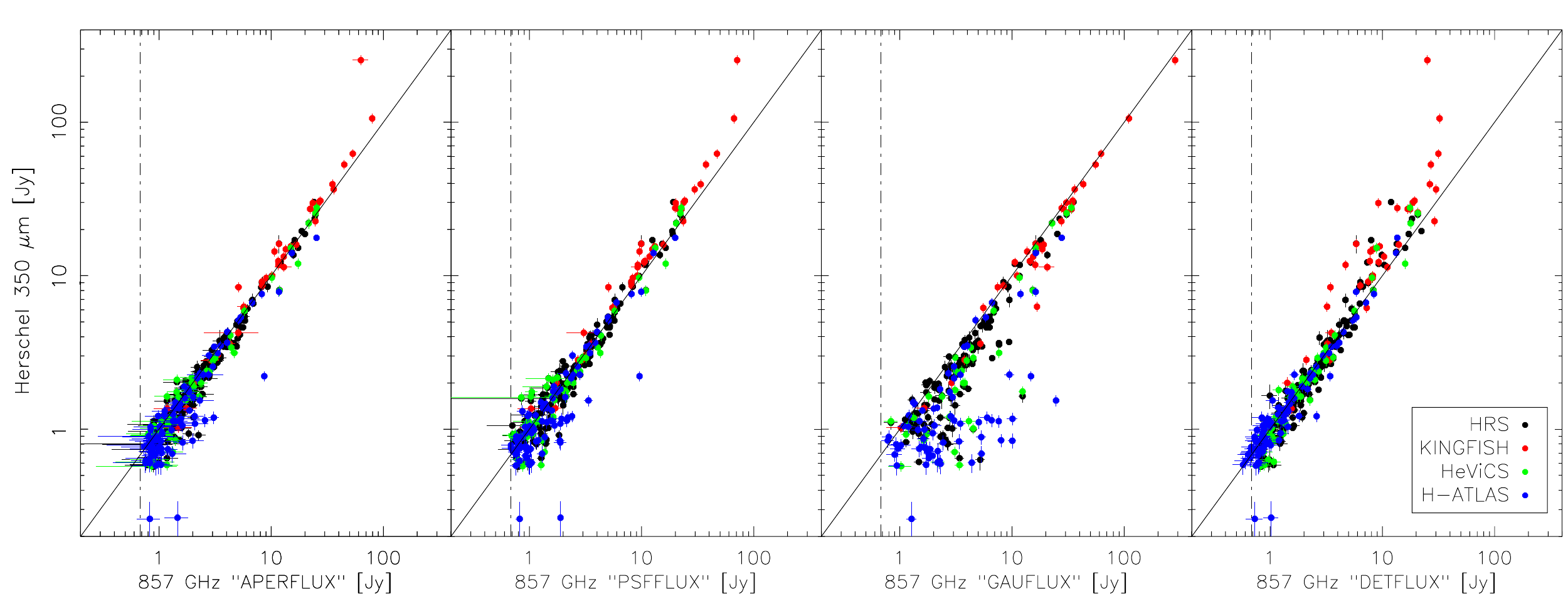}
\end{center}
\caption{Comparison between \Herschel\ and \Planck\ flux densities at 545\,GHz (\emph{top}) and 857\,GHz (\emph{bottom}). All four PCCS flux densities estimates are shown, from left to right, APERFLUX, PSFFLUX, GAUFLUX, and DETFLUX. The \Herschel\ 500\,\micron\ data have been extrapolated to 550\,\micron\ (545\,GHz) assuming a spectral index of 2.7. The vertical dashed line shows the 90\,\% completeness level of the PCCS. The diagonals show the line of equality between the flux densities.}
\label{fig:photo_herschel}
\end{figure*}

Figure~\ref{fig:photo_herschel} shows the comparison between \Planck\ flux densities at 545 and 857\,GHz and four \Herschel\ catalogues, HRS \citep{boselli10}, KINGFISH \citep{kennicut11}, HeViCS \citep{davies12}, and H-ATLAS \citep{eales10}. Inter-calibration offsets between the two instruments were corrected prior to comparison \citep{planck2013-p02b, planck2013-p03f}. To compare with 545\,GHz flux densities, the \Herschel\ 500\,\micron\ data have been extrapolated to 550\,\micron\ (545\,GHz) assuming a spectral index of 2.7, which is the mean value found for the matched objects. At 350\,\micron\ (857\,GHz) no correction has been applied since the \Herschel\ and \Planck\ filters are nearly the same.

At low flux densities, the smallest dispersion is achieved by the DETFLUX photometry because the filtering process removes structure not associated with compact sources. At high flux densities, the brightest objects in the KINGFISH survey are extended galaxies that are resolved by \Planck\, so their flux densities are underestimated by DETFLUX, APERFLUX and PSFFLUX. GAUFLUX accounts for the size of the source and is therefore able to estimate the flux density correctly. For extended sources like these, we recommend the use of GAUFLUX (see also Sect.~\ref{sec:cautionary}).

All these results show that the flux density measurements in the PCCS are in reasonable agreement with those obtained at ground-based observatories or with higher resolution instruments like SCUBA and those of \Herschel. That agreement, in turn, means that the solid angles of \Planck\ beams are understood to comparable accuracy.

\subsection{Comparison between internal and external validation}

To check the consistency of the two validation processes, we extend the HFI internal validation to 70\,GHz and compare with the results of the external validation.  Simulations were constructed at 70\,GHz as outlined in Sect.~\ref{sec:int_val} and the injected sources were extracted using the HFI--MHW extraction algorithm.  The simulations passed the internal consistency tests discussed in Sect. \ref{sec:int_val}, allowing us to determine the reliability using simulation reliability estimate, as was the case for  100--217\,GHz.

\begin{figure}
\begin{center}
\hspace{-1.5cm} \includegraphics[width=1.1\columnwidth]{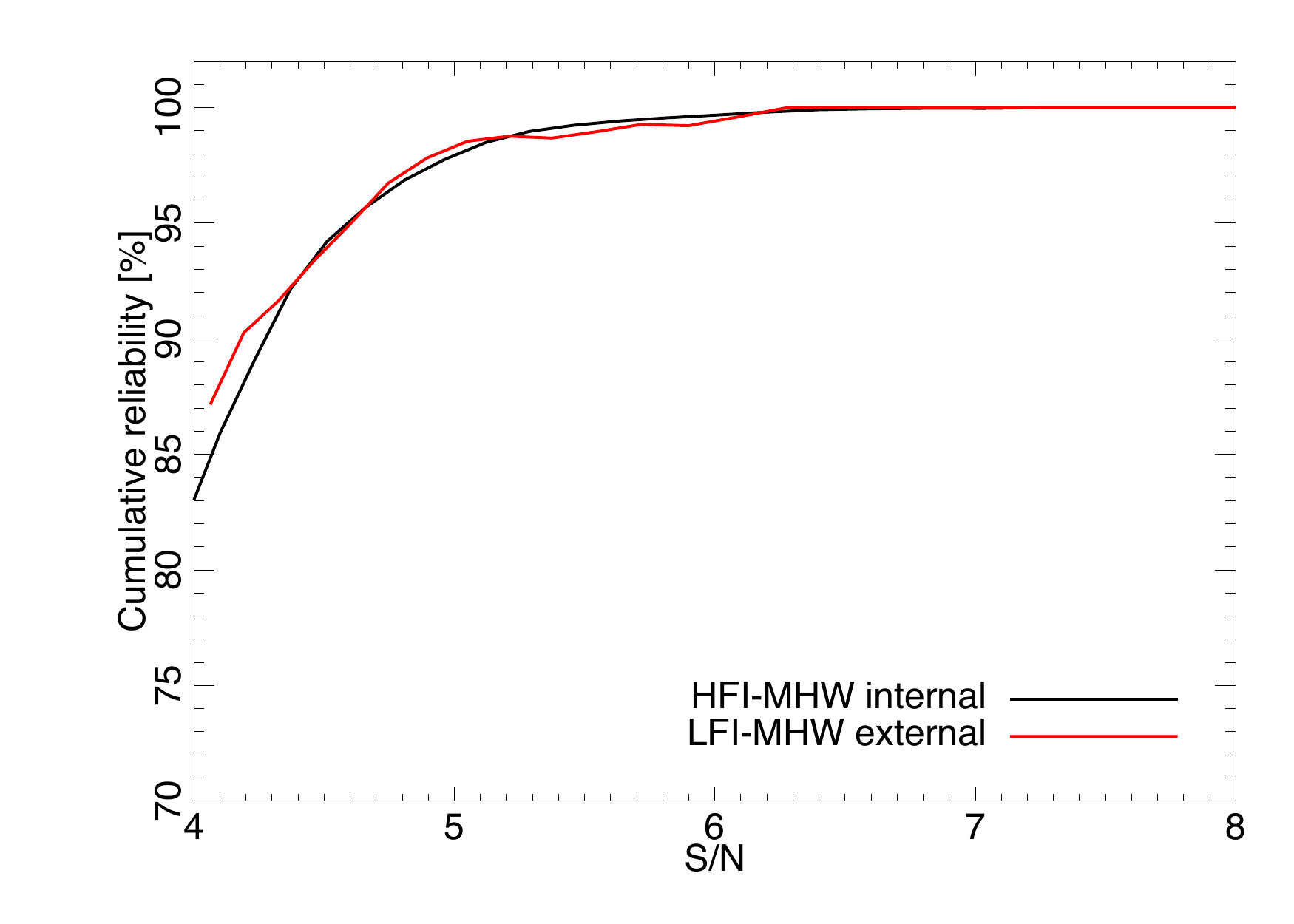} \\
\hspace{-1.5cm} \includegraphics[width=1.1\columnwidth]{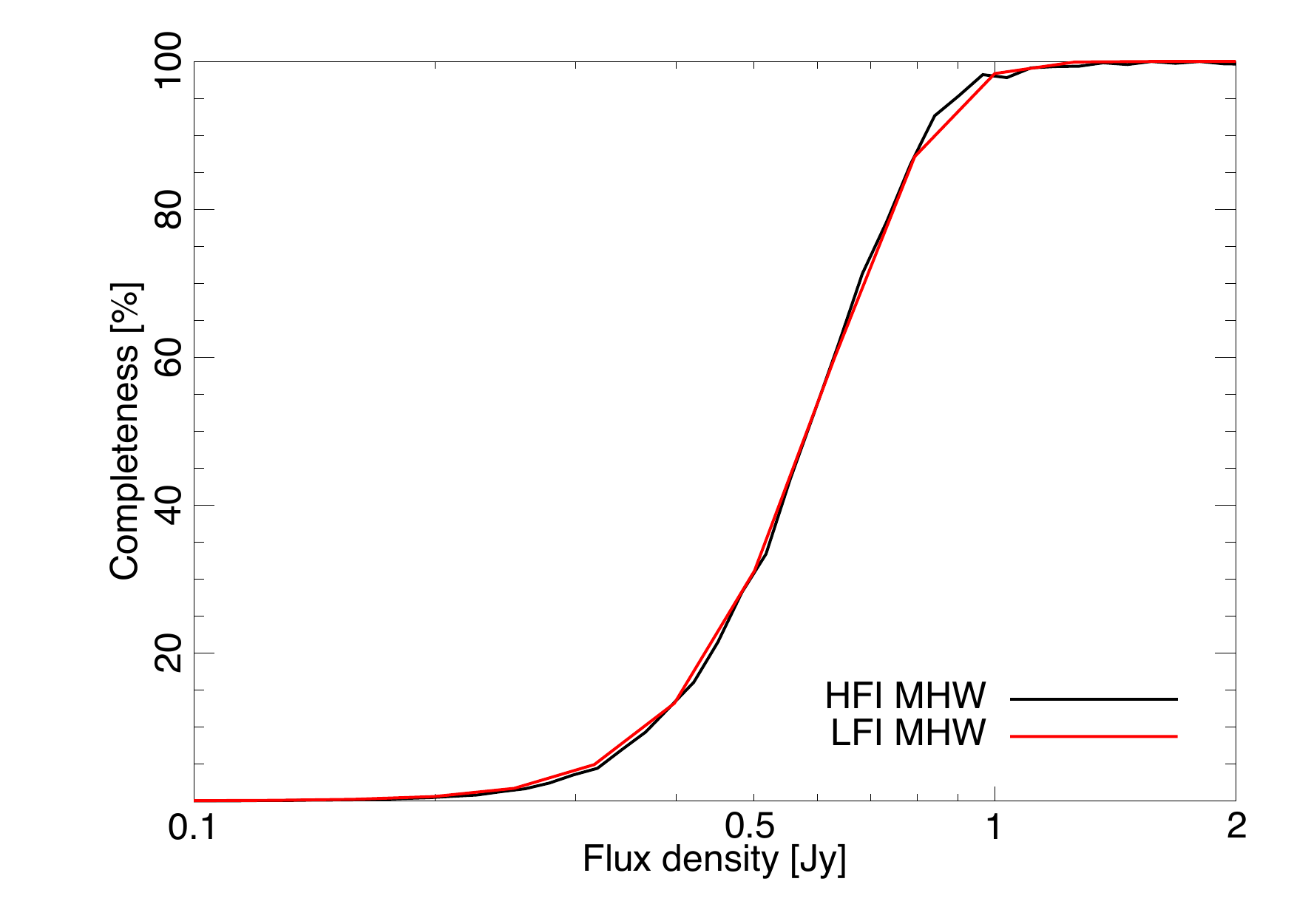}
\caption{Cumulative reliability (\emph{top panel}) and differential completeness (\emph{bottom panel}) of the HFI--MHW and LFI--MHW catalogues at 70\,GHz as determined by their respective internal and external validation procedures. }
\label{fig:valid_compare}
\end{center}
\end{figure}

Figure~\ref{fig:valid_compare} shows the completeness and reliability for the HFI--MHW and LFI--MHW catalogues as estimated using their respective validations at 70\,GHz.  We compare the external validation of the LFI--MHW catalogue with the internal validation of the HFI--MHW catalogue. Both the reliability and the completeness determined from each of the validations are in good agreement.

\subsection{Impact of Galactic cirrus at high frequency}
\label{sec:confusion}

\begin{figure}
\begin{center}
\includegraphics[width=\columnwidth]{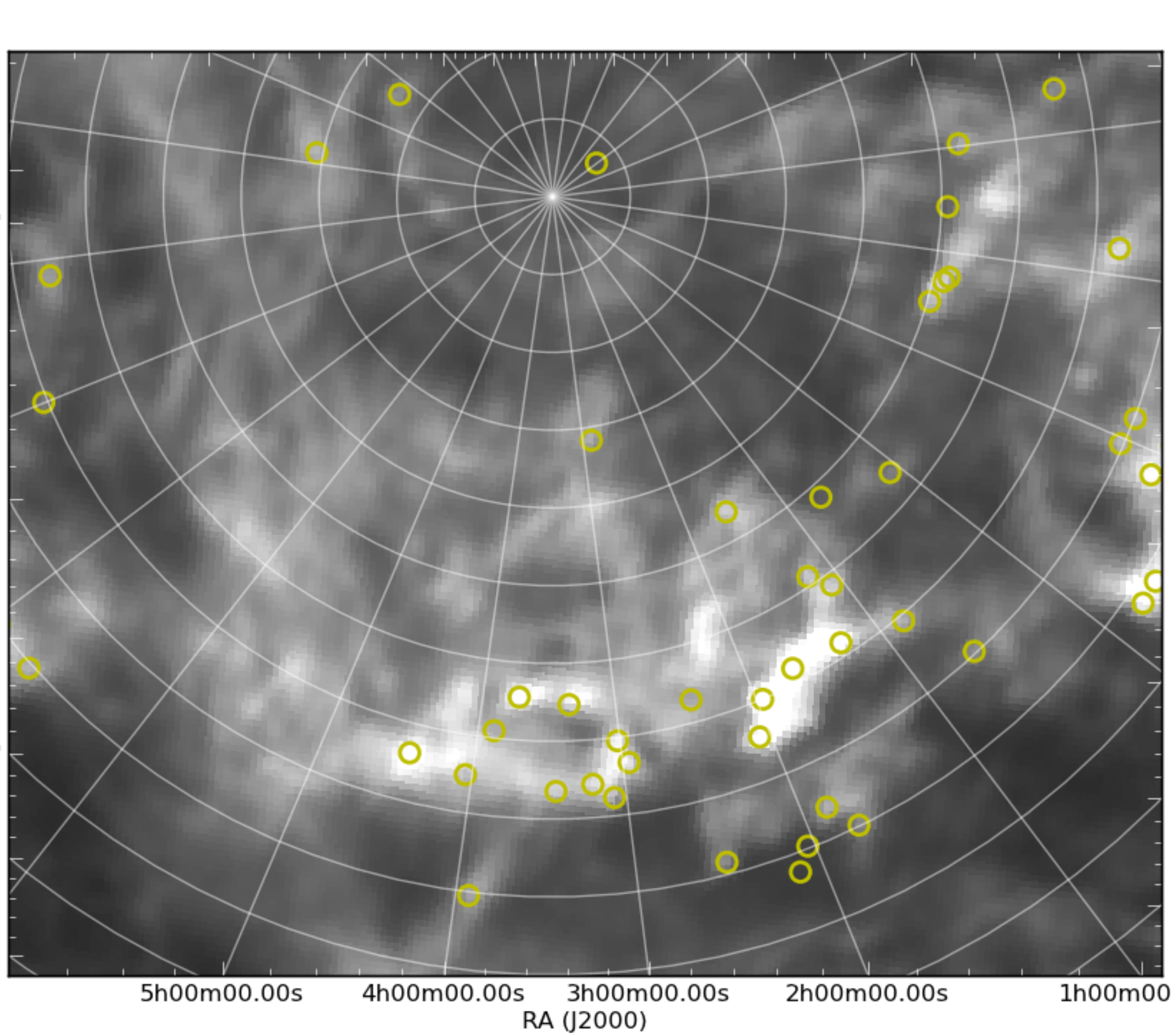} \\ 
\includegraphics[width=\columnwidth]{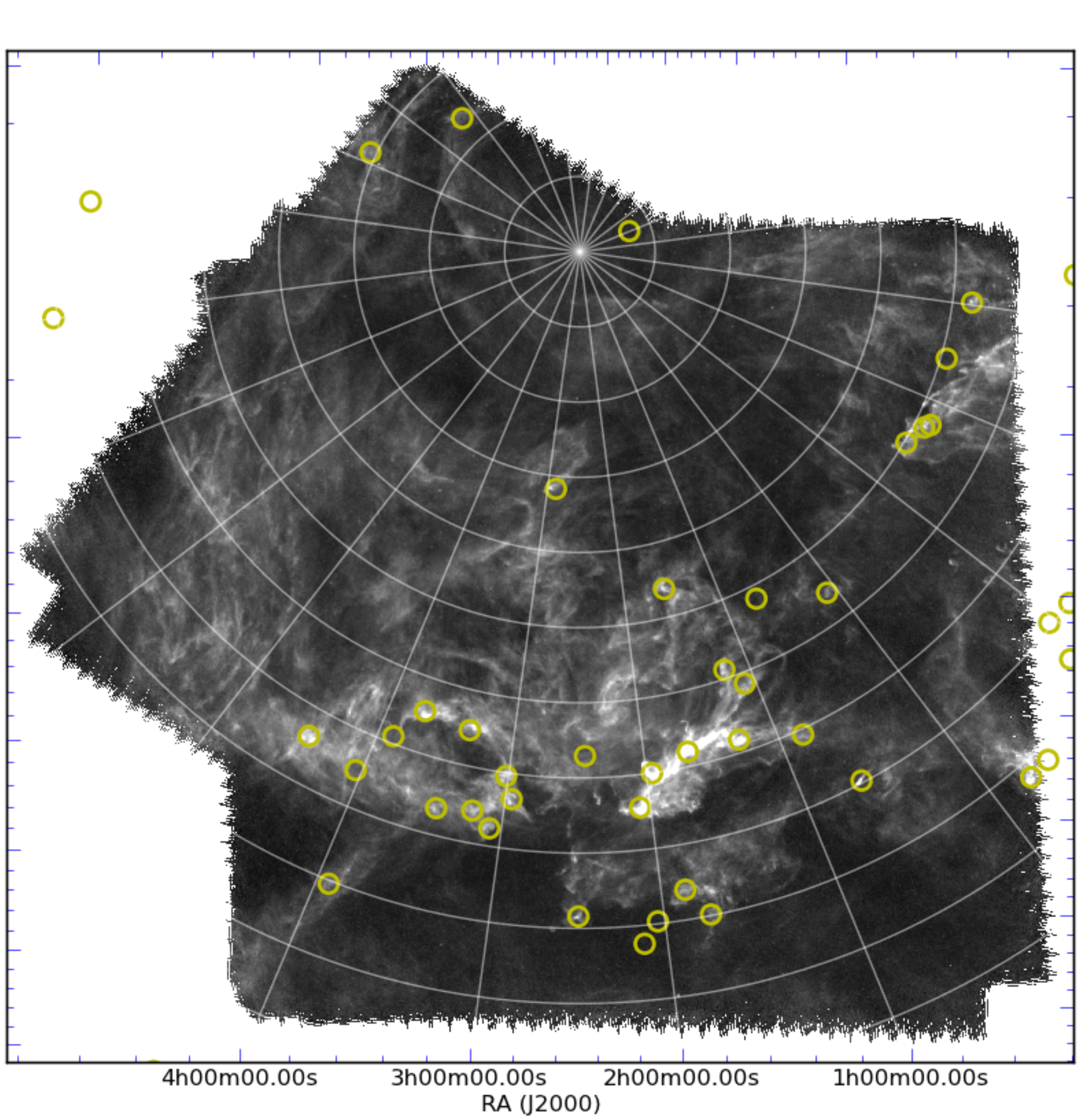} 
\end{center}
\caption{The Polaris field observed by \Planck\ (\emph{top}) and \Herschel\ (\emph{bottom}) at 857\,GHz (350\,\micron). Structures that appear to be compact sources to \Planck, shown with yellow circles, are revealed to be cirrus knots when observed at higher resolution. They are located in regions with bright backgrounds, which provides a proxy for identifying them.  The declination grid has spacing of 30 arc-minutes.}
\label{fig:polaris_field}
\end{figure}

The intensity fluctuations in the \Planck\ high-frequency maps are dominated by faint star-forming galaxies and Galactic cirrus \citep{condon74,hacking87,franceschini89,helou90,toffolatti98,dole03,negrello04,dole06}. The filamentary structure of Galactic cirrus at small angular scales (from a few tens of arcseconds up to a few tens of arcminutes or a degree) is often visible as knots in \Planck\ maps. These sources, which appear compact in  the \Planck\ maps, appear as filamentary structures when viewed by high-resolution instruments such as \Herschel/SPIRE.  An example is the Polaris field (see Fig.~\ref{fig:polaris_field}), where \Herschel\ does not detect sources above the extragalactic density counts, but \Planck\ detects a sharply increasing number of sources with lower interstellar brightness that are coincident with filaments.

Using the few \Herschel\ fields available, we are able to establish a statistical evaluation of how the spurious source density behaves. We consider real sources to be compact structures that are not part of the interstellar quasi-stationary turbulent cascade. The other apparent sources are artefacts of the detection algorithms on the general interstellar structure and depend strongly on the angular resolution used. They are not useful as sources in a catalogue.

To control the spurious detections induced by the cirrus filaments, we apply higher \snr\ thresholds in the Galactic zone for 353, 545, and 857\,GHz (see Table \ref{tab:pccs}).  These thresholds remove the bulk of the spurious sources identified in the \Herschel\ SPIRE fields in this zone, while preserving the majority of the extragalactic compact objects.

For the extragalactic zone, we note that there is a local threshold in brightness that we estimate to be approximately 3--5\,\MJysr\ at 857\,GHz, above which the probability of cirrus-induced spurious detections increases.  This is not used to threshold the catalogue, but could be used as a further control of spurious detections.

In some areas the situation is more complicated.  \Herschel\ does detect ``real'' Galactic (protostellar) sources in filaments in brighter regions \citep[like the Aquila rift,][]{andre10}. These sources often are not fully unresolved but are embedded in an envelope and the filamentary structure. These sources usually lie in sky regions of much higher brightnesses, and are located within the Galactic zone.

\begin{figure}
\begin{center}
\includegraphics[angle=90,width=\columnwidth]{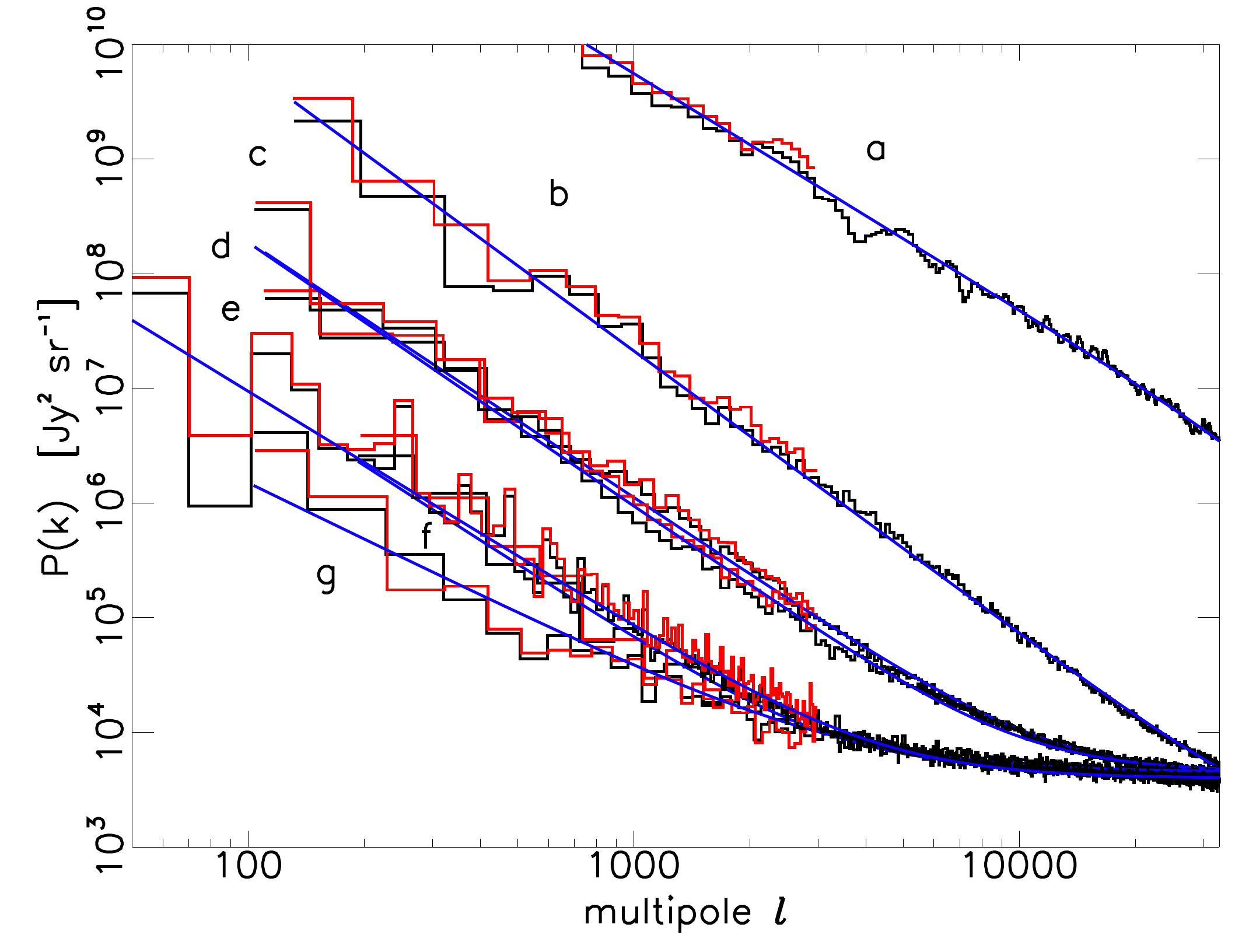}
\end{center}
\caption{Power spectra of six fields observed by both \Planck\ (red) and \Herschel\ (black). Fits to the spectra are shown in blue. There is a good agreement between \Planck\ and \Herschel\ in the common multipole range (typically $\ell < 3000$). Fields are, from top to bottom: (a)~Aquila; (b)~Polaris; (c)~Spider; (d)~Draco; (e)~Gama; (f)~FLS; and (g)~XMM-LSS. No real Galactic sources are expected in fields (b) -- (g), only extragalactic sources (correlated and Poisson components) and cirrus at larger angular scales. Real Galactic sources are detected, however, in (a)~\citep{andre10}: the power spectrum is orders of magnitude above the other fields, demonstrating the need to separate the Galactic from extragalactic zones, and the use of the background brightness as a proxy to estimate the cirrus contamination.}
\label{fig:cirrus_ps}
\end{figure}

We suggest a local definition of the presence of ``real'' Galactic sources: the power spectra of the maps at 857\,GHz retain their power law behaviour all the way from large scales measured by \Planck\ to the smallest scales measured by \Herschel\ (with a very good overlap), with the flat part of the power spectrum after noise removal being at the level set by extragalactic sources. The power spectra of the fields considered in this analysis are shown in Fig.~\ref{fig:cirrus_ps}.

\section{Characteristics of the PCCS}\label{sec:characteristics}

\subsection{Sensitivity and positional uncertainties}

\begin{figure}
\begin{center}
\includegraphics[width=\columnwidth]{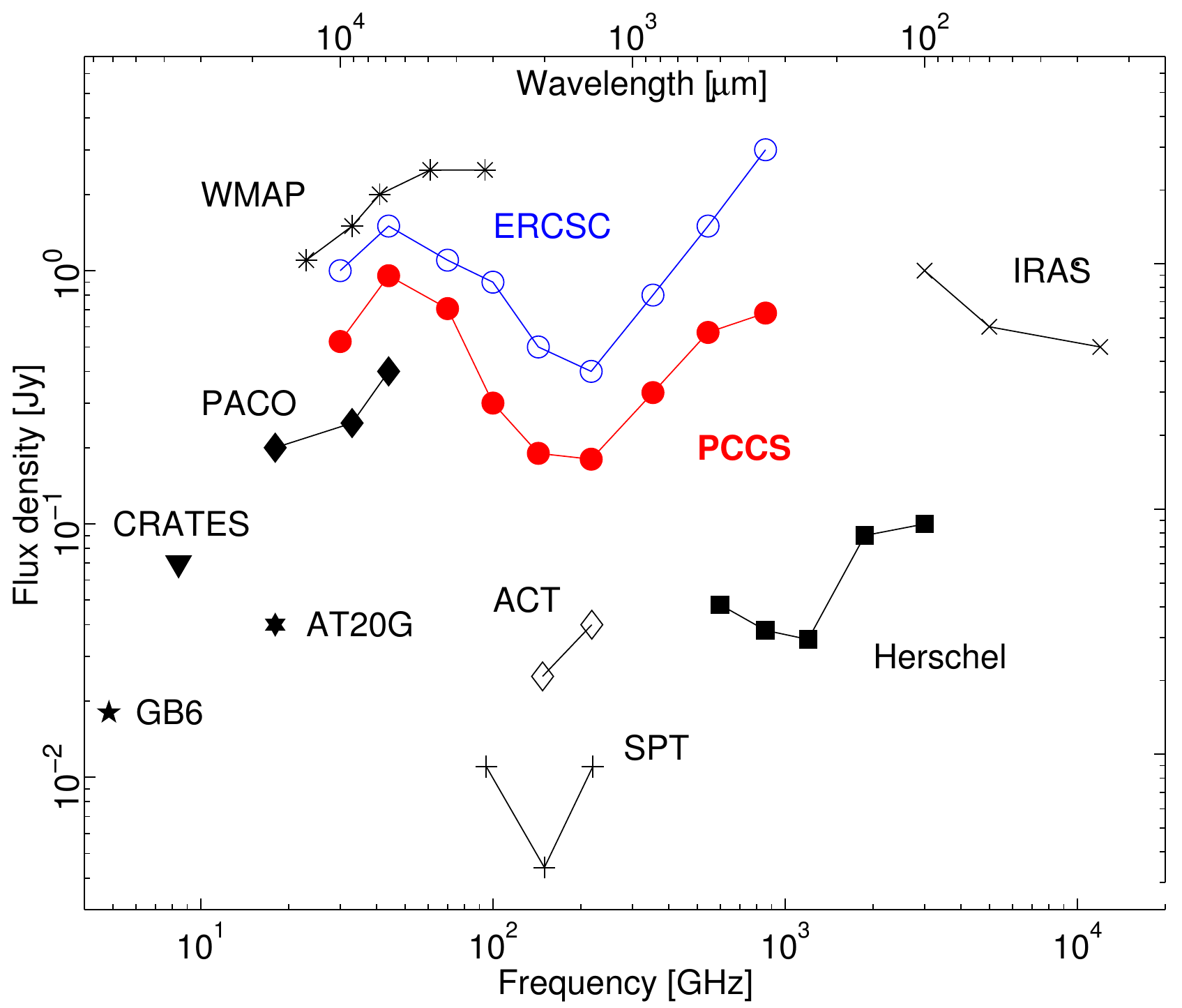}
\caption{The PCCS completeness level outside the Galactic plane (see Table \ref{tab:pccs}) is shown relative to other wide area surveys. The ERCSC completeness levels have been obtained from \citet{planck2011-6.1} up to 70\,GHz and \citet{planck2012-VII} for the other channels, while the \textit{WMAP} ones are from \citet{gnuevo08} up to 41\,GHz and \citet{lanz13} for 61 and 94\,GHz. The sensitivity levels for \Herschel\ SPIRE and PACS instruments are from \citet{clements10} and \citet{rigby11}, respectively. The other wide area surveys shown as a comparison are: GB6 \citep{gregory96}, CRATES \citep{healey07}, AT20G \citep{murphy10}, PACO \citep{bonavera11}, SPT \citep{Moc13}, ACT \citep{Mar13} and IRAS \citep{beichman88}.
\label{fig:sensitivity}}
\end{center}
\end{figure}

Table~\ref{tab:pccs} shows the effective beam FWHM, the minimum flux density (after excluding the faintest 10\,\% of sources) and the 90\,\% completeness level of all nine lists in the PCCS. As an illustration, Fig.~\ref{fig:sensitivity} shows the completeness level of PCCS at high Galactic latitude ($|{b}| > 30^\circ$) relative to the previous ERCSC and other wide area surveys at comparable frequencies. It is clear from this comparison that the sensitivity of the PCCS is a significant improvement on that of the ERCSC (see Sect.~\ref{sec:ercsc}) and that both catalogues are more complete then the \textit{WMAP} ones. Note that the PCCS detection limit increases inside the Galactic plane.

Figure~\ref{fig:qa_real} shows how the sensitivity of the catalogues varies across the sky due to the scanning strategy (the minimum noise is at the ecliptic poles where the sky is observed many times) and due to the effect of Galactic emission (near the Galactic plane and in particular Galactic regions).

The positional accuracy of the ERCSC was confirmed to be better than FWHM/5 \citep{planck2011-1.10,bonavera11}. In the case of the PCCS we have found similar results, as expected, since we have made corrections for two types of pointing errors that affected the ERCSC \citep{planck2011-1.10}. The first was due to time-dependent, thermally-driven misalignment between the star tracker and the telescope \citep{planck2013-p01}. The second was due to uncorrected stellar aberration across the focal plane. The misalignment resulting from stellar aberration is of the same magnitude as the positional uncertainties, and hence was not apparent in the ERCSC.

As explained in Sect.~\ref{sec:ext_val}, by comparing the positions derived with the detection method used to build the PCCS with the PACO sample \citep{massardi11,bonavera11,bonaldi13}, we have estimated the distribution of the pointing uncertainties up to 353\,GHz. In the case of 545 and 857\,GHz we derived the same quantities from the comparison with \Herschel\ sources. The median values of these distributions are reported in Table~\ref{tab:pccs}. The estimated positional uncertainties are below FWHM/5. These results are in good agreement with the values derived from the internal validation (see Table \ref{tab:qa}).

\subsection{Statistical properties of the PCCS}
\label{sec:stats}

Table~\ref{tab:pccs2} shows the numbers of sources internally matched within PCCS by finding them in neighbouring channels. It shows the number $N_{\rm both}$ of sources matched \textit{both} above and below in frequency (e.g., sources at 100\,GHz found in both the 70 and 143\,GHz catalogues), the number $N_{\rm either}$ matched \textit{either} above or below in frequency (a less stringent criterion), and the fraction of sources so matched. A source is considered to be matched if the positions are closer than the larger FWHM of the two channels. A catalogue was extracted from the IRIS 100\,\micron\ map~\citep{mdeschenes05} using the MHW2 pipeline, and that is used as the neighbouring channel above 857\,GHz. The IRIS mask, which removes around 2.1\,\% of the sky, was applied to the 857\,GHz catalogue before doing this comparison, and this reduces the number of sources to 24119, a decrease of about 1\,\%. The number of matches obtained for the 857\,GHz channel only includes sources outside the IRIS mask. For the 30\,GHz channel, the matches were evaluated using only the channel above, 44\,GHz. The low percentage of internal matches of the 30\,GHz channel results from two factors: the generally negative spectral index of the sources at these frequencies and the relatively low sensitivity of the 44\,GHz receivers. In fact, when the sensitivity of one of the neighbouring channels is worse, the percentage of matched sources is lower,  as is the case between 70 and 100\,GHz and between 217 and 353\,GHz.

\begin{figure*}
\begin{center}
\includegraphics[width=\textwidth]{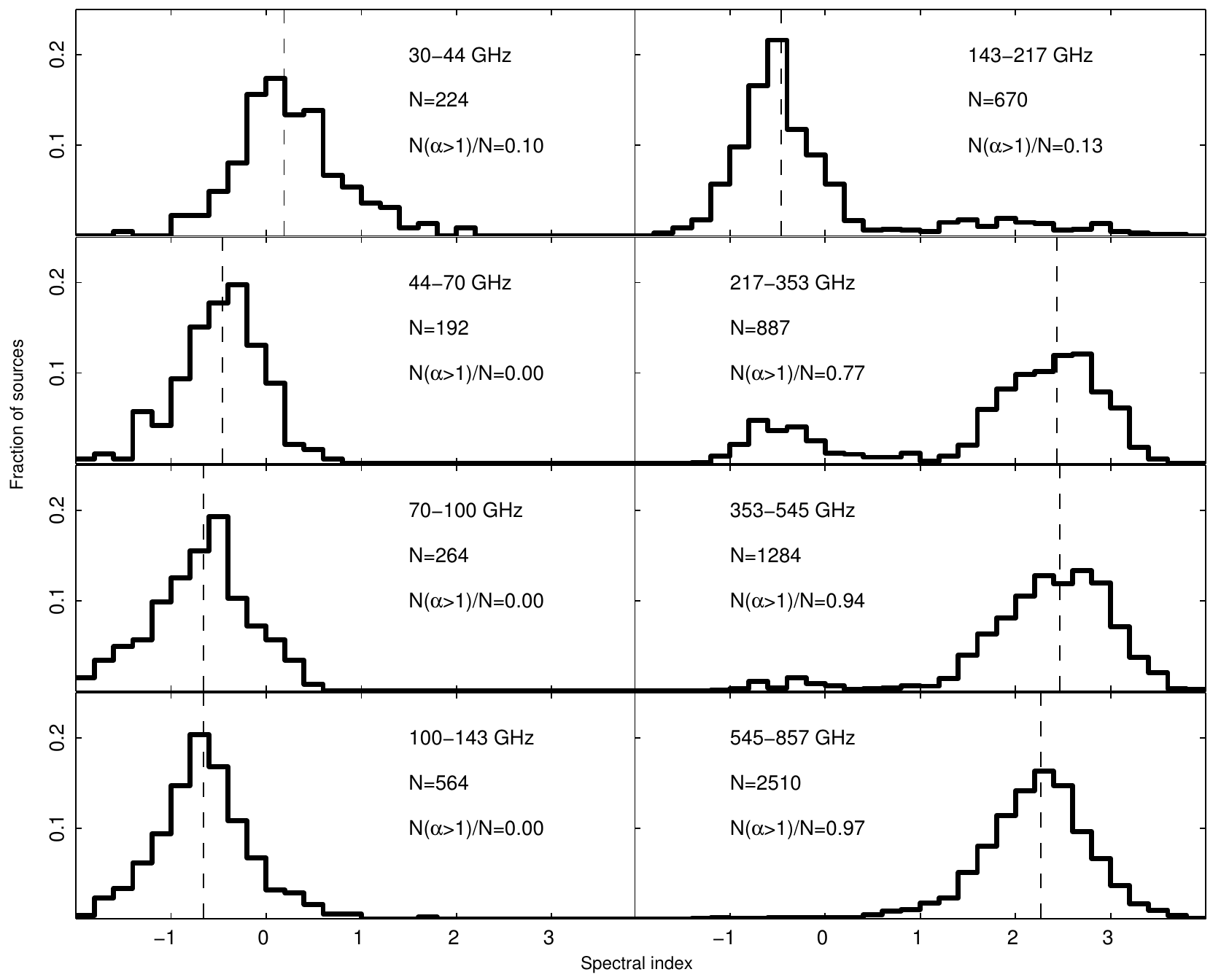}
\caption{Spectral indices of PCCS sources matched between contiguous channels with $|b|>30^\circ$. Each panel also shows the number of sources and the fraction with $\alpha>1$. The median values are indicated by vertical dashed lines.}
\label{fig:spec_index}
\end{center}
\end{figure*}

Figure~\ref{fig:spec_index} shows histograms of the spectral indices obtained from the matches between contiguous channels. As expected, the high frequency channels (545 and 857\,GHz) are dominated ($>90\,\%$) by dusty galaxies and the low frequency ones are dominated ($>95\,\%$) by synchrotron sources. In addition, two striking results initially obtained making use of the ERCSC are also seen in Figure~\ref{fig:spec_index}: i) the difference between the median values of the spectral indices below 70\,GHz indicates that there is a significant steepening in blazar spectra as demonstrated in \citet{planck2011-6.1}; ii) the high frequency counts (at least for frequencies $\leq217$\,GHz) of extragalactic sources are dominated at the bright end by synchrotron emitters, not dusty galaxies \citep{planck2012-VII}.

The deeper completeness levels and, as a consequence, the higher number of sources compared with the ERCSC (see next section), will allow the extension of previous studies to more sources and to fainter flux densities. However, this is beyond the scope of this paper and will be addressed in future publications.

\begin{table}
\begingroup
\newdimen\tblskip \tblskip=5pt
  \caption{Summary of sources matched between neighbouring channels. The number of sources with matches in {\it both} the lower-frequency and the higher-frequency channels is $N_{\rm both}$, and the number of sources with matches in {\it one or both} of the adjacent channels is $N_{\rm either}$.     \label{tab:pccs2}}
\nointerlineskip
\vskip -3mm
\footnotesize
\setbox\tablebox=\vbox{
   \newdimen\digitwidth
   \setbox0=\hbox{\rm 0}
   \digitwidth=\wd0
   \catcode`*=\active
   \def*{\kern\digitwidth}
   \newdimen\signwidth
   \setbox0=\hbox{+}
   \signwidth=\wd0
   \catcode`!=\active
   \def!{\kern\signwidth}
\halign to \hsize{\hbox to 0.8in{#\leaderfil}\tabskip=1em plus 0.2em&\hfil#\hfil&\hfil#\hfil&\hfil#\hfil&\hfil#\hfil\tabskip=0pt\cr
\noalign{\doubleline}
\omit& \raisebox{-3pt}{Number}& \multispan2{\hfil Number matched\hfil}& \raisebox{-3pt}{Fraction}\cr
\noalign{\vskip -5pt}
\omit&&\multispan2{\hrulefill}\cr
\noalign{\vskip -1pt}
\omit \hfil Channel\hfil& of sources& $N_{\rm both}$& $N_{\rm either}$& matched\cr
\noalign{\vskip 3pt\hrule\vskip 5pt}
    *30$^{\rm a}$& *1256& \ldots& **629& 50.1\,\% \cr
    *44& **731& *530& **664&  90.8\,\% \cr
    *70& **939& *552& **815& 86.8\,\% \cr
    100& *3850& *772& *2758&  71.6\,\% \cr
    143& *5675& 2454& *4645& 81.9\,\% \cr
    217& 16070& 3351& 10624& 66.1\,\% \cr
    353& 13613& 8029& 12079& 88.7\,\% \cr
    545& 16933& 9382& 14535& 85.8\,\% \cr
    857$^{b}$& 24381& 6904& 18061& 74.9\,\% \cr
 \noalign{\vskip 3pt\hrule\vskip 3pt}}}
\endPlancktable
\tablenote {{\rm a}} The $30$\,GHz channel is only matched with the 44\,GHz channel above.\par
\tablenote {{b}} The $857$\,GHz channel is matched above with a catalogue extracted from the IRIS maps using the HFI--MHW. Both catalogues were cut with the IRIS mask prior to matching.\par 
\endgroup
\end{table}

\subsection{Comparison with the \Planck\ ERCSC}\label{sec:ercsc}

The Early Release Compact Source Catalogue is a catalogue of high-reliability sources, both Galactic and extragalactic, detected over the full sky, in the first \Planck\ all-sky survey. One of the primary goals of the ERCSC was to provide an early catalogue of sources for follow-up observations with existing facilities, in particular \Herschel, while they were still in their cryogenic operational phase. The PCCS differs from the ERCSC both in the data and the philosophy.

\begin{figure*}
\begin{center}
\includegraphics[width=\textwidth]{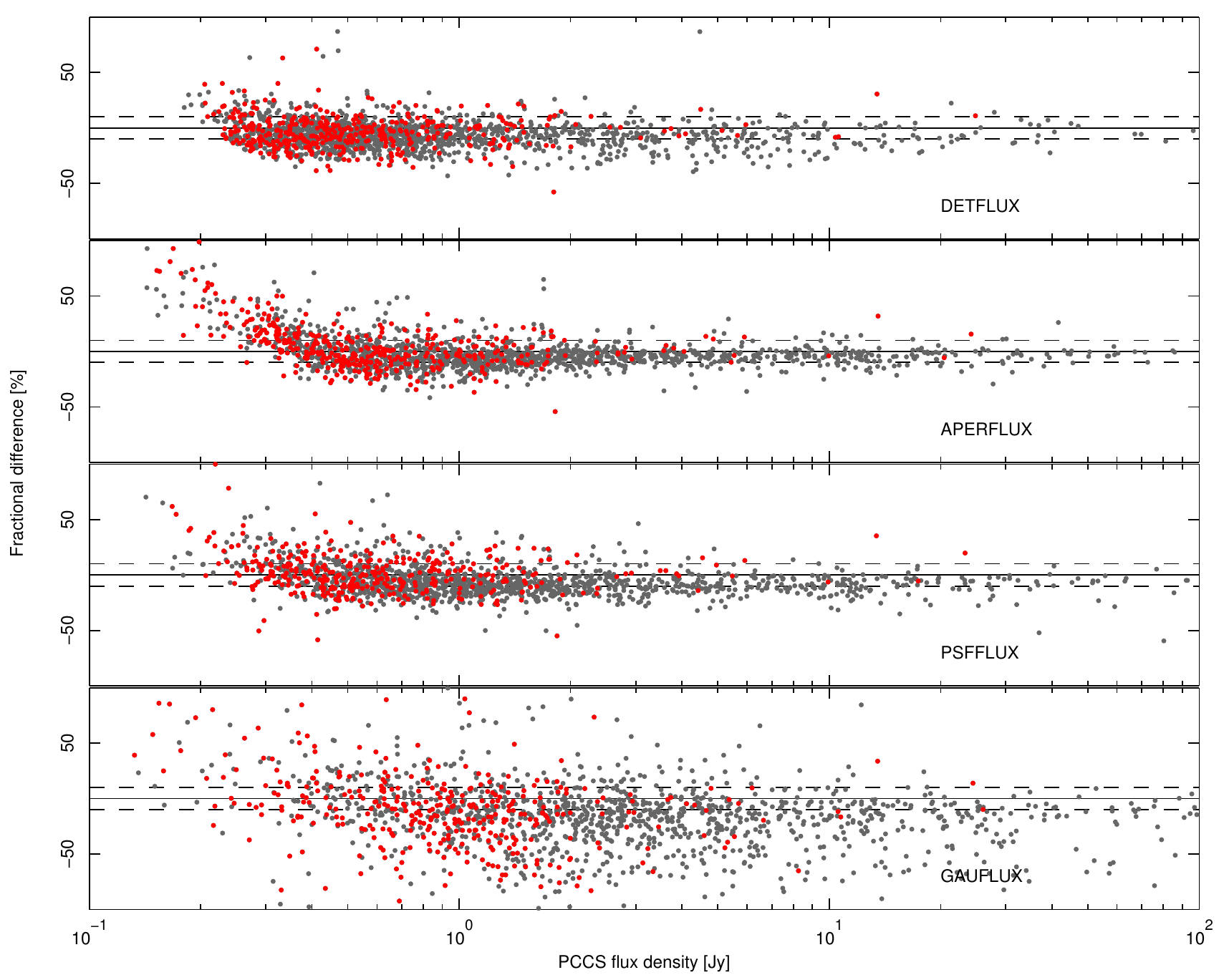}
\caption{Comparison of ERCSC and PCCS photometries at 143\,GHz. Grey points correspond to common sources below $|{b}|<30^{\circ}$ while red ones show the common ones for $|{b}|>30^{\circ}$. Dashed lines indicates the $\pm10\,\%$ uncertainty level.
\label{fig:photo_ercsc}}
\end{center}
\end{figure*}

The data used to build the ERCSC consisted of one complete survey and 60\,\% of the second survey included in the maps. The data used for the PCCS consists of two complete surveys and 60\,\% of the third survey. Moreover, our knowledge of the instruments has improved during this time, and this translates into a better calibration and quality of the maps, and better characterization of the beams \citep{planck2013-p02,planck2013-p03}. The beam size and shapes are crucial to obtaining precise measurements of the flux densities. The change in beam sizes between those used for ERCSC and the present values used for the PCCS is $\sim2$\,\% in the LFI channels and $\sim8$\,\% in the HFI ones. Figure~\ref{fig:photo_ercsc} shows a comparison at 143\,GHz between the photometries from ERCSC and PCCS. Similar results are obtained for all the other channels.

The primary goal of the ERCSC, to provide a reliable catalogue, was successfully accomplished. The goal of the PCCS is to increase the completeness of the catalogue while  maintaining a good global reliability ($>80$\,\% by construction). This has led to the higher number of detections per channel (a factor 2--4 more sources) and better sensitivity achieved by the PCCS (see also Fig.~\ref{fig:sensitivity} for a direct comparison between the PCCS and ERCSC completeness levels).

\section{The PCCS: access, content and usage}\label{sec:content}

The PCCS is available from the ESA \Planck\ Legacy Archive\footnote{\url{http://pla.esac.esa.int/pla/pla.jnlp}}. The source lists contain 24 columns. The 857\,GHz source list has six additional columns that consist of the band-filled aperture flux densities and associated uncertainties in the three adjacent frequency channels, 217--545\,GHz, for each source detected at 857\,GHz.

\subsection{Catalogue content and usage}
\label{sec:usage}

Detailed information about the catalogue content and format can be found in the Explanatory Supplement \citep{planck2013-p28} and in the FITS files headers. Here we summarize the most important features of the catalogues. The key columns in the catalogues are:
\begin{itemize}
\item Source identification: NAME (string).
\item Position: GLON and GLAT contain the Galactic coordinates, and RA and DEC the equatorial coordinates (J2000).
\item Flux density: the four estimates of flux density (DETFLUX, APERFLUX, PSFFLUX, and GAUFLUX) in mJy, and their associated uncertainties (with the \_ERR suffix).
\item Source extension: the EXTENDED flag is set to 1 if a source is extended. See the definition below.
\item Cirrus indicator: the CIRRUS\_N column contains a cirrus indicator for the HFI channels. See the definition below.
\item External validation: the EXT\_VAL contains a summary of the external validation for the LFI channels. See the definition below.
\item Identification with ERCSC: the ERCSC column indicates the name of the ERCSC counterpart, if there is one, at this channel.
\end{itemize}

A source is classified as extended if
\begin{equation}
\mathrm{FWHM}_\mathrm{eff} \geq 1.5 \,\mathrm{FWHM}_\mathrm{nom},
\end{equation}
where $\mathrm{FWHM}_\mathrm{nom}$ is the nominal beam size for the selected channel and the quantity $\mathrm{ FWHM}_\mathrm{eff}$ is the geometric mean of the major and minor FWHM values from the Gaussian fit,
\begin{equation}
\mathrm{FWHM}_\mathrm{eff}= \sqrt{\mathrm{FWHM}_1\,\mathrm{FWHM}_2}.
\end{equation}
In the upper HFI bands, sources that are extended tend to be associated with structure in the Galactic interstellar medium although individual nearby galaxies are also extended sources as seen by \Planck~\citep[see][]{planck2011-6.4a}. The choice of the threshold, 1.5 times the beam width, is motivated by the accuracy with which source profiles can be measured from maps where the point spread function is critically sampled (1\farcm7 pixel scale for a FHWM of $\sim$4\arcmin). Naturally, faint sources for which the Gaussian fitting failed do not have the EXTENDED flag set.

Sources in the HFI channels have a cirrus indicator, CIRRUS\_N. This is the number of sources detected at 857\,GHz (using a uniform \snr\ threshold of 5) within a 1\degr\  radius of the source. Many 857\,GHz detections at this \snr\ threshold in the Galactic region will be from cirrus knots, so it provides a useful indicator of the presence of cirrus.

The EXT\_VAL column summarizes the cross-matching with external catalogues. For the LFI channels this is the set of radio catalogues used in the external validation (see Sect.~\ref{sec:ext_val}). For HFI channels it is the catalogue extracted from the IRIS map (see Sect.~\ref{sec:stats}). The EXT\_VAL flag takes the value 0, 1, or 2:
\begin{enumerate}
\item[0:] The source has no clear counterpart in any of the external catalogues and it has not been detected in other \Planck\ channels.
\item[1:] The source has no clear counterpart in any of the external catalogues, but it has been detected in other \Planck\ channels.
\item[2:] For the LFI channels, the source has a clear counterpart in the radio catalogues. For the HFI channels, the source either has a clear counterpart in the radio catalogues or in both the IRIS catalogue and all the higher \Planck\ channels.
\end{enumerate}
This flag provides valuable information about the reliability of individual sources: those flagged as EXT\_VAL= 2 are already known, those with EXT\_VAL = 1 have been detected in other \Planck\ channels and are therefore potentially new sources, and those with EXT\_VAL = 0 appear in only a single channel, and thus are more likely to be spurious. For the LFI channels, the matrix filters \citep{herranz09} are used to determine whether a source has been detected in other \Planck\ channels. For the HFI channels, the cross-matching is carried out a posteriori from the catalogues (see Sect.~\ref{sec:stats}).

As described in Sect.~\ref{sec:photometry}, four measures of flux density are provided in units of mJy. For extended sources, both DETFLUX and PSFFLUX are likely to produce underestimates of the true source flux density. Furthermore, at faint flux densities corresponding to low \snr, the PSF and GAUSSIAN fits may fail. This would be represented either by a negative flux density or by a significant difference between the GAUFLUX and DETFLUX values. In general, for bright extended sources, we recommend using the GAUFLUX and GAUFLUX\_ERR values although even these might be biased high if the source is located in a region of complex, diffuse foreground emission. Uncertainties in the flux density measured by each technique are reflected in the corresponding \_ERR column.

The median positional uncertainty, given in Table \ref{tab:pccs}, is only a statistical estimate for each band. Individual sources could have a larger positional offset depending on the local background rms and \snr. As this quantity has been obtained by comparison with external data sets it also takes into account any astrometric offset in the maps.

\subsection{Cautionary notes on the use of catalogues}
\label{sec:cautionary}

In this section, we remind readers of the preliminary nature of the PCCS and list some cautionary notes for users of the catalogue.  The PCCS is based solely on the nine frequency maps derived from the nominal mission, which ended in November 2010.  The HFI instrument continued to operate stably for another 14 months after the end of the nominal mission, and the LFI instrument is expected to complete an additional 5.4 full-sky surveys not included in the PCCS.  Thus the PCCS is based on only a fraction of \Planck\ data: approximately 1/3 in the case of LFI. The observations following the nominal mission will also allow for better control of systematic errors, which in turn is likely to improve the quality and accuracy of a later, more complete catalogue of \Planck\ sources based on the entire mission. Our understanding of the instrument (effective beam size, for instance) has improved since the ERCSC was issued and will surely continue to improve.  Likewise, we can expect further improvements in the use of ground-based and other facilities to validate properties of the catalogue, such as flux density scales.  Note the improvement over validation efforts for the ERCSC, and the extension of external validation to 143 and 217\,GHz (Sect. 3.2.2).  It is also reasonable to expect further refinements in the algorithms used to detect sources and to measure their properties.  Finally, the PCCS does not address, as future catalogues will, the issue of polarization.

As noted earlier, the aim of the PCCS is to provide as complete a list as possible of \Planck\ sources with a reasonable degree of reliability. The criteria used to include or exclude candidate sources differ from channel to channel and in different parts of the sky; they also are based on different S/N levels.  These differences were consequences of our desire to make the catalogue as complete as possible, yet maintain $> 80$\% reliability.  These differences have to be taken into account when using the PCCS for statistical studies.  On the other hand, we have endeavoured to ensure that the flux density scales of the various channel catalogues are consistent.  They appear to be at the few percent level \citep[see discusion in][]{planck2013-p01a}.

We now turn to several specific cautions and comments for users of the PCCS.

\paragraph{Variability:} At radio frequencies, up to and including 217\,GHz, many of the extragalactic sources are highly variable. A small fraction of them vary even on time scales of a few hours based on observed changes in the flux density as a source drifts through the different \Planck\ horns \citep{planck2013-p02,planck2013-p03}. Follow-up observations of these sources might show significant differences in flux density compared to the values in the data products. Although the maps used for the PCCS are based on 2.6 sky coverages, the PCCS provides only a single average flux density estimate over all \Planck\ data samples that were included in the maps and does not contain any measure of the variability of the sources from survey to survey.

\paragraph{Contamination from CO:} At infrared/submillimetre frequencies (100\,GHz and above), the \Planck\ bandpasses straddle energetically significant CO lines. The effect is the greatest at 100\,GHz, where the line contributes strongly to the flux densities of Galactic sources.  Aperture photometry at PCCS candidate locations using the type-3 CO map derived from \Planck\ component separation \citep{planck2013-p03a} shows that 91\% of detections within the 
85\% Galactic dust mask have CO contamination of their APERFLUX estimates at the level of 20\% or higher.  Outside this mask, this contamination drops to 6\%.  Follow-up observations of these Galactic sources, especially those associated with Galactic star-forming regions, at a similar frequency but different bandpass, should correct for the contribution of line emission to the measured continuum flux density of the source.

\paragraph{Photometry:} Each source has multiple estimates of flux density, DETFLUX, APERFLUX, GAUFLUX and PSFFLUX, as defined above. The appropriate photometry to be used depends on the nature of the source. For sources that are unresolved at the spatial resolution of \Planck, APERFLUX and DETFLUX are most appropriate. Even in this regime, PSF or Gaussian fits of faint sources fail and consequently these have a PSFFLUX/GAUFLUX value of NaN ("Not a Number"). For bright resolved sources, GAUFLUX might be most appropriate although GAUFLUX appears to overestimate the flux density of the sources close to the Galactic plane due to an inability to fit for the contribution of the Galactic background at the spatial resolution of the data. For the 353--857\,GHz channels, the complex nature of the diffuse emission and the relative undersampling of the beam produces a bias in DETFLUX, we therefore recommend that APERFLUX is used instead (see Fig.~\ref{fig:photo_herschel}).

\paragraph{Calibration:} The absolute calibration uncertainties of \Planck\ are $<1$\,\% for 30--217\,GHz and are $<1.2$\% at 353\,GHz.  For 545 and 857\,GHz, the absolute calibration uncertainty is $<10$\% \citep{planck2013-p02, planck2013-p02b,planck2013-p03, planck2013-p03f}.  For these two channels the calibration uncertainty is an appreciable systematic error on the photometry, which is not included in the internal validation (as it was not simulated) or the external comparison with \Herschel\ photometry (see Sect. \ref{sec:highfreq}) as the inter-calibration between HFI and SPIRE was corrected prior to comparison.

\paragraph{Colour correction:} The flux density estimates have not been colour corrected. Colour corrections are described in \citet{planck2013-p02} and \citet{planck2013-p03}. The typical amplitude of the correction is below $4$\,\% for synchrotron-like spectra in the LFI channels and below $15$\,\% for thermal emission in the  HFI channels. Colour corrections can be calculated using \textit{Planck} unit conversion and colour correction software (UC\_CC), included in the data release \citep{planck2013-p03d}, which uses  the instrumental bandpasses \citep{planck2013-p28}.

\paragraph{Cirrus/ISM:} A significant fraction of the sources detected in the upper HFI bands could be associated with Galactic interstellar medium features or cirrus. The value of CIRRUS\_N in the catalogue can be used to flag sources that might be clustered together and thereby associated with ISM structure. Candidate ISM features can also be selected by choosing objects with EXTENDED = 1 although nearby Galactic and extragalactic sources that are extended at \Planck\ spatial resolution will meet this criterion too.
The 857\,GHz brightness proxy described in Sect.~\ref{sec:confusion} can also be used as indicator of cirrus contamination.

\section{Conclusions}\label{sec:conclusions}

The PCCS lists sources extracted from the \Planck\ nominal mission data in each of its nine frequency bands. By construction its reliability is $>80\,\%$ and a special effort was made to use simple selection procedures in order to facilitate statistical analyses. With a common detection method for all the channels and the additional three photometric estimates, spectral analysis can also be performed safely. The deeper completeness levels and, as a consequence, the higher number of sources compared with the ERCSC, will allow the extension of previous studies to more sources and to fainter flux densities. The PCCS is the natural evolution of the ERCSC, but both lack polarization and multi-frequency information. Future releases will take advantage of the full mission data and they will contain information on properties of sources not available in this release, including polarization and variability, and association of sources detected in different bands.

This paper describes the construction and properties of this preliminary catalogue.  We have not attempted to exploit the PCCS for science purposes, preferring instead to leave this to future papers and to the wider community.

\begin{acknowledgements}
The development of \Planck\ has been supported by: ESA; CNES and CNRS/INSU-IN2P3-INP (France); ASI, CNR, and INAF (Italy); NASA and DoE (USA); STFC and UKSA (UK); CSIC, MICINN, JA and RES (Spain); Tekes, AoF and CSC (Finland); DLR and MPG (Germany); CSA (Canada); DTU Space (Denmark); SER/SSO (Switzerland); RCN (Norway); SFI (Ireland); FCT/MCTES (Portugal); and PRACE (EU). A description of the \Planck\ Collaboration and a list of its members, including the technical or scientific activities in which they have been involved, can be found at \url{http://www.sciops.esa.int/index.php?project=planck&page=Planck_Collaboration}. This research has made use of Aladin.

We are deeply grateful to Rick Perley and the U.S. National Radio Astronomy Observatory for participating in the joint Planck-VLA observations.  We also thank Steve Eales and the H-ATLAS team (Eales et al, 2010 PASP 122 499E) for sharing their SPIRE catalogs prior to publication,
allowing a comparison of flux densities between HFI and SPIRE.
\end{acknowledgements}

\bibliographystyle{aa}
\bibliography{Planck_bib,pccs}

\begin{thebibliography}{78}
\expandafter\ifx\csname natexlab\endcsname\relax\def\natexlab#1{#1}\fi

\bibitem[{{AMI Consortium} {et~al.}(2012){AMI Consortium}, {Perrott}, {Green},
  {Davies}, {Franzen}, {Grainge}, {Hobson}, {Hurley-Walker}, {Lasenby},
  {Olamaie}, {Pooley}, {Rodr{\'{\i}}guez-Gonz{\'a}lvez}, {Saunders}, {Scaife},
  {Schammel}, {Scott}, {Shimwell}, {Titterington}, \& {Waldram}}]{AMI2012}
{AMI Consortium}, {Perrott}, Y.~C., {Green}, D.~A., {et~al.} 2012, \mnras, 421,
  L6

\bibitem[{{Andr{\'e}} {et~al.}(2010){Andr{\'e}}, {Men'shchikov}, {Bontemps},
  {K{\"o}nyves}, {Motte}, {Schneider}, {Didelon}, {Minier}, {Saraceno},
  {Ward-Thompson}, {di Francesco}, {White}, {Molinari}, {Testi}, {Abergel},
  {Griffin}, {Henning}, {Royer}, {Mer{\'{\i}}n}, {Vavrek}, {Attard},
  {Arzoumanian}, {Wilson}, {Ade}, {Aussel}, {Baluteau}, {Benedettini},
  {Bernard}, {Blommaert}, {Cambr{\'e}sy}, {Cox}, {di Giorgio}, {Hargrave},
  {Hennemann}, {Huang}, {Kirk}, {Krause}, {Launhardt}, {Leeks}, {Le Pennec},
  {Li}, {Martin}, {Maury}, {Olofsson}, {Omont}, {Peretto}, {Pezzuto}, {Prusti},
  {Roussel}, {Russeil}, {Sauvage}, {Sibthorpe}, {Sicilia-Aguilar}, {Spinoglio},
  {Waelkens}, {Woodcraft}, \& {Zavagno}}]{andre10}
{Andr{\'e}}, P., {Men'shchikov}, A., {Bontemps}, S., {et~al.} 2010, \aap, 518,
  L102

\bibitem[{{Beichman} {et~al.}(1988){Beichman}, {Neugebauer}, {Habing}, {Clegg},
  \& {Chester}}]{beichman88}
{Beichman}, C.~A., {Neugebauer}, G., {Habing}, H.~J., {Clegg}, P.~E., \&
  {Chester}, T.~J., eds. 1988, {Infrared astronomical satellite (IRAS) catalogs
  and atlases. Volume 1: Explanatory supplement}, Vol.~1

\bibitem[{{Bennett} {et~al.}(2013){Bennett}, {Larson}, {Weiland}, {Jarosik},
  {Hinshaw}, {Odegard}, {Smith}, {Hill}, {Gold}, {Halpern}, {Komatsu}, {Nolta},
  {Page}, {Spergel}, {Wollack}, {Dunkley}, {Kogut}, {Limon}, {Meyer}, {Tucker},
  \& {Wright}}]{Bennett2013}
{Bennett}, C.~L., {Larson}, D., {Weiland}, J.~L., {et~al.} 2013, \apjs, 208, 20

\bibitem[{{Bonaldi} {et~al.}(2013){Bonaldi}, {Bonavera}, {Massardi}, \& {De
  Zotti}}]{bonaldi13}
{Bonaldi}, A., {Bonavera}, L., {Massardi}, M., \& {De Zotti}, G. 2013, \mnras,
  428, 1845

\bibitem[{{Bonavera} {et~al.}(2011){Bonavera}, {Massardi}, {Bonaldi},
  {Gonz{\'a}lez-Nuevo}, {de Zotti}, \& {Ekers}}]{bonavera11}
{Bonavera}, L., {Massardi}, M., {Bonaldi}, A., {et~al.} 2011, \mnras, 416, 559

\bibitem[{{Boselli} {et~al.}(2010){Boselli}, {Eales}, {Cortese}, {Bendo},
  {Chanial}, {Buat}, {Davies}, {Auld}, {Rigby}, {Baes}, {Barlow}, {Bock},
  {Bradford}, {Castro-Rodriguez}, {Charlot}, {Clements}, {Cormier}, {Dwek},
  {Elbaz}, {Galametz}, {Galliano}, {Gear}, {Glenn}, {Gomez}, {Griffin}, {Hony},
  {Isaak}, {Levenson}, {Lu}, {Madden}, {O'Halloran}, {Okamura}, {Oliver},
  {Page}, {Panuzzo}, {Papageorgiou}, {Parkin}, {Perez-Fournon}, {Pohlen},
  {Rangwala}, {Roussel}, {Rykala}, {Sacchi}, {Sauvage}, {Schulz}, {Schirm},
  {Smith}, {Spinoglio}, {Stevens}, {Symeonidis}, {Vaccari}, {Vigroux},
  {Wilson}, {Wozniak}, {Wright}, \& {Zeilinger}}]{boselli10}
{Boselli}, A., {Eales}, S., {Cortese}, L., {et~al.} 2010, \pasp, 122, 261

\bibitem[{{Chen} {et~al.}(2013){Chen}, {Rachen}, {L{\'o}pez-Caniego},
  {Dickinson}, {Pearson}, {Fuhrmann}, {Krichbaum}, \& {Partridge}}]{Chen13}
{Chen}, X., {Rachen}, J.~P., {L{\'o}pez-Caniego}, M., {et~al.} 2013, \aap, 553,
  A107

\bibitem[{{Clemens} {et~al.}(2013){Clemens}, {Negrello}, {De Zotti},
  {Gonzalez-Nuevo}, {Bonavera}, {Cosco}, {Guarese}, {Boaretto}, {Salucci},
  {Baccigalupi}, {Clements}, {Danese}, {Lapi}, {Mandolesi}, {Partridge},
  {Perrotta}, {Serjeant}, {Scott}, \& {Toffolatti}}]{Clemens13}
{Clemens}, M.~S., {Negrello}, M., {De Zotti}, G., {et~al.} 2013, \mnras

\bibitem[{{Clements} {et~al.}(2010){Clements}, {Rigby}, {Maddox}, {Dunne},
  {Mortier}, {Pearson}, {Amblard}, {Auld}, {Baes}, {Bonfield}, {Burgarella},
  {Buttiglione}, {Cava}, {Cooray}, {Dariush}, {de Zotti}, {Dye}, {Eales},
  {Frayer}, {Fritz}, {Gardner}, {Gonzalez-Nuevo}, {Herranz}, {Ibar}, {Ivison},
  {Jarvis}, {Lagache}, {Leeuw}, {Lopez-Caniego}, {Negrello}, {Pascale},
  {Pohlen}, {Rodighiero}, {Samui}, {Serjeant}, {Sibthorpe}, {Scott}, {Smith},
  {Temi}, {Thompson}, {Valtchanov}, {van der Werf}, \& {Verma}}]{clements10}
{Clements}, D.~L., {Rigby}, E., {Maddox}, S., {et~al.} 2010, \aap, 518, L8

\bibitem[{{Condon}(1974)}]{condon74}
{Condon}, J.~J. 1974, \apj, 188, 279

\bibitem[{{Crawford} {et~al.}(2010){Crawford}, {Switzer}, {Holzapfel},
  {Reichardt}, {Marrone}, \& {Vieira}}]{Crawford10}
{Crawford}, T.~M., {Switzer}, E.~R., {Holzapfel}, W.~L., {et~al.} 2010, \apj,
  718, 513

\bibitem[{{Dale} {et~al.}(2005){Dale}, {Bendo}, {Engelbracht}, {Gordon},
  {Regan}, {Armus}, {Cannon}, {Calzetti}, {Draine}, {Helou}, {Joseph},
  {Kennicutt}, {Li}, {Murphy}, {Roussel}, {Walter}, {Hanson}, {Hollenbach},
  {Jarrett}, {Kewley}, {Lamanna}, {Leitherer}, {Meyer}, {Rieke}, {Rieke},
  {Sheth}, {Smith}, \& {Thornley}}]{dale05}
{Dale}, D.~A., {Bendo}, G.~J., {Engelbracht}, C.~W., {et~al.} 2005, \apj, 633,
  857

\bibitem[{{Davies} {et~al.}(2013){Davies}, {Bianchi}, {Baes}, {Boselli},
  {Ciesla}, {Clemens}, {Davis}, {De Looze}, {Alighieri}, {Fuller}, {Fritz},
  {Hunt}, {Serra}, {Smith}, {Verstappen}, {Vlahakis}, {Xilouris}, {Bomans},
  {Hughes}, {Garcia-Appadoo}, \& {Madden}}]{davies12}
{Davies}, J.~I., {Bianchi}, S., {Baes}, M., {et~al.} 2013, \mnras, 428, 834

\bibitem[{{de Zotti} {et~al.}(2005){de Zotti}, {Ricci}, {Mesa}, {Silva},
  {Mazzotta}, {Toffolatti}, \& {Gonz{\'a}lez-Nuevo}}]{dezotti05}
{de Zotti}, G., {Ricci}, R., {Mesa}, D., {et~al.} 2005, \aap, 431, 893

\bibitem[{{Delabrouille} {et~al.}(2013){Delabrouille}, {Betoule}, {Melin},
  {Miville-Desch{\^e}nes}, {Gonzalez-Nuevo}, {Le Jeune}, {Castex}, {de Zotti},
  {Basak}, {Ashdown}, {Aumont}, {Baccigalupi}, {Banday}, {Bernard}, {Bouchet},
  {Clements}, {da Silva}, {Dickinson}, {Dodu}, {Dolag}, {Elsner}, {Fauvet},
  {Fa{\"y}}, {Giardino}, {Leach}, {Lesgourgues}, {Liguori},
  {Mac{\'{\i}}as-P{\'e}rez}, {Massardi}, {Matarrese}, {Mazzotta}, {Montier},
  {Mottet}, {Paladini}, {Partridge}, {Piffaretti}, {Prezeau}, {Prunet},
  {Ricciardi}, {Roman}, {Schaefer}, \& {Toffolatti}}]{psm}
{Delabrouille}, J., {Betoule}, M., {Melin}, J.-B., {et~al.} 2013, \aap, 553,
  A96

\bibitem[{{Dole} {et~al.}(2003){Dole}, {Lagache}, \& {Puget}}]{dole03}
{Dole}, H., {Lagache}, G., \& {Puget}, J.-L. 2003, \apj, 585, 617

\bibitem[{{Dole} {et~al.}(2006){Dole}, {Lagache}, {Puget}, {Caputi},
  {Fern{\'a}ndez-Conde}, {Le Floc'h}, {Papovich}, {P{\'e}rez-Gonz{\'a}lez},
  {Rieke}, \& {Blaylock}}]{dole06}
{Dole}, H., {Lagache}, G., {Puget}, J.-L., {et~al.} 2006, \aap, 451, 417

\bibitem[{{Dunne} {et~al.}(2000){Dunne}, {Eales}, {Edmunds}, {Ivison},
  {Alexander}, \& {Clements}}]{dunne00}
{Dunne}, L., {Eales}, S., {Edmunds}, M., {et~al.} 2000, \mnras, 315, 115

\bibitem[{{Eales} {et~al.}(2010){Eales}, {Dunne}, {Clements}, {Cooray}, {de
  Zotti}, {Dye}, {Ivison}, {Jarvis}, {Lagache}, {Maddox}, {Negrello},
  {Serjeant}, {Thompson}, {van Kampen}, {Amblard}, {Andreani}, {Baes},
  {Beelen}, {Bendo}, {Benford}, {Bertoldi}, {Bock}, {Bonfield}, {Boselli},
  {Bridge}, {Buat}, {Burgarella}, {Carlberg}, {Cava}, {Chanial}, {Charlot},
  {Christopher}, {Coles}, {Cortese}, {Dariush}, {da Cunha}, {Dalton}, {Danese},
  {Dannerbauer}, {Driver}, {Dunlop}, {Fan}, {Farrah}, {Frayer}, {Frenk},
  {Geach}, {Gardner}, {Gomez}, {Gonz{\'a}lez-Nuevo}, {Gonz{\'a}lez-Solares},
  {Griffin}, {Hardcastle}, {Hatziminaoglou}, {Herranz}, {Hughes}, {Ibar},
  {Jeong}, {Lacey}, {Lapi}, {Lawrence}, {Lee}, {Leeuw}, {Liske},
  {L{\'o}pez-Caniego}, {M{\"u}ller}, {Nandra}, {Panuzzo}, {Papageorgiou},
  {Patanchon}, {Peacock}, {Pearson}, {Phillipps}, {Pohlen}, {Popescu},
  {Rawlings}, {Rigby}, {Rigopoulou}, {Robotham}, {Rodighiero}, {Sansom},
  {Schulz}, {Scott}, {Smith}, {Sibthorpe}, {Smail}, {Stevens}, {Sutherland},
  {Takeuchi}, {Tedds}, {Temi}, {Tuffs}, {Trichas}, {Vaccari}, {Valtchanov},
  {van der Werf}, {Verma}, {Vieria}, {Vlahakis}, \& {White}}]{eales10}
{Eales}, S., {Dunne}, L., {Clements}, D., {et~al.} 2010, \pasp, 122, 499

\bibitem[{{Eddington}(1940)}]{eddington40}
{Eddington}, Sir, A.~S. 1940, \mnras, 100, 354

\bibitem[{{Franceschini} {et~al.}(1989){Franceschini}, {Toffolatti}, {Danese},
  \& {de Zotti}}]{franceschini89}
{Franceschini}, A., {Toffolatti}, L., {Danese}, L., \& {de Zotti}, G. 1989,
  \apj, 344, 35

\bibitem[{{Fu} {et~al.}(2012){Fu}, {Jullo}, {Cooray}, {Bussmann}, {Ivison},
  {P{\'e}rez-Fournon}, {Djorgovski}, {Scoville}, {Yan}, {Riechers}, {Aguirre},
  {Auld}, {Baes}, {Baker}, {Bradford}, {Cava}, {Clements}, {Dannerbauer},
  {Dariush}, {De Zotti}, {Dole}, {Dunne}, {Dye}, {Eales}, {Frayer}, {Gavazzi},
  {Gurwell}, {Harris}, {Herranz}, {Hopwood}, {Hoyos}, {Ibar}, {Jarvis}, {Kim},
  {Leeuw}, {Lupu}, {Maddox}, {Mart{\'{\i}}nez-Navajas}, {Micha{\l}owski},
  {Negrello}, {Omont}, {Rosenman}, {Scott}, {Serjeant}, {Smail}, {Swinbank},
  {Valiante}, {Verma}, {Vieira}, {Wardlow}, \& {van der Werf}}]{fu12}
{Fu}, H., {Jullo}, E., {Cooray}, A., {et~al.} 2012, \apj, 753, 134

\bibitem[{{Gonz{\'a}lez-Nuevo} {et~al.}(2006){Gonz{\'a}lez-Nuevo},
  {Arg{\"u}eso}, {L{\'o}pez-Caniego}, {Toffolatti}, {Sanz}, {Vielva}, \&
  {Herranz}}]{gnuevo06}
{Gonz{\'a}lez-Nuevo}, J., {Arg{\"u}eso}, F., {L{\'o}pez-Caniego}, M., {et~al.}
  2006, \mnras, 369, 1603

\bibitem[{{Gonz{\'a}lez-Nuevo} {et~al.}(2008){Gonz{\'a}lez-Nuevo}, {Massardi},
  {Arg{\"u}eso}, {Herranz}, {Toffolatti}, {Sanz}, {L{\'o}pez-Caniego}, \& {de
  Zotti}}]{gnuevo08}
{Gonz{\'a}lez-Nuevo}, J., {Massardi}, M., {Arg{\"u}eso}, F., {et~al.} 2008,
  \mnras, 384, 711

\bibitem[{{G{\'o}rski} {et~al.}(2005){G{\'o}rski}, {Hivon}, {Banday},
  {Wandelt}, {Hansen}, {Reinecke}, \& {Bartelmann}}]{gorski05}
{G{\'o}rski}, K.~M., {Hivon}, E., {Banday}, A.~J., {et~al.} 2005, \apj, 622,
  759

\bibitem[{{Gregory} {et~al.}(1996){Gregory}, {Scott}, {Douglas}, \&
  {Condon}}]{gregory96}
{Gregory}, P.~C., {Scott}, W.~K., {Douglas}, K., \& {Condon}, J.~J. 1996,
  \apjs, 103, 427

\bibitem[{{Griffin} {et~al.}(2010){Griffin}, {Abergel}, {Abreu}, {Ade},
  {Andr{\'e}}, {Augueres}, {Babbedge}, {Bae}, {Baillie}, {Baluteau}, {Barlow},
  {Bendo}, {Benielli}, {Bock}, {Bonhomme}, {Brisbin}, {Brockley-Blatt},
  {Caldwell}, {Cara}, {Castro-Rodriguez}, {Cerulli}, {Chanial}, {Chen},
  {Clark}, {Clements}, {Clerc}, {Coker}, {Communal}, {Conversi}, {Cox},
  {Crumb}, {Cunningham}, {Daly}, {Davis}, {de Antoni}, {Delderfield}, {Devin},
  {di Giorgio}, {Didschuns}, {Dohlen}, {Donati}, {Dowell}, {Dowell}, {Duband},
  {Dumaye}, {Emery}, {Ferlet}, {Ferrand}, {Fontignie}, {Fox}, {Franceschini},
  {Frerking}, {Fulton}, {Garcia}, {Gastaud}, {Gear}, {Glenn}, {Goizel},
  {Griffin}, {Grundy}, {Guest}, {Guillemet}, {Hargrave}, {Harwit}, {Hastings},
  {Hatziminaoglou}, {Herman}, {Hinde}, {Hristov}, {Huang}, {Imhof}, {Isaak},
  {Israelsson}, {Ivison}, {Jennings}, {Kiernan}, {King}, {Lange}, {Latter},
  {Laurent}, {Laurent}, {Leeks}, {Lellouch}, {Levenson}, {Li}, {Li},
  {Lilienthal}, {Lim}, {Liu}, {Lu}, {Madden}, {Mainetti}, {Marliani}, {McKay},
  {Mercier}, {Molinari}, {Morris}, {Moseley}, {Mulder}, {Mur}, {Naylor},
  {Nguyen}, {O'Halloran}, {Oliver}, {Olofsson}, {Olofsson}, {Orfei}, {Page},
  {Pain}, {Panuzzo}, {Papageorgiou}, {Parks}, {Parr-Burman}, {Pearce},
  {Pearson}, {P{\'e}rez-Fournon}, {Pinsard}, {Pisano}, {Podosek}, {Pohlen},
  {Polehampton}, {Pouliquen}, {Rigopoulou}, {Rizzo}, {Roseboom}, {Roussel},
  {Rowan-Robinson}, {Rownd}, {Saraceno}, {Sauvage}, {Savage}, {Savini},
  {Sawyer}, {Scharmberg}, {Schmitt}, {Schneider}, {Schulz}, {Schwartz},
  {Shafer}, {Shupe}, {Sibthorpe}, {Sidher}, {Smith}, {Smith}, {Smith},
  {Spencer}, {Stobie}, {Sudiwala}, {Sukhatme}, {Surace}, {Stevens}, {Swinyard},
  {Trichas}, {Tourette}, {Triou}, {Tseng}, {Tucker}, {Turner}, {Vaccari},
  {Valtchanov}, {Vigroux}, {Virique}, {Voellmer}, {Walker}, {Ward}, {Waskett},
  {Weilert}, {Wesson}, {White}, {Whitehouse}, {Wilson}, {Winter}, {Woodcraft},
  {Wright}, {Xu}, {Zavagno}, {Zemcov}, {Zhang}, \& {Zonca}}]{Griffin10}
{Griffin}, M.~J., {Abergel}, A., {Abreu}, A., {et~al.} 2010, \aap, 518, L3

\bibitem[{{Hacking} {et~al.}(1987){Hacking}, {Condon}, \& {Houck}}]{hacking87}
{Hacking}, P., {Condon}, J.~J., \& {Houck}, J.~R. 1987, \apjl, 316, L15

\bibitem[{{Healey} {et~al.}(2007){Healey}, {Romani}, {Taylor}, {Sadler},
  {Ricci}, {Murphy}, {Ulvestad}, \& {Winn}}]{healey07}
{Healey}, S.~E., {Romani}, R.~W., {Taylor}, G.~B., {et~al.} 2007, \apjs, 171,
  61

\bibitem[{{Helou} \& {Beichman}(1990)}]{helou90}
{Helou}, G. \& {Beichman}, C.~A. 1990, in Liege International Astrophysical
  Colloquia, Vol.~29, Liege International Astrophysical Colloquia, ed.
  B.~{Kaldeich}, 117--123

\bibitem[{{Herranz} {et~al.}(2013){Herranz}, {Gonz{\'a}lez-Nuevo}, {Clements},
  {De Zotti}, {Lopez-Caniego}, {Lapi}, {Rodighiero}, {Danese}, {Fu}, {Cooray},
  {Baes}, {Bendo}, {Bonavera}, {Carrera}, {Dole}, {Eales}, {Ivison}, {Jarvis},
  {Lagache}, {Massardi}, {Micha{\l}owski}, {Negrello}, {Rigby}, {Scott},
  {Valiante}, {Valtchanov}, {Van der Werf}, {Auld}, {Buttiglione}, {Dariush},
  {Dunne}, {Hopwood}, {Hoyos}, {Ibar}, \& {Maddox}}]{herranz13}
{Herranz}, D., {Gonz{\'a}lez-Nuevo}, J., {Clements}, D.~L., {et~al.} 2013,
  \aap, 549, A31

\bibitem[{{Herranz} {et~al.}(2009){Herranz}, {L{\'o}pez-Caniego}, {Sanz}, \&
  {Gonz{\'a}lez-Nuevo}}]{herranz09}
{Herranz}, D., {L{\'o}pez-Caniego}, M., {Sanz}, J.~L., \& {Gonz{\'a}lez-Nuevo},
  J. 2009, \mnras, 394, 510

\bibitem[{{Herranz} \& {Sanz}(2008)}]{herranz08}
{Herranz}, D. \& {Sanz}, J.~L. 2008, IEEE Journal of Selected Topics in Signal
  Processing, 2, 727

\bibitem[{{Kennicutt} {et~al.}(2011){Kennicutt}, {Calzetti}, {Aniano},
  {Appleton}, {Armus}, {Beir{\~a}o}, {Bolatto}, {Brandl}, {Crocker}, {Croxall},
  {Dale}, {Meyer}, {Draine}, {Engelbracht}, {Galametz}, {Gordon}, {Groves},
  {Hao}, {Helou}, {Hinz}, {Hunt}, {Johnson}, {Koda}, {Krause}, {Leroy}, {Li},
  {Meidt}, {Montiel}, {Murphy}, {Rahman}, {Rix}, {Roussel}, {Sandstrom},
  {Sauvage}, {Schinnerer}, {Skibba}, {Smith}, {Srinivasan}, {Vigroux},
  {Walter}, {Wilson}, {Wolfire}, \& {Zibetti}}]{kennicut11}
{Kennicutt}, R.~C., {Calzetti}, D., {Aniano}, G., {et~al.} 2011, \pasp, 123,
  1347

\bibitem[{{Kurinsky} {et~al.}(2013){Kurinsky}, {Sajina}, {Partridge}, {Myers},
  {Chen}, \& {L{\'o}pez-Caniego}}]{kurinsky13}
{Kurinsky}, N., {Sajina}, A., {Partridge}, B., {et~al.} 2013, \aap, 549, A133

\bibitem[{{Lanz} {et~al.}(2013){Lanz}, {Herranz}, {L{\'o}pez-Caniego},
  {Gonz{\'a}lez-Nuevo}, {de Zotti}, {Massardi}, \& {Sanz}}]{lanz13}
{Lanz}, L.~F., {Herranz}, D., {L{\'o}pez-Caniego}, M., {et~al.} 2013, \mnras,
  428, 3048

\bibitem[{{Leach} {et~al.}(2008){Leach}, {Cardoso}, {Baccigalupi}, {Barreiro},
  {Betoule}, {Bobin}, {Bonaldi}, {Delabrouille}, {de Zotti}, {Dickinson},
  {Eriksen}, {Gonz{\'a}lez-Nuevo}, {Hansen}, {Herranz}, {Le Jeune},
  {L{\'o}pez-Caniego}, {Mart{\'{\i}}nez-Gonz{\'a}lez}, {Massardi}, {Melin},
  {Miville-Desch{\^e}nes}, {Patanchon}, {Prunet}, {Ricciardi}, {Salerno},
  {Sanz}, {Starck}, {Stivoli}, {Stolyarov}, {Stompor}, \& {Vielva}}]{leach08}
{Leach}, S.~M., {Cardoso}, J.-F., {Baccigalupi}, C., {et~al.} 2008, \aap, 491,
  597

\bibitem[{{L{\'o}pez-Caniego} {et~al.}(2007){L{\'o}pez-Caniego},
  {Gonz{\'a}lez-Nuevo}, {Herranz}, {Massardi}, {Sanz}, {De Zotti},
  {Toffolatti}, \& {Arg{\"u}eso}}]{caniego07}
{L{\'o}pez-Caniego}, M., {Gonz{\'a}lez-Nuevo}, J., {Herranz}, D., {et~al.}
  2007, \apjs, 170, 108

\bibitem[{{L{\'o}pez-Caniego} {et~al.}(2006){L{\'o}pez-Caniego}, {Herranz},
  {Gonz{\'a}lez-Nuevo}, {Sanz}, {Barreiro}, {Vielva}, {Arg{\"u}eso}, \&
  {Toffolatti}}]{caniego06}
{L{\'o}pez-Caniego}, M., {Herranz}, D., {Gonz{\'a}lez-Nuevo}, J., {et~al.}
  2006, \mnras, 370, 2047

\bibitem[{{Marsden} {et~al.}(2013){Marsden}, {Gralla}, {Marriage}, {Switzer},
  {Partridge}, {Massardi}, {Morales}, {Addison}, {Bond}, {Crichton}, {Das},
  {Devlin}, {Dunner}, {Hajian}, {Hilton}, {Hincks}, {Hughes}, {Irwin},
  {Kosowsky}, {Menanteau}, {Moodley}, {Niemack}, {Page}, {Reese}, {Schmitt},
  {Sehgal}, {Sievers}, {Staggs}, {Swetz}, {Thornton}, \& {Wollack}}]{Mar13}
{Marsden}, D., {Gralla}, M., {Marriage}, T.~A., {et~al.} 2013, ArXiv:1306.2288

\bibitem[{{Massardi} {et~al.}(2011){Massardi}, {Bonaldi}, {Bonavera},
  {L{\'o}pez-Caniego}, {de Zotti}, \& {Ekers}}]{massardi11}
{Massardi}, M., {Bonaldi}, A., {Bonavera}, L., {et~al.} 2011, \mnras, 415, 1597

\bibitem[{{Massardi} {et~al.}(2009){Massardi}, {L{\'o}pez-Caniego},
  {Gonz{\'a}lez-Nuevo}, {Herranz}, {de Zotti}, \& {Sanz}}]{massardi09}
{Massardi}, M., {L{\'o}pez-Caniego}, M., {Gonz{\'a}lez-Nuevo}, J., {et~al.}
  2009, \mnras, 392, 733

\bibitem[{{Mitra} {et~al.}(2011){Mitra}, {Rocha}, {G{\'o}rski}, {Huffenberger},
  {Eriksen}, {Ashdown}, \& {Lawrence}}]{mitra2010}
{Mitra}, S., {Rocha}, G., {G{\'o}rski}, K.~M., {et~al.} 2011, \apjs, 193, 5

\bibitem[{{Miville-Desch{\^e}nes} \& {Lagache}(2005)}]{mdeschenes05}
{Miville-Desch{\^e}nes}, M.-A. \& {Lagache}, G. 2005, \apjs, 157, 302

\bibitem[{{Mocanu} {et~al.}(2013){Mocanu}, {Crawford}, {Vieira}, {Aird},
  {Aravena}, {Austermann}, {Benson}, {B{\'e}thermin}, {Bleem}, {Bothwell},
  {Carlstrom}, {Chang}, {Chapman}, {Cho}, {Crites}, {de Haan}, {Dobbs},
  {Everett}, {George}, {Halverson}, {Harrington}, {Hezaveh}, {Holder},
  {Holzapfel}, {Hoover}, {Hrubes}, {Keisler}, {Knox}, {Lee}, {Leitch},
  {Lueker}, {Luong-Van}, {Marrone}, {McMahon}, {Mehl}, {Meyer}, {Mohr},
  {Montroy}, {Natoli}, {Padin}, {Plagge}, {Pryke}, {Rest}, {Reichardt}, {Ruhl},
  {Sayre}, {Schaffer}, {Shirokoff}, {Spieler}, {Spilker}, {Stalder},
  {Staniszewski}, {Stark}, {Story}, {Switzer}, {Vanderlinde}, \&
  {Williamson}}]{Moc13}
{Mocanu}, L.~M., {Crawford}, T.~M., {Vieira}, J.~D., {et~al.} 2013,
  ArXiv:1306.3470

\bibitem[{{Murphy} {et~al.}(2010){Murphy}, {Sadler}, {Ekers}, {Massardi},
  {Hancock}, {Mahony}, {Ricci}, {Burke-Spolaor}, {Calabretta}, {Chhetri}, {de
  Zotti}, {Edwards}, {Ekers}, {Jackson}, {Kesteven}, {Lindley}, {Newton-McGee},
  {Phillips}, {Roberts}, {Sault}, {Staveley-Smith}, {Subrahmanyan}, {Walker},
  \& {Wilson}}]{murphy10}
{Murphy}, T., {Sadler}, E.~M., {Ekers}, R.~D., {et~al.} 2010, \mnras, 402, 2403

\bibitem[{{Negrello} {et~al.}(2013){Negrello}, {Clemens}, {Gonzalez-Nuevo}, {De
  Zotti}, {Bonavera}, {Cosco}, {Guarese}, {Boaretto}, {Serjeant}, {Toffolatti},
  {Lapi}, {Bethermin}, {Castex}, {Clements}, {Delabrouille}, {Dole},
  {Franceschini}, {Mandolesi}, {Marchetti}, {Partridge}, \&
  {Sajina}}]{negrello13}
{Negrello}, M., {Clemens}, M., {Gonzalez-Nuevo}, J., {et~al.} 2013, \mnras,
  429, 1309

\bibitem[{{Negrello} {et~al.}(2004){Negrello}, {Magliocchetti}, {Moscardini},
  {De Zotti}, {Granato}, \& {Silva}}]{negrello04}
{Negrello}, M., {Magliocchetti}, M., {Moscardini}, L., {et~al.} 2004, \mnras,
  352, 493

\bibitem[{{Perley} \& {Butler}(2013)}]{Per13}
{Perley}, R.~A. \& {Butler}, B.~J. 2013, \apjs, 204, 19

\bibitem[{{Planck Collaboration ES}(2013)}]{planck2013-p28}
{Planck Collaboration ES}. 2013, {The Explanatory Supplement to the Planck 2013
  results} ({ESA})

\bibitem[{{Planck Collaboration I}(2013)}]{planck2013-p01}
{Planck Collaboration I}. 2013, Submitted to \aap, [arXiv:astro-ph/1303.5062]

\bibitem[{{Planck Collaboration II}(2013)}]{planck2013-p02}
{Planck Collaboration II}. 2013, Submitted to \aap, [arXiv:astro-ph/1303.5063]

\bibitem[{{Planck Collaboration III}(2013)}]{planck2013-p02a}
{Planck Collaboration III}. 2013, Submitted to \aap, [arXiv:astro-ph/1303.5064]

\bibitem[{{Planck Collaboration Int. VII}(2013)}]{planck2012-VII}
{Planck Collaboration Int. VII}. 2013, \aap, 550, A133

\bibitem[{{Planck Collaboration Int. XIV}(2013)}]{planck2013-XIV}
{Planck Collaboration Int. XIV}. 2013, {Planck intermediate results. XIV. Dust
  emission at millimetre wavelengths in the Galactic plane} ({Submitted to
  \aap, [arXiv:astro-ph/1307.6815]})

\bibitem[{{Planck Collaboration IV}(2013)}]{planck2013-p02d}
{Planck Collaboration IV}. 2013, Submitted to \aap, [arXiv:astro-ph/1303.5065]

\bibitem[{{Planck Collaboration IX}(2013)}]{planck2013-p03d}
{Planck Collaboration IX}. 2013, Submitted to \aap, [arXiv:astro-ph/1303.5070]

\bibitem[{{Planck Collaboration V}(2013)}]{planck2013-p02b}
{Planck Collaboration V}. 2013, Submitted to \aap, [arXiv:astro-ph/1303.5066]

\bibitem[{{Planck Collaboration VI}(2013)}]{planck2013-p03}
{Planck Collaboration VI}. 2013, Submitted to \aap, [arXiv:astro-ph/1303.5067]

\bibitem[{{Planck Collaboration VII}(2011)}]{planck2011-1.10}
{Planck Collaboration VII}. 2011, \aap, 536, A7

\bibitem[{{Planck Collaboration VII}(2013)}]{planck2013-p03c}
{Planck Collaboration VII}. 2013, Submitted to \aap, [arXiv:astro-ph/1303.5068]

\bibitem[{{Planck Collaboration VIII}(2013)}]{planck2013-p03f}
{Planck Collaboration VIII}. 2013, Submitted to \aap,
  [arXiv:astro-ph/1303.5069]

\bibitem[{{Planck Collaboration X}(2013)}]{planck2013-p03e}
{Planck Collaboration X}. 2013, Submitted to \aap, [arXiv:astro-ph/1303.5071]

\bibitem[{{Planck Collaboration XI}(2013)}]{planck2013-p01a}
{Planck Collaboration XI}. 2013, In preparation

\bibitem[{{Planck Collaboration XII}(2013)}]{planck2013-p06}
{Planck Collaboration XII}. 2013, Submitted to \aap, [arXiv:astro-ph/1303.5072]

\bibitem[{{Planck Collaboration XIII}(2011)}]{planck2011-6.1}
{Planck Collaboration XIII}. 2011, \aap, 536, A13

\bibitem[{{Planck Collaboration XIII}(2013)}]{planck2013-p03a}
{Planck Collaboration XIII}. 2013, Submitted to \aap,
  [arXiv:astro-ph/1303.5073]

\bibitem[{{Planck Collaboration XIV}(2011)}]{planck2011-6.2}
{Planck Collaboration XIV}. 2011, \aap, 536, A14

\bibitem[{{Planck Collaboration XIV}(2013)}]{planck2013-pip88}
{Planck Collaboration XIV}. 2013, Submitted to \aap, [arXiv:astro-ph/1303.5074]

\bibitem[{{Planck Collaboration XIX}(2011)}]{planck2011-7.0}
{Planck Collaboration XIX}. 2011, \aap, 536, A19

\bibitem[{{Planck Collaboration XV}(2011)}]{planck2011-6.3a}
{Planck Collaboration XV}. 2011, \aap, 536, A15

\bibitem[{{Planck Collaboration XV}(2013)}]{planck2013-p08}
{Planck Collaboration XV}. 2013, Submitted to \aap, [arXiv:astro-ph/1303.5075]

\bibitem[{{Planck Collaboration XVI}(2011)}]{planck2011-6.4a}
{Planck Collaboration XVI}. 2011, \aap, 536, A16

\bibitem[{{Planck Collaboration XXIII}(2011)}]{planck2011-7.7b}
{Planck Collaboration XXIII}. 2011, \aap, 536, A23

\bibitem[{{Planck Collaboration XXIX}(2013)}]{planck2013-p05a}
{Planck Collaboration XXIX}. 2013, Submitted to \aap,
  [arXiv:astro-ph/1303.5089]

\bibitem[{{Rigby} {et~al.}(2011){Rigby}, {Maddox}, {Dunne}, {Negrello},
  {Smith}, {Gonz{\'a}lez-Nuevo}, {Herranz}, {L{\'o}pez-Caniego}, {Auld},
  {Buttiglione}, {Baes}, {Cava}, {Cooray}, {Clements}, {Dariush}, {de Zotti},
  {Dye}, {Eales}, {Frayer}, {Fritz}, {Hopwood}, {Ibar}, {Ivison}, {Jarvis},
  {Panuzzo}, {Pascale}, {Pohlen}, {Rodighiero}, {Serjeant}, {Temi}, \&
  {Thompson}}]{rigby11}
{Rigby}, E.~E., {Maddox}, S.~J., {Dunne}, L., {et~al.} 2011, \mnras, 415, 2336

\bibitem[{{Toffolatti} {et~al.}(1998){Toffolatti}, {Argueso Gomez}, {de Zotti},
  {Mazzei}, {Franceschini}, {Danese}, \& {Burigana}}]{toffolatti98}
{Toffolatti}, L., {Argueso Gomez}, F., {de Zotti}, G., {et~al.} 1998, \mnras,
  297, 117

\end{thebibliography}

\appendix

\section{Photometry}\label{sec:appendix}

This appendix describes in detail the photometry methods used in the PCCS.

\subsection{Aperture Photometry}
\label{sec:aperflux}

The aperture photometry is evaluated by centring a circular aperture on the position of the source. An annulus around this aperture is used to evaluate the background. In the absence of noise, the observed flux density of the source, $S_{\rm{obs}}$, may be written as:
\begin{equation}
S_{\mathrm{obs}}=\left( S_{\mathrm{ap}} - S_{\mathrm{an}}  \left(\frac{k_0^2}{k_2^2-k_1^2}\right)\right) \; ,
\end{equation}
where $k_0$ is the radius of the aperture, $k_1$ and $k_2$ are the inner and outer radii of the annulus, and, $S_{\mathrm{ap}}$ and $S_{\mathrm{an}}$ are the flux densities of the source in the aperture and annulus.  Both $S_{\mathrm{ap}}$ and $S_{\mathrm{ann}}$ may be written in terms of the true flux density of the source, $S_{\mathrm{true}}$. This gives the following relationship between the observed and true flux densities of the source:
\begin{equation}
\label{eqn:obstrue_eqn}
S_{\mathrm{obs}}=\left( \frac{\Omega_{k_0}}{\Omega} - \left(\frac{\Omega_{k_2}-\Omega_{k_1}}{\Omega} \right) \left(\frac{k_0^2}{k_2^2-k_1^2}\right)\right) S_{\mathrm{true}} \; ,
\end{equation}
where ${\Omega}$ is the solid angle of the beam,  and $\Omega_{k_0}$,  $\Omega_{k_1}$, and $\Omega_{k_2}$ are the beam solid angles out to the radii of $k_0$, $k_1$ and $k_2$. This provides the correction factor to be applied to the observed flux density, which accounts for both the flux density of the source missing from the aperture and that removed through background subtraction.

Assuming a circularly symmetric Gaussian beam and that $k_0$, $k_1$ and $k_2$ are given in units of the FWHM, equation~\ref{eqn:obstrue_eqn} may be written as:
\begin{equation}
S_{\mathrm{obs}}=\left(1 - \left(\frac{1}{2}\right)^{4k_0^2} -\left( \left(\frac{1}{2}\right)^{4k_1^2} - \left(\frac{1}{2}\right)^{4k_2^2}\right) \frac{k_0^2}{k_2^2-k_1^2}\right) S_{\mathrm{true}} \; .
\end{equation}
We used a radius of 1 FWHM for the aperture, $k_0 = 1$, and the annulus is located immediately outside of the aperture and has a width of 1 FWHM, $k_1 = 1$ and $k_2 = 2$.

The beams however are not exactly Gaussian so the effective FWHM is used to determine the radii of the aperture and annulus, and the correction factor is evaluated using:
\begin{equation}
S_{\mathrm{obs}}=\left(\frac{4\Omega_{\mathrm{FWHM1}}-\Omega_{\mathrm{FWHM2}}}{3\Omega }\right) S_{\mathrm{true}} \; ,
\end{equation}
where $\Omega_{\mathrm{FWHM1}}$ and $\Omega_{\mathrm{FWHM2}}$ are the beam solid angles within radii of 1 and 2 times the effective FWHM.

The noise level per pixel is estimated from the variance of the pixels that lie in the annulus, hence the uncertainties in the estimates of the background and the flux density within the aperture may be evaluated, allowing the uncertainty on $S_{\mathrm{obs}}$ to be calculated. The diffuse sky emission is a source of uncertainty in the photometry, thus it contributes a component to the ``noise'' that is correlated between pixels. Given that the exact degree of correlation is not known and is likely to vary with position on the sky, a correction factor to account for the correlated noise is evaluated by performing aperture photometry nearby, in regions without detected sources. Its value is such that it scales the residuals normalized by the uncertainties to a Gaussian of unit variance.

\subsection{PSF Photometry}
\label{sec:psfflux}

The PSF photometry is obtained by fitting a model of the PSF to the map at the position of the source. The PSF is obtained from the effective beam~\citep{planck2013-p02, planck2013-p03}. The model of the source is
\begin{equation}
\vec{m} = A \vec{P} + C \; ,
\end{equation}
where $\vec{P}$ is the PSF at the position of the source, $A$ is the amplitude of the source and $C$ is a the (constant) background. The best-fit values of the parameters $\beta = (A, C)$ are found by minimising the $\chi^2$ between the model and the data, $\vec{d}$,
\begin{equation}
\chi^2(\beta) = (\vec{d} - \vec{m}(\beta))^{\rm T} \tens{N}^{-1} (\vec{d} - \vec{m}(\beta)) \; ,
\label{eqn:phot_chisq}
\end{equation}
where $\tens{N}$ is the covariance matrix of the noise. The noise is assumed to be uncorrelated between pixels and proportional to the inverse of the number of hits in each pixel. The overall normalization of the noise is adjusted by setting $\chi^2 = 1$ at the best-fit value of the parameters. This has the effect of inflating the uncertainties to account for any mismatch between the modelled PSF and the true shape of the source in the map. The uncertainties on the parameters are computed from the curvature of the $\chi^2$. The best-fit amplitude and its uncertainty are converted to units of flux density using the area of the PSF.

\subsection{Gaussian Fit Photometry}
\label{sec:gauflux}

The Gaussian fit photometry is obtained by fitting a 2-dimensional Gaussian to the map at the position of the source. The model consists of a elliptical Gaussian centred at the position of the source plus a linear background,
\begin{equation}
m(\vec{x}) = A \exp\left[-\vec{x}^{\rm T}\tens{Q}^{-1}\vec{x}/2\right] + \vec{B}\cdot\vec{x} + C \; ,
\end{equation}
where $A$ is the amplitude of the source, \tens{Q} is the covariance matrix of the elliptical Gaussian profile, and $\vec{B}$ and $C$ are the background parameters. It is assumed that the source is at the origin of the coordinates $\vec{x}$. The components of $\tens{Q}$ can be expressed as a function of the semi-axes $a$ and $b$, and an orientation angle $\theta$ as
\begin{equation}
\tens{Q}^{-1}=\vec{R}^{\rm T} \vec{C}^{-1} \vec{R} \; ,
\end{equation}
where \vec{R} is the rotation matrix,
\begin{equation}
\vec{R}=\left[ \begin{array}{cc} \cos\theta & -\sin\theta \\ \sin\theta & \cos\theta \end{array} \right] \; ,
\end{equation}
and
\begin{equation}
\vec{C}^{-1}=\left[ \begin{array}{cc} 1/a^2 & 0 \\ 0 & 1/b^2 \end{array} \right] \; .
\end{equation}
There are seven parameters to fit
\begin{equation}
\beta = [A, B_1, B_2, C, a, b, \theta] \; .
\end{equation}
The model is fitted to the data by minimising the $\chi^2$~(\ref{eqn:phot_chisq}) between a pixelized version of the model $\vec{m}$ and the data $\vec{d}$. The uncertainties on the parameters are given by the diagonal elements of the covariance matrix of the fit. Assuming that the elliptical Gaussian model is a good approximation to the real source profile, the amplitude of the source and its uncertainty are converted to units of flux density using the area of the Gaussian.

\raggedright

\end{document}